\documentclass[useAMS,usenatbib]{mn2e}
          
\usepackage{graphicx}

\newif\ifAMStwofonts                        
\newcommand{\lsimeq}{{_<\atop^{\sim}}}
\newcommand{\gsimeq}{{_>\atop^{\sim}}}

\begin{document}

\title[Galaxy and AGN Evolution in the IR]{Modelling Galaxy and AGN Evolution in the IR: Black Hole Accretion versus Star-Formation Activity}
\author[C. Gruppioni, F. Pozzi, G. Zamorani and C. Vignali]{C. Gruppioni$^{(1)}$\thanks{E-mail: carlotta.gruppioni@oabo.inaf.it}, F. Pozzi$^{(2)}$, G. Zamorani$^{(1)}$ and C. Vignali$^{(2)}$\\
$^{(1)}$INAF: Osservatorio Astronomico di Bologna, via Ranzani 1, I--40127 Bologna, Italy.\\
$^{(2)}$Dipartimento di Astronomia, Universit\`a di Bologna, via Ranzani 1, I--40127 Bologna, Italy}

   \date{Accepted 2011 May 3.  Received 2011 May 3; in original form 2010 November 12}

\pagerange{\pageref{firstpage}--\pageref{lastpage}} \pubyear{2011}

\maketitle

\label{firstpage}

\begin{abstract}
We present a new backward evolution model for galaxies and AGNs in the infrared (IR). What is new in this model is the separate study of the evolutionary properties of the different IR populations (i.e. spiral galaxies, starburst galaxies, low-luminosity AGNs, ``unobscured'' type 1 AGNs and ``obscured'' type 2 AGNs) defined through a detailed analysis of the spectral energy distributions (SEDs) of large samples of IR selected sources. The evolutionary parameters have been constrained by means of all the available observables from surveys in the mid- and far-IR (source counts, redshift and luminosity distributions, luminosity functions). By decomposing the SEDs 
representative of the three AGN classes into three distinct components (a stellar component emitting most of its power
in the optical/near-IR, an AGN component due to hot dust heated by the central
black hole peaking in the mid-IR, and a starburst component dominating the far-IR spectrum) we have disentangled the AGN contribution to the monochromatic and total IR luminosity emitted by the different populations
considered in our model from that due to star-formation activity. We have then obtained an estimate of the total IR luminosity density (and star-formation density -- SFD --  produced by IR galaxies) and the first ever estimate of the black hole mass accretion
density (BHAR) from the IR. The derived evolution of the BHAR is in agreement with estimates from X-rays, though the BHAR values we derive from IR are slightly higher than the X-ray ones. Finally, we have simulated source counts, redshift distributions and SFD and BHAR that we expect to obtain with the future cosmological Surveys in the mid-/far-IR that will be performed with {\em JWST-MIRI} and {\em SPICA-SAFARI}.
\end{abstract}


\section{Introduction}
\label{authorf_sec:intro}
In the past years strong observational evidence of high rates of evolution for infrared (IR) galaxies has been obtained by means of
two independent findings: the detection of a large amount of energy contained in the Cosmic Infrared Background 
(CIRB, Hauser \& Dwek 2001), and the number counts from several deep cosmological surveys (from 15 $\mu$m to 850 $\mu$m)
largely exceeding the no-evolution expectations. Both results agree in requiring a strong increase in the IR energy density between 
the present time and $z$$\sim$1--2.
The discovery of the CIRB, which is interpreted as the integrated emission from dust present 
in galaxies, has offered new perspectives on our understanding of galaxy formation and evolution, since it provides a key constraint 
on the history of star-formation (SF) and accretion in the Universe.
The resolution of the CIRB into individual sources has been one of the 
main goals for the {\em Infrared Space Observatory} ({\em ISO}; Kessler et al. 1996) and the {\em Spitzer Space Telescope} (Werner et al. 2004). 
The deep cosmological surveys carried out by {\em ISO} and {\em Spitzer} 
allowed new insights into the IR population contributing to the CIRB, showing source counts in excess with respect to
no-evolution predictions and revealing the existence of a population of distant, dusty IR-bright galaxies that are missed by optical surveys. 
The major contributors to the CIRB are Luminous (L$_{8-1000\mu m} > 10^{11}$ L$_{\odot}$, LIRGs) and Ultra-Luminous IR Galaxies 
(L$_{8-1000\mu m} > 10^{12}$ L$_{\odot}$, ULIRGs): they have been discovered by the {\em Infrared Astronomical Satellite} ({\em IRAS}; Neugebauer 1984) 
as rare objects in the local Universe radiating most of their energy in the IR, but have been successively found by {\em ISO}, {\em Spitzer} and also 
{\em SCUBA} on the JCMT in the sub-millimeter (sub-mm), to become more and more important as the redshift increases.  In particular, the so-called 
``sub-mm galaxies'' (SMGs) detected by {\em SCUBA} (i.e. Smail, Ivison \& Blain 1997) and subsequently confirmed by {\em Spitzer} (i.e. Pope et al. 2006), 
which emit a significant fraction 
of their rest-frame bolometric luminosity in the far-infrared (FIR)/sub-mm and seem to be mostly at $z\sim2-3$, represent a direct evidence for that.    
The existence of this strongly evolving population of dusty, massive galaxies that form the bulk of their stars at high redshifts  (SMGs and high-z (U)LIRGs are found 
to be already massive galaxies at $z\sim2$; i.e. Dye et al. 2008), is an example of anti-hierarchical behaviour, showing how crucial this
IR-bright, dust-obscured galaxy population is to understand galaxy formation and evolution.

The new results provided by the {\em Balloon-borne Large Aperture Sub-millimeter Telescope} ({\em BLAST}; Pascale et al. 2008) and the {\em Herschel Space Observatory} (Pilbratt et al. 2010) in the FIR/sub-mm domain (e.g., Patanchon et al. 2009; Bethermin et al. 2010; Berta et al. 2010; Oliver et al. 2010; Gruppioni et al. 2010),
together with the availability of the new space facilities in the coming years, such as 
{\em James Webb Space Telescope} ({\em JWST}; Gardner et al. 2006) and furtherly the {\em SPace Infrared telescope for Cosmology and Astrophysics}
({\em SPICA}; Nakagawa et al. 2009),
open a new perspective to study in detail the population of IR galaxies beyond $z=1-2$, requiring new models 
to explain the high rate of evolution observed at lower redshifts. 
Two main weaknesses exist in most of the current models for IR sources: 1) the failure in reproducing the observed redshift distributions; 2) the
severe underestimate of the contribution from Active Galactic Nuclei (AGNs).
The new models should take into account that IR galaxies can host both SF and AGN 
activity (i.e. Lutz et al. 1998) and that in the far-infrared (FIR) even the emission from Seyfert and Quasars can be dominated by SF
(Lutz et al. 2004; Schweitzer et al. 2006; Shao et al. 2010). 
We can no longer neglect the AGN contribution in modelling and interpreting IR data, but we
should rather understand and quantify the AGN presence within IR galaxies and its connection with SF activity.
In particular, a key cosmological question that needs to be
answered by models and future observations regards the role of AGN in galaxy formation and evolution. In fact,
the seminal discovery that all massive galaxies in the local Universe harbour super-massive black holes 
(SMBH; $M_{BH}>10^6 M_{\odot}$) implies that all massive galaxies have hosted AGN at some time 
during their life (i.e. Magorrian et al. 1998). Recent works in the X-rays, optical and mid-infrared (MIR) have shown that many  heavily obscured
AGNs escape detection even in the deepest optical and X-ray observations, but, as expected from their high level of obscuration,
reveal themselves in the IR (i.e. Daddi et al. 2007; Fiore et al. 2008).
When planning future IR surveys with new facilities and/or when trying to interpret new data results, it is therefore necessary 
to consider models that properly take into account the presence of AGNs within a significant fraction of the IR population, 
in order to be able to answer the still open questions about galaxy formation and evolution, like i.e. What drives the evolution of the massive, dusty 
distant galaxy population? What feedback/interplay exists between the AGN and star-forming phase and how 
does it relate to cosmic downsizing?

We have developed a new backward evolution model fitting all the main constraints provided by the IR/sub-mm surveys (from 15 $\mu$m to 500 $\mu$m), where
we have taken particular care in properly identifying and modelling the AGN contribution at different luminosities and redshifts. Our model starts from the
classical approach consisting of evolving a local luminosity function (LLF) in luminosity and/or density with redshift. In particular, we have decomposed the LLF into different 
populations of galaxies and AGNs, each of them evolving independently. What is really new in this model is the way we have distinguished between the different IR populations: 
the separation into classes is based on a detailed Spectral Energy Distribution (SED) study performed 
on a large spectroscopic sample of MIR selected sources (Gruppioni et al. 2008). The shape of the SED provides a more meaningful physical characterisation of  different populations than e.g. their luminosity, which is commonly used to separate IR sources into different classes (i.e. LIGs, ULIGs, HyLIGs).
We have constrained our model by means of the MIR (i.e. from {\em ISO} and {\em Spitzer}) and FIR (i.e. from {\em Herschel}) data. In particular, as constraints in the FIR we have considered the very recent results of the PACS Evolutionary Probe (PEP) Survey (Berta et al. 2010; Gruppioni et al. 2010), of the {\em Herschel}-ATLAS (H-ATLAS) Survey (Eales et al. 2010; Clements et al. 2010) and of the {\em Herschel} Multi-tiered Extra-galactic (HerMES) Survey (Oliver et al. 2010; Vaccari et al. 2010). The model evolutions and luminosity functions have then been used to derive the evolution with redshift of the star-formation and accretion densities from IR luminosity. To this purpose, we have analysed the relative contributions of AGN and star-formation activity to the IR luminosity of the populations considered in the model by means of a SED decomposition obtained with the Fritz et al. (2006) code. 

The paper is structured as follows. In Section \ref{authorf_sec:model} we provide a detailed description of our model, including the population separation, the considered MIR and FIR data constraints,
the LLF parameters and the evolutionary paths. Section \ref{authorf_sec:results} reports on the results of our model in the MIR and FIR/sub-mm bands, while in Section~\ref{authorf_sec:agn}
we discuss the contribution of AGNs to the total IR luminosity and its evolution, deriving also
an estimate of the Black Hole accretion rate density (BHAR) and star-formation density (SFD) evolution with redshift. In Section~\ref{authorf_sec:pred} the model predictions for Surveys with the {\em Mid-InfraRed Instrument} ({\em MIRI}; Wright et al. 2004) on board of {\em JWST} and with the {\em SPICA FAR-Infrared Instrument} ({\em SAFARI}; Swinyard et al. 2009) are discussed, while 
our conclusions are presented in Section~\ref{authorf_sec:concl}.

\section{The Model}
\label{authorf_sec:model}
The model starts from the classical approach of evolving a local luminosity function (LLF) with redshift, but considers five different IR populations evolving independently. As starting LLF we consider a parametrisation of the 12-$\mu$m one derived by Fang et al. (1998), consistently with the Rush, Malkan \& Spinoglio (RMS, 1993) LLF, decomposed into single population LLFs as to reproduce the total observed LLF. 
As parametrisation for the LLF ($\Phi(L)$, defined as the differential LF per decade in luminosity), we adopt the form first derived by Saunders et al. (1990) to describe the 60-$\mu$m LF of IRAS galaxies, found to be too broad to be described by a standard Schechter function. Saunders et al. (1990) found a better representation for the IR LF with $\Phi(L)$ given by the function:
\begin{equation}
\Phi(L)=\Phi^{\star}\left(\frac{L}{L^{\star}}\right)^{1-\alpha} exp\left[-\frac{1}{2\sigma^2}log_{10}^2\left(1+\frac{L}{L^{\star}}\right)\right]
\end{equation}
which behaves as a power law for L$<<$L$^{\star}$ and as a Gaussian in $logL$ for L$>>$L$^{\star}$.

The relative fractions of sources within the different classes are defined on the basis of a detailed broad-band SED-fitting analysis performed on a large sample of MIR (15-$\mu$m) selected 
galaxies, all with spectroscopic redshift and classification (Gruppioni et al. 2008). The MIR sample on which we have originally based the source classification consists of 203 sources from the 
ELAIS-S1 survey at 15 $\mu$m (Lari et al. 2001; Gruppioni et al. 2002), for which a detailed optical spectroscopical analysis has been performed and presented by La Franca et al. (2004).  
Based on the SED-fitting technique Gruppioni et al. (2008) have classified the MIR sources, identifying AGN signatures in about 50\% of them. Similar AGN fractions in IR surveys have been recently found either in local samples (i.e. Smith et al. 2008; Goulding and Alexander 2009) or at cosmological distances through different identification techniques (i.e. Brand et al. 2006; Treister et al. 2006).
The parametrisation of the single LLFs, as well as the evolution parameters for each population have been obtained
through a minimisation algorithm aimed at finding the best-fitting parameters to simultaneously reproduce the redshift and luminosity distributions observed for the five populations of the Gruppioni et al. (2008) sample (with the same optical and IR completeness corrections and redshift limits of the sample data applied to the model), the 15-$\mu$m to 500-$\mu$m source counts (as described in the following Sections). 

\begin{figure}
  \begin{center}
    \includegraphics[width=8 cm]{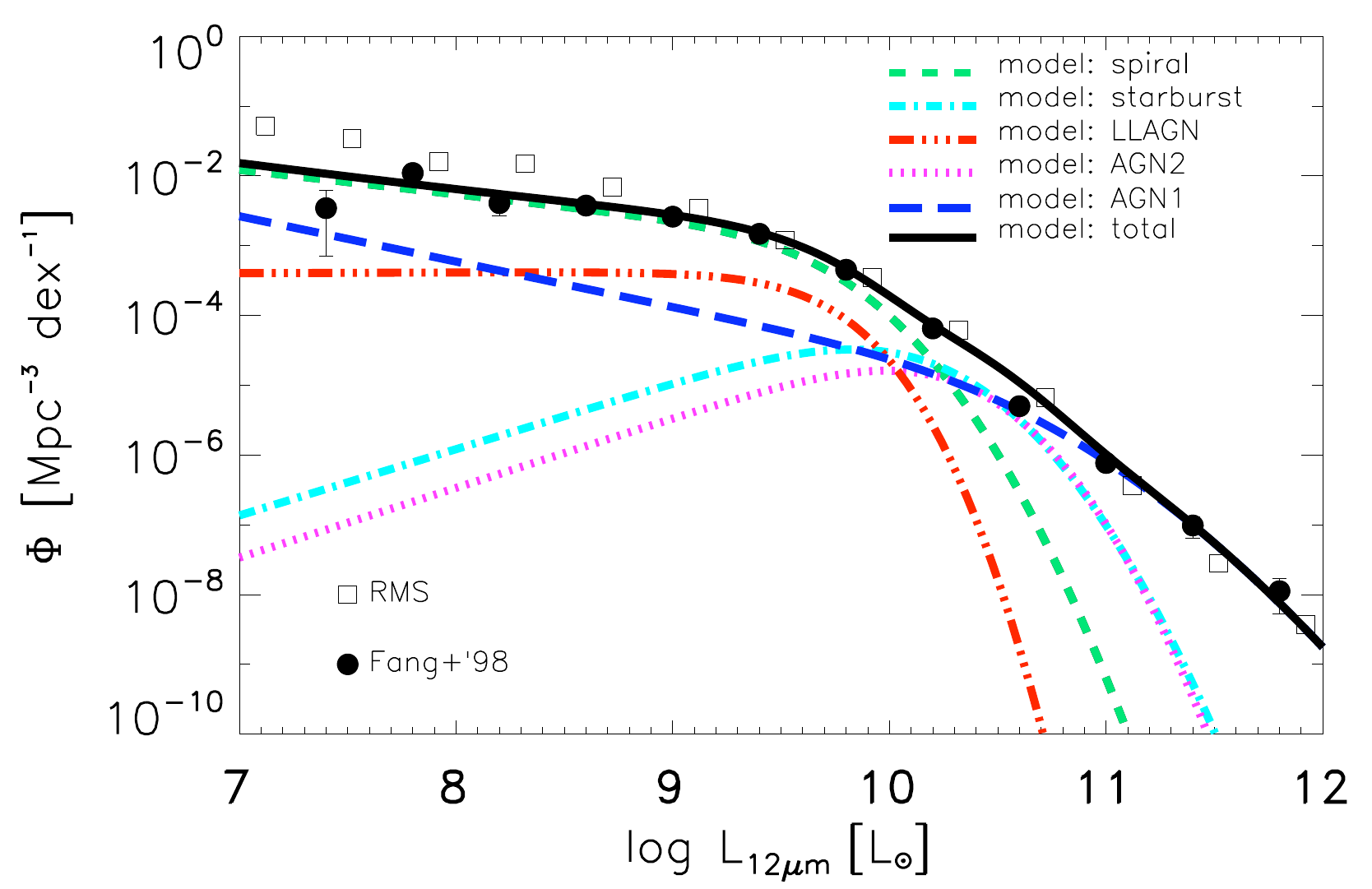}
      \end{center}
  \caption{Model LLF: {\tt spiral} galaxy (green dashed), {\tt starburst} galaxy (cyan dot-dashed), {\tt LLAGN} (red dot-dot-dot dashed), {\tt AGN2} (magenta dotted), {\tt AGN1} (blue long-dashed), total (black solid). The different populations are described in Section~\ref{authorf_sec:pop}. 
The observed Fang et al. (1998) and RMS data points are overplotted as black filled circles and open squares respectively.} 
\label{authorf_fig:llf}
\end{figure}

\begin{figure}
  \begin{center}
    \includegraphics[width=8 cm]{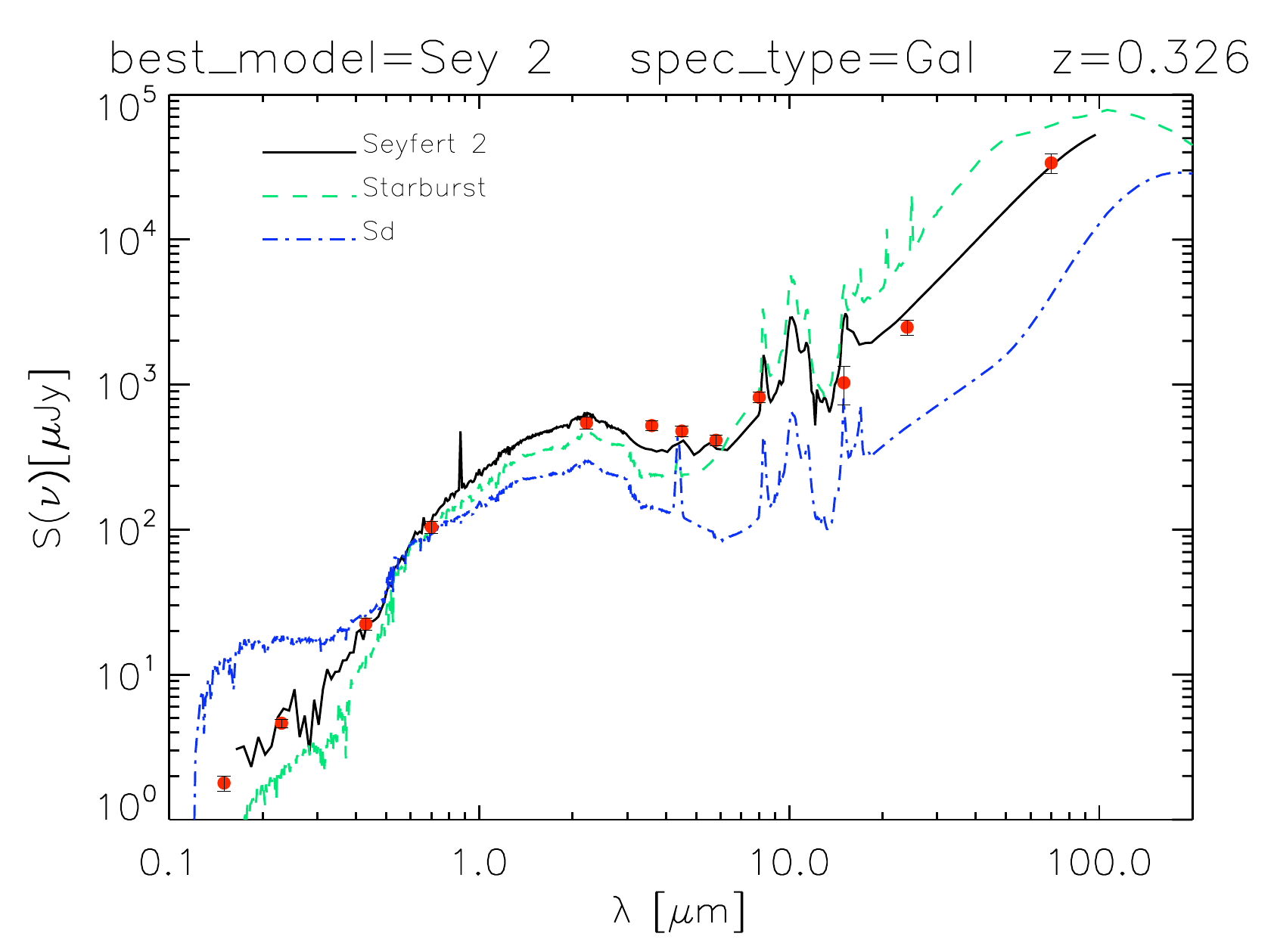}
      \end{center}
  \caption{Example of an observed SED (red filled circles) of a MIR source spectroscopically classified as {\tt galaxy}, compared to three different template SEDs (from Polletta et al. 2007): Seyfert 2 (black solid line, best-fit), Starburst Galaxy (green dashed) and Sd Spiral Galaxy (blue dot-dashed), normalised to the optical band.} 
\label{authorf_fig:Sey2}
\end{figure}
In Figure \ref{authorf_fig:llf} we show the LLF decomposition into the five different populations (whose parameters are reported in Table~\ref{authorf_tab:llf}), overplotted to the Fang et al. (1998) and RMS 12-$\mu$m observed data points.

\begin{table}
 \caption{15-$\mu$m LLF Parameters}
\begin{tabular}{|l|c|c|r|r|}
\hline \hline
 SED class &  log$_{10}$$(L^{\star}(0))$ & $\Phi_0^{\star}$  & $\alpha$  &  $\sigma$ \\    \hline \hline 
{\tt spiral} & 9.45 &  1.7$\times$10$^{-3}$ &  1.35 &  0.30 \\
{\tt starburst} & 9.70 & 5.0$\times$10$^{-5}$ & 0.05 & 0.31 \\
{\tt LLAGN}  & 9.76 &  4.3$\times$10$^{-4}$ &  0.99 & 0.18 \\
{\tt AGN2} & 9.95 & 3.0$\times$10$^{-5}$ & 0.001  &  0.27 \\
{\tt AGN1}  & 9.80 &  4.0$\times$10$^{-5}$ & 1.65 &  0.60 \\
\hline \hline
\end{tabular}
\label{authorf_tab:llf}
\end{table}

\subsection{The Infrared Populations}
\label{authorf_sec:pop}
As mentioned in the previous section, the classification of the IR sources into different populations in our model is based on the results of a broad-band SED-fitting analysis for one of the largest available spectroscopic samples of MIR-selected galaxies and AGNs at intermediate redshifts ($z<1.5$) performed by Gruppioni et al. (2008). 
The sample, consisting of 72\% of the 15-$\mu$m
ELAIS-SWIRE Survey in S1 (Lari et al. 2001; Gruppioni et al. 2002), contains 203 extragalactic sources, all with measured spectroscopic redshift (La Franca et al. 2004). The sample of 203 15-$\mu$m sources is composed by a deeper sample in the central S1$\_$5 area, which is 99\% spectroscopically complete to $R=21.6$, plus a shallower sample in the rest of the field, which is 97\% spectroscopically complete to $R=20.5$. The same optical/15-$\mu$m limits of the data have been applied to the model when comparing the model to the data in ELAIS-S1 (see the single populations source counts, redshift and luminosity distributions plotted in Figures 5 and 6).  
Most of these sources have full multi-wavelength coverage from the far-UV ({\em GALEX}; Martin et al. 2005) to the FIR ({\em Spitzer}) and lie in the redshift range $0.1 < z < 1.3$. This large sample allowed us for the first time to characterise the 
spectral properties of sources responsible for the strong evolution observed in the MIR and to construct an observational library of templates for IR galaxies and AGNs at intermediate $z$. The
observed SEDs have been interpreted by performing a fit with several local template SEDs representative of different classes of IR galaxies and AGNs (Polletta et al. 2007), comparing the resulting
SED classification with the spectroscopic one. The considered templates included three elliptical galaxies of different ages, one lenticular, seven spirals, three starbursts, three type 1 QSOs, one type 2 QSO, Seyfert 1, 1.8 and 2 and two composite ULIRGs, containing both starburst and AGN component, in the wavelength range between 0.1 and 1000 $\mu$m. Of the original sample of 203 sources 41\% is well reproduced by galaxy templates (S0 to Sdm, no ellipticals), 6.5\% by starburst templates, 35\% by Seyfert~2/Seyfert~1.8 templates, 5.5\% by templates typical of ``obscured'' AGNs, characterised by large column densities of obscuring material (i.e. Markarian 231 or IRAS19254) and 12\% by ``unobscured'' AGN templates. 
Note that the two SEDs reproducing those of the the ``obscured'' AGN population are  empirical templates created to
fit the SEDs of the heavily obscured broad absorption-line (BAL) QSO Markarian~231 (Berta 2005) and 
the Seyfert 2 galaxy IRAS~19254$-$7245 South (Berta et al. 2003). These two SEDs are similar in shape, containing a
powerful starburst component, mainly responsible for their FIR emission,
and an AGN component that contributes to the MIR (Farrah et al. 2003), and reproduce the SEDs of ``obscured'' AGNs regardless of their optical spectra (i.e. broad or narrow lines in the optical; Gruppioni et al. 2008).
The fraction of objects showing different levels of AGN activity in their SEDs is significantly higher ($\sim$53\%) than that derived from the optical spectroscopy ($\sim$29\%; La Franca et al. 2004), particularly because of the identification of AGN activity in the SEDs of objects spectroscopically classified as galaxies. This might be partially due to the fact that the spectroscopic classification can be somewhat unreliable because of dilution of AGN signatures due to the host galaxy light in the optical band. It is likely that in most of these objects the AGN is either obscured or of low-luminosity, and thus it does not dominate the energetic output at any wavelength, except in the MIR, showing up just in the range where the host galaxy SED has a minimum. 
Similar results have been obtained from the same detailed broad-band SED-fitting analysis performed on a sample of 100- and 160-$\mu$m {\em Herschel} sources in the PEP GOODS-N field (Gruppioni et al. 2010). In the fields where we have performed detailed SED-fitting analyses, we have done further investigations on the available X-ray images to check whether a correlation between our SED classification and the X-ray detection and luminosity exists or not. Indeed, we find that the X-ray detected sources are mainly objects classified as AGNs on the SED-fitting basis, with their luminosities being higher than those of the few detected galaxy-SED ones. In particular, from a match between the positions of IR selected sources classified by Gruppioni et al. (2008; 2010), with those of the X-ray sources in ELAIS-S1 (XMM: Puccetti et al. 2005; Chandra: Vignali et al. in preparation) and in GOODS-N (Chandra: Alexander et al. 2005) respectively, we found that, indeed, the fraction of matches for the Seyfert2/1.8 SED-classified objects is significantly higher than those found for objects classified as starburst and spiral galaxies on the basis of their SEDs. In particular, in ELAIS-S1 we found 41\% of XMM+Chandra detections for the Seyfert2/Seyfert1.8 sources against 15\% for the starburst+spiral galaxies, while in the GOODS-N 
we found 25\% of Chandra detections for the Seyfert 2's (with $<$L(2-8keV)$>$$\sim$10$^{43}$ erg s$^{-1}$) against 12\% and 4\% for the starburst ($<$L(2-8keV)$>$$\sim$10$^{42.5}$ erg s$^{-1}$) and spiral galaxies ($<$L(2-8keV)$>$$\sim$10$^{42}$ erg s$^{-1}$) respectively. 
The Seyfert 2/Seyfert 1.8-SED class is likely to be composed by a mixture of optical Seyfert 2 and LINERs: we believe most of them being low luminosity AGNs, not necessarily obscured ones.\\
In figure \ref{authorf_fig:Sey2} we show an example of the IR sources whose observed SED is best-fitted by a Seyfert 2/Seyfert 1.8 template SED, but whose optical spectrum is typical of a spiral/star-forming galaxy. Note that, as discussed in detail in Section~\ref{authorf_sec:agn}, the Seyfert 2/Seyfert 1.8 SEDs are dominated by star formation in all the MIR/FIR bands, except in the range 3--10 $\mu$m, where the
warm dust heated by the AGN shows up. Therefore, these objects can be considered as star-forming galaxies, but in our analysis we prefer to keep them as a separate class as a reminder that they are likely to contain an AGN that might be important in other bands (i.e. X-rays).
\begin{figure*}
  \begin{center}
    \includegraphics[width=15cm]{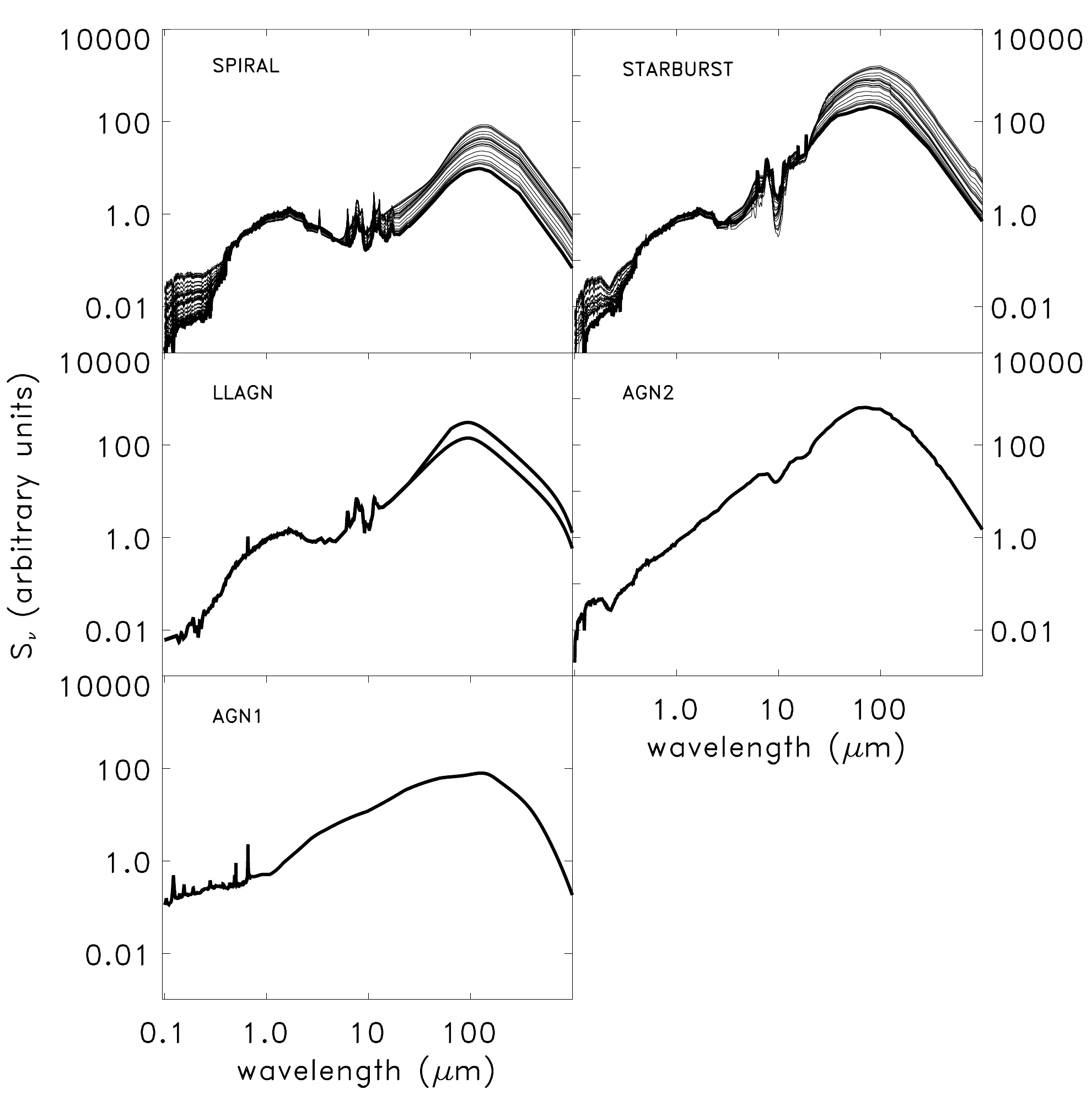}
      \end{center}
  \caption{Template SEDs considered for the five IR populations. {\em Top left:} {\tt spiral} SEDs, evolving with L$_{15\mu m}$ from an Sa spiral to an Sdm one, with fine interpolations in between. {\em Top right:} {\tt starburst} SEDs evolving from the moderate starburst M82 to the extreme starburst Arp220. {\em Middle left:} Seyfert 2 SEDs (original Polletta's and modified -- e.g. with a higher FIR bump -- as to reproduce a fraction of {\em Herschel} sources showing an excess in the FIR with respect to the original template), assumed as representative of the {\tt LLAGN} population. {\em Middle right:} Markarian 231 SED, assumed as template for the {\tt AGN2} population. {\em Bottom right:} TQSO1 SED, assumed for the {\tt AGN1} population.} 
\label{authorf_fig:templ}
\end{figure*}

In the model we have grouped the sources into five ``broad'' SED classes, according to their SED-fitting classification. The five populations considered in the model are: 
normal spiral galaxies ({\tt spiral}: Sa, Sb, Sc, Sd, Sdm SEDs), galaxies powered by star-formation ({\tt starburst}: M82, NGC6090, IRAS22491, IRAS20551, Arp220 SEDs), objects containing a low luminosity AGN ({\tt LLAGN}: Seyfert 2, Seyfert 1.8 SEDs), obscured AGNs ({\tt AGN2}: Markarian 231, IRAS19254 SEDs), and unobscured AGNs ({\tt AGN1}: QSO1, TQSO1, BQSO1 SEDs). 

\noindent For the {\tt spiral} population we have assumed a SED evolving with luminosity from an Sa spiral (L$_{15\mu m}$$<$$10^9$ L$_\odot$) to an Sdm
spiral (L$_{15\mu m}\gsimeq 10^{10}$ L$_\odot$), while for the {\tt starburst} population the SEDs vary from a moderate starburst galaxy (NGC6090, L$_{15\mu m}$$\simeq$10$^{10}$ L$_\odot$), up to an extreme one 
(Arp220, L$_{15\mu m}$$>$10$^{11.9}$L$_\odot$), with fine interpolations between these known templates. These choices follow the conclusions of Gruppioni et al. (2008) that the {\tt galaxy} and {\tt starburst}
SEDs evolve with $z$ (and/or L), from early- to late-type and from moderate to extreme starbursts respectively. \\
For the AGN-dominated populations ({\tt AGN1} and {\tt AGN2}), a single SED has been considered in the model as representative of the whole class: in particular, we have assumed the TQSO1 and Markarian 231 templates (see Polletta et a. 2007 for a description) respectively. 
The choice of a single template SED as representative of each AGN-dominated population in the model is motivated by the fact that the bulk of sources ($>$90\%) in each class is very well fitted by the selected templates, with apparently no clear trend with $z$ or $L$, as instead clearly observed for spiral and starburst galaxies.
For the same reason, for the {\tt LLAGN} population we have assumed just two templates: the original Polletta et al. (2007) Seyfert 2 one, reproducing most of the ELAIS-S1 {\tt LLAGN}, and a modified Seyfert 2 template, with a higher FIR bump (at $\lambda$$>$24 $\mu$m), needed to fit $\sim$40\% of the {\tt LLAGN} SEDs observed by {\em Herschel} in the GOODS-N field, showing an excess in the FIR (see Section~\ref{authorf_sec:results} and Gruppioni et al. 2010 for details). The two {\tt LLAGN} templates are used in the model in the same proportions as observed in the PEP GOODS-N data sample.\\
As we will show in Section~\ref{authorf_sec:agn}, all the template SEDs containing an AGN are "composite" SEDs, where AGN and starburst coexist, with the two components having different relative levels of importance in different wavelength ranges and for different SED classes. However, in the current work we have analysed the evolution of the different SED-classes as a whole, assuming that the AGN and starburst components influence each other and co-evolve within the same object. It is very hard to say how the evolution of the two components can differ: this would require an extremely accurate analysis and exquisite quality data to decompose the observed SED of each source and study the evolution of the AGN and starburst component separately. While such an analysis would be matter for a future work (Pozzi et al., in preparation), here we study the evolution of different SED populations, as resulted from a detailed SED-fitting anaysis.\\  
In Figure \ref{authorf_fig:templ} we show the template SEDs used for the five IR populations considered in our model. 

\subsection{Observational Constraints from {\em ISO}, {\em Spitzer} and {\em Herschel} Surveys}
\label{authorf_sec:constr}
The five populations considered in the model are characterised by different evolutionary properties. The evolution parameters are defined by minimising the differences between data and model expectations, simultaneously for the single population luminosity and redshift distributions and for the differential source counts (total and for each population) considering all the data from IR surveys available in the literature (from 15 $\mu$m to 500 $\mu$m). All the source counts from different surveys at the same wavelength have been combined by binning in flux density and averaging each data point weighted by its formal error (inverse of the squared error). The combined source counts, shown as grey shaded areas in Figures 5, 7, 8 and 9, have then been used to constrain our model.\\
The MIR data considered for constraining the global source counts are from several {\em ISO} and {\em Spitzer} surveys: at 15 $\mu$m from ELAIS-S1 (Gruppioni et al. 2002), HDF-N, HDF-S, Marano Firback, Marano Deep, and Marano Ultradeep (Elbaz et al. 1999), ultradeep lensed (Metcalfe et al. 2003), Lockman Deep and Shallow (Rodighiero et al. 2004); at 24 $\mu$m from GOODS (Papovich et al. 2004) and SWIRE (Shupe et al. 2008). The 15-$\mu$m source counts have also been constrained separately for the different populations using the ELAIS-S1 data. The redshift and luminosity distributions have been constrained by using the ELAIS-S1 spectroscopic redshifts (La Franca et al. 2004) at 15-$\mu$m and the GOODS-S and -N (Rodighiero et al. 2010) and COSMOS (Le Floc'h et al. 2009) spectroscopic and photometric redshifts at different flux density levels at 24 $\mu$m. Using the La Franca et al. (2004) spectroscopic sample and the SED-based classification of Gruppioni et al. (2008) we have been able to separately constrain the redshift and luminosity distributions of each population at 15 $\mu$m, while at 24 $\mu$m the comparison has been performed on the total distributions only, since the 24 $\mu$m samples available had no classified objects through SED-fitting as we would need for a direct match by populations. \\
In the FIR we have considered as constraints to source counts the recent results from {\em Herschel}, as well as previous results from {\em ISO} and {\em Spitzer}: at 70 $\mu$m from {\em Spitzer} (CDF-S and BOOTES: Dole et al. 2004; GOODS: Frayer et al. 2006a; FLS: Frayer et al. 2006b; COSMOS: Frayer et al. 2009; GOODS/FIDEL, COSMOS and SWIRE: Bethermin et al. 2010a) and from the shorter wavelength photometer (PACS; Poglitsch et al. 2010) of {\em Herschel} (PEP GOODS-S: Berta et al. in preparation);
at 100 $\mu$m from {\em ISO} (Lockman Hole: Rodighiero \& Franceschini 2004; ELAIS: Heraudeau et al. 2004) and from {\em Herschel-PACS} (PEP GOODS-N, GOODS-S, Lockman Hole, COSMOS: Berta et al. 2010); at 160 $\mu$m from {\em ISO} (FIRBACK: Dole et al. 2001), {\em Spitzer} (CDF-S and BOOTES: Dole et al. 2004; FLS: Frayer et al. 2006b; COSMOS: Frayer et al. 2009; GOODS/FIDEL, COSMOS and SWIRE: Bethermin et al. 2010a) and {\em Herschel-PACS} (PEP GOODS-N, GOODS-S, Lockman Hole, COSMOS: Berta et al. 2010). The global redshift distributions of the PEP sources at 100 $\mu$m and 160 $\mu$m have also been considered as constrain for our model.
In the sub-mm we have considered the source counts from the recent Surveys with the longer wavelength instrument (SPIRE; Griffin et al. 2010) on {\em Herschel}: at 250 $\mu$m, 350 $\mu$m and 500 $\mu$m from HerMES (Oliver et al. 2010) and H-ATLAS (Clements et al. 2010).

\subsection{Evolution}
\label{authorf_sec:evol}
For the {\tt spiral}, {\tt starburst}, {\tt LLAGN} and {\tt obscAGN} populations, both luminosity and density evolution are required in order to fit the observables, while for {\tt unobsAGN} no density evolution is needed, just luminosity. The shape of the evolution functions considered here is not the commonly used $(1+z)^k$ with a z$_{break}$, but it is a more physical representation
of the evolution, peaking at a given redshift and decreasing at higher redshifts, with different peaking redshifts and decreasing/increasing skew rates for the different populations. We have modelled the evolution with $z$ of the luminosity $L$ and density $\rho$ by means of a variant of the ``skew-normal distribution'' function, which is an extension of the normal (Gaussian) probability distribution, allowing for the presence of skewness (i.e. Azzalini et al. 1985). In particular, we have considered the following shape and parametrization for the evolution:
\begin{eqnarray}
log_{10}(L^\star(z))=log_{10}(L^\star(0)) + \frac{A_L}{\omega \sqrt{2\pi}} e^{-z^2/2\omega^2}\times erf\left(\kappa \frac{z}{\omega}\right) \label{equation:evoll} \\
log_{10}(\Phi^\star(z))=log_{10}(\Phi^\star(0)) + \frac{A_{\Phi}}{\omega \sqrt{2\pi}} e^{-z^2/2\omega^2}\times erf\left(\kappa \frac{z}{\omega}\right) \label{equation:evolr}
\end{eqnarray} 
\begin{table}
 \caption{Evolution Parameters}
\begin{tabular}{|l|c|c|c|c|}
\hline \hline
 SED class &  $A_L$ & $A_{\Phi}$  & $\omega$  &  $\kappa$ \\    \hline \hline 
{\tt spiral} & 1.0 &  2.5 &  2.5 &   5.0\\
{\tt starburst} & 12.0 & 8.0 & 3.5 & 3.0 \\
{\tt LLAGN}  & 8.5 & 6.5  &  2.2 & 1.8 \\
{\tt AGN2} & 19.0 & 16.0&  2.8 &  0.8 \\
{\tt AGN1}  & 17.8 &  $-$ &  4.6 &  3.1 \\
\hline \hline
\end{tabular}
\label{authorf_tab:evol}
\end{table}
\noindent where $erf(x)=\frac{2}{\sqrt{\pi}}\int^{x}_{0}e^{-t^2}dt$ is the ``error function'', $A_L$ and $A_{\Phi}$ are the normalizations for the evolution of $L^\star(z)$ and $\Phi^\star(z)$ respectively and $\kappa$ (known as {\em shape parameter}, because it regulates the shape of the function) and $\omega$ (scale factor) in different combinations define the skewness of the function and the redshift location of the peak. In order to limit the number of free parameters in the fit, for each population we have assumed the same values of $\kappa$ and $\omega$ for both luminosity and density evolution. 
\begin{figure}
  \begin{center}
    \includegraphics[width=8 cm]{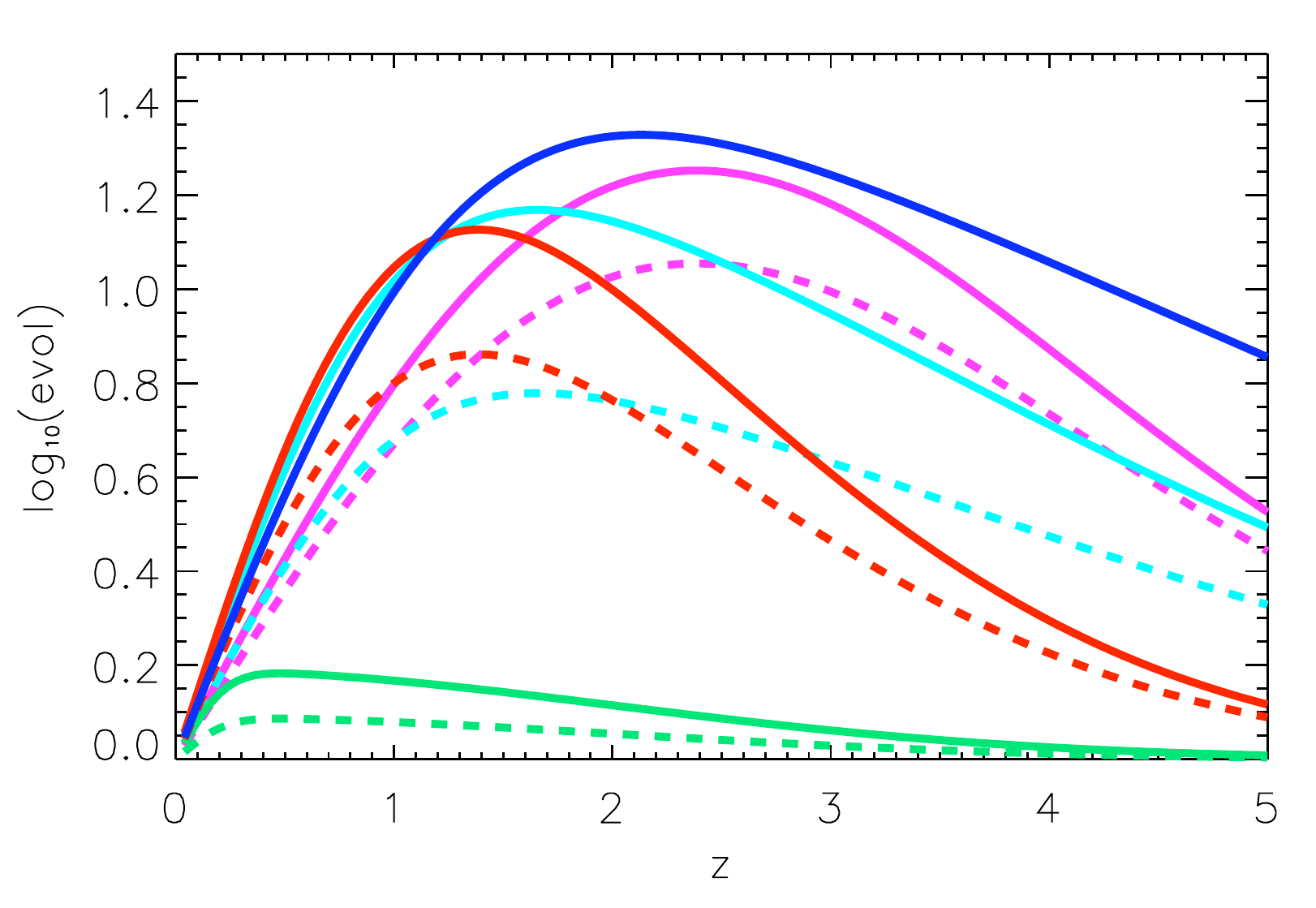}
      \end{center}
  \caption{The luminosity (solid lines) and density (dashed lines) evolution curves, described in Equations~\ref{equation:evoll} and \ref{equation:evolr},  for the five populations considered in the model (green: {\tt spiral}, cyan: {\tt starburst}, red: {\tt LLAGN}, magenta: {\tt AGN2}, blue: {\tt AGN1}) that best reproduce all the available IR data.} 
\label{authorf_fig:evol}
\end{figure}

By considering all the available IR observational constraints, we have found as best estimates for the parameters describing the evolution of the five populations (equations \ref{equation:evoll} and \ref{equation:evolr}) the values reported in Table~\ref{authorf_tab:evol}.
\begin{figure*}
  \begin{center}
    \includegraphics[width=10 cm,height=7cm]{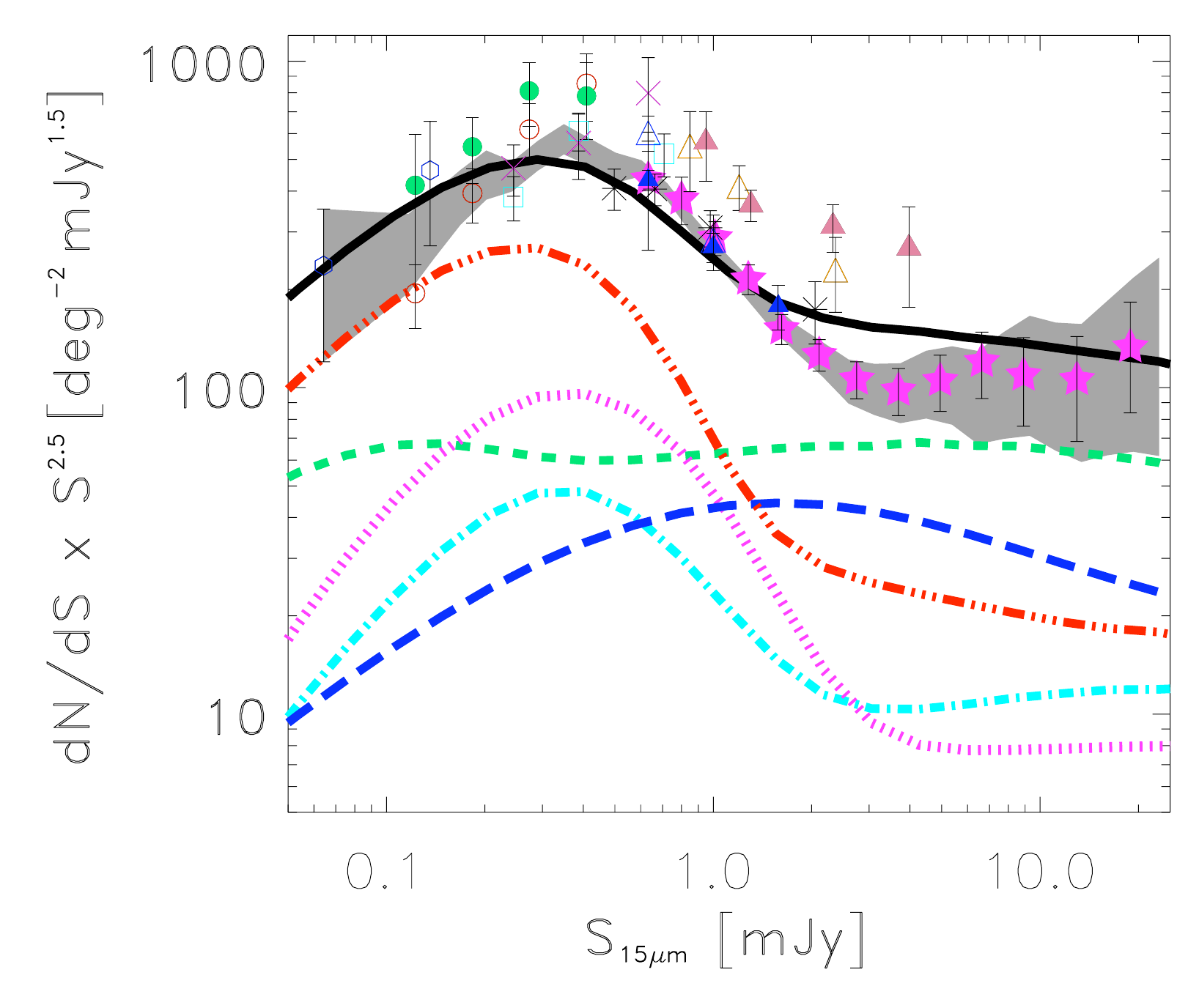}
    \includegraphics[width=14 cm,height=8cm]{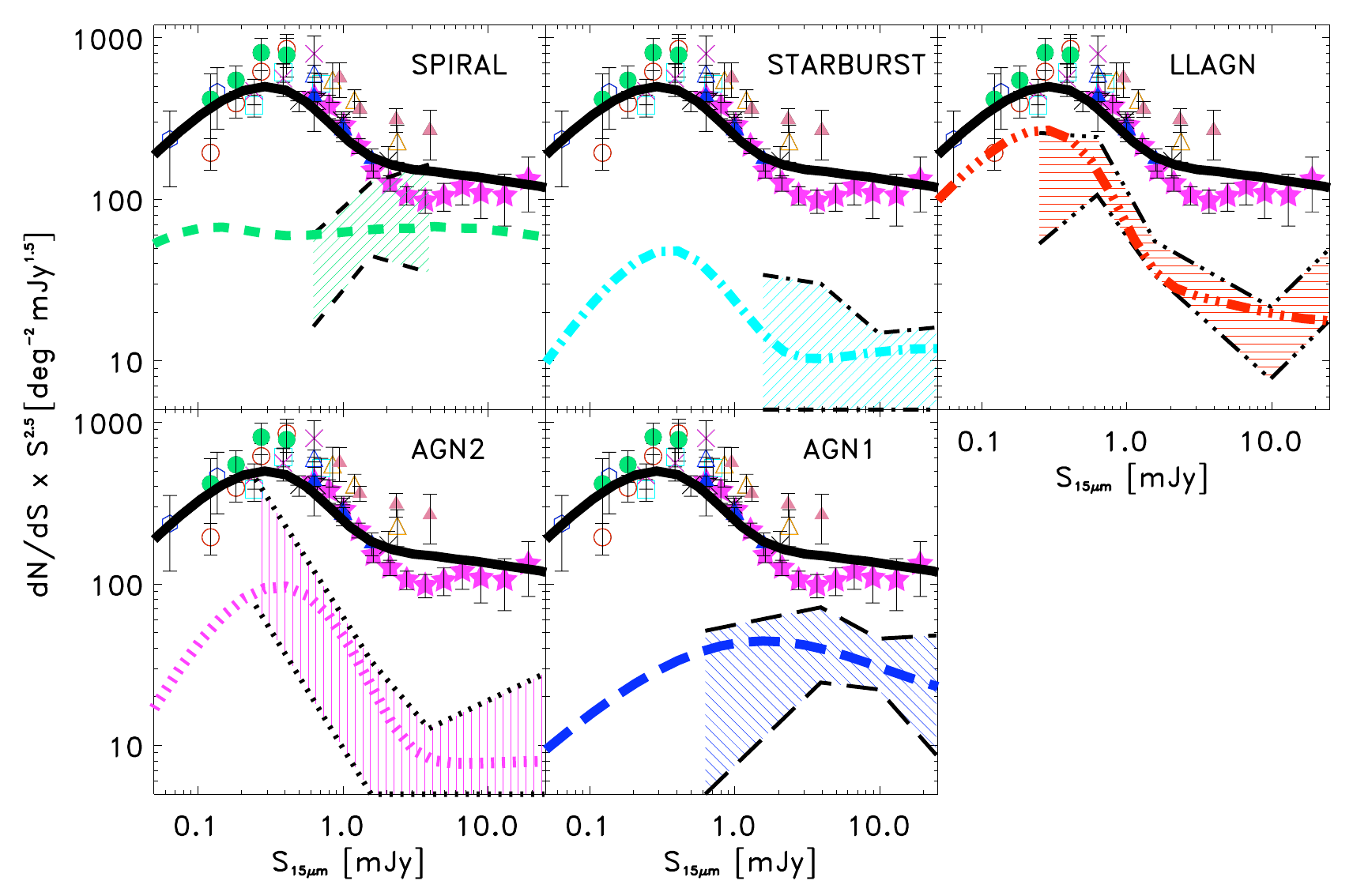}        
       \end{center}
  \caption{Differential source counts normalised to Euclidean at 15-$\mu$m. Several survey data are shown (red open circles: HDF-N, green filled circles: HDF-S, cyan open squares: Marano Firback, asterisks: Marano Deep, magenta diagonal crosses: Marano Ultradeep, pink filled triangles: Lockman Shallow, orange open triangles: Lockman Deep from Elbaz et al. 1999; magenta filled stars: ELAIS-S1 from Gruppioni et al. 2002; blue open hexagons: Abell2390 ultradeep lensed field from Metcalfe et al. 2003; blue filled triangles: Lockman Shallow, blue open triangles: Lockman Deep from Rodighiero et al. 2004) and compared to our model expectations. The grey shaded area represents the uncertainty region of the weighted averaged counts from all the different surveys. The different contributions from the five populations are plotted in different colours (black solid: total; green dashed: {\tt spiral} galaxies; cyan dot-dashed: {\tt starburst} galaxies; red dot-dot-dot-dashed: {\tt LLAGN}; magenta dotted: {\tt AGN2}; blue long-dashed: {\tt AGN1}), all together ({\em top panel}), and separately by population in the five {\em bottom panels} (where the coloured dashed area represents the 1$\sigma$ uncertainty region of the ELAIS-S1 observed counts for that population).} 
\label{authorf_fig:cntmir}
\end{figure*}
\begin{figure}
  \begin{center}
\includegraphics[width=8 cm,height=6cm]{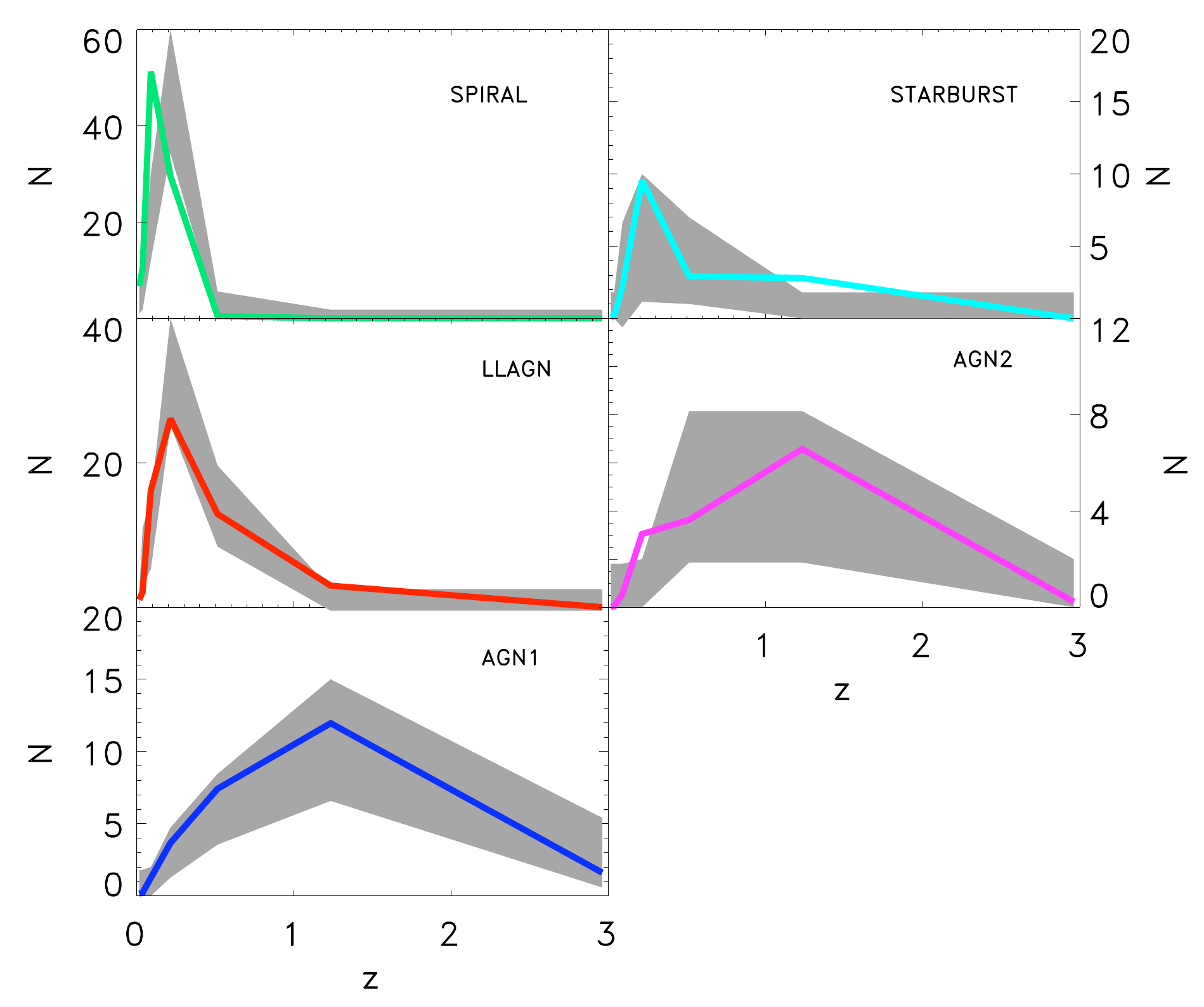}  
\includegraphics[width=8 cm, height=6cm]{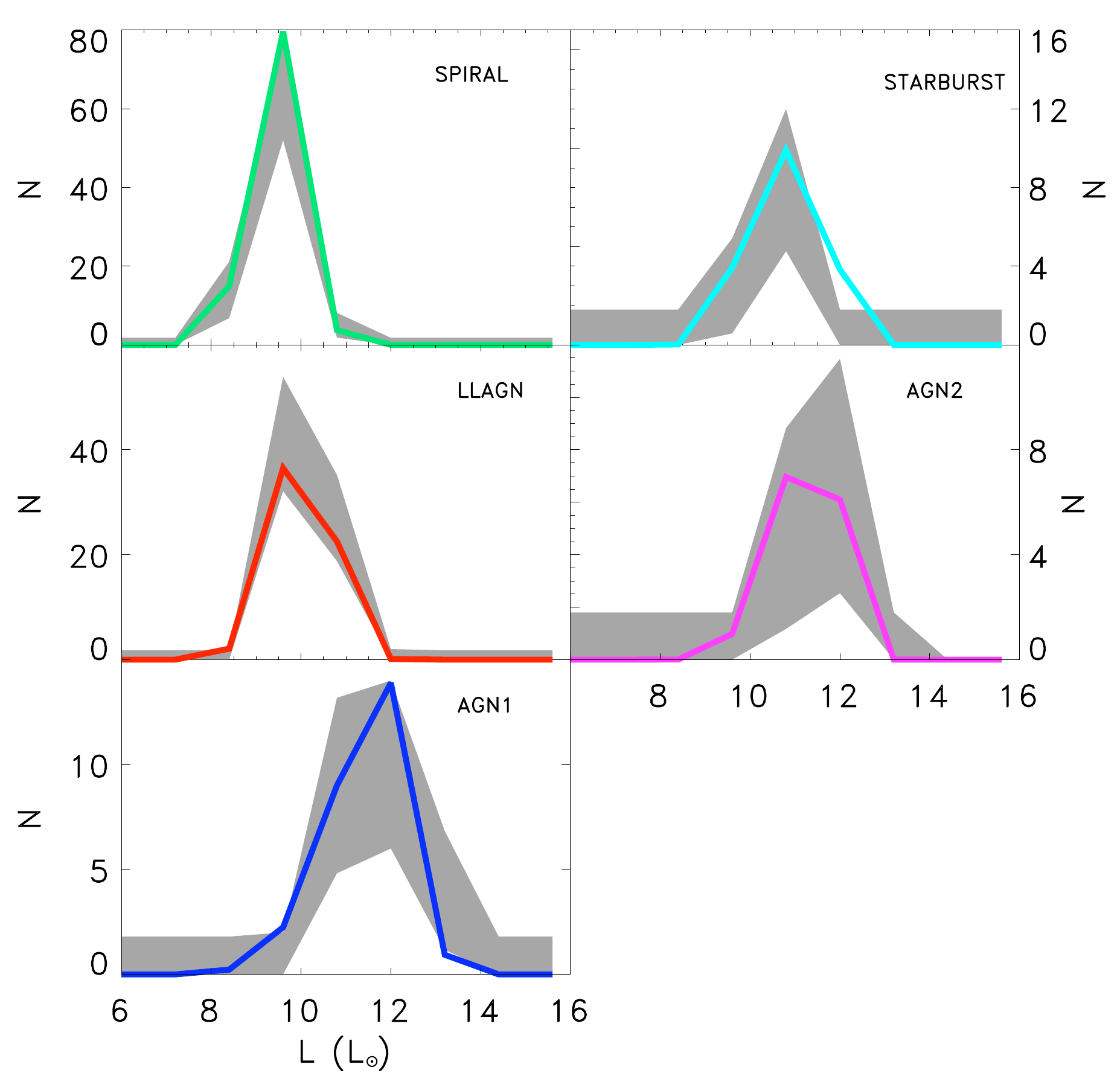} 
       \end{center}
  \caption{Redshift ({\em upper panels}) and luminosity ({\em lower panels}) distributions of the ELAIS-S1 15-$\mu$m sources down to a flux density S$_{15\mu m}$$=$0.5 mJy. The data relative to the five populations in the ELAIS-S1 Survey (Gruppioni et al. 2008) are shown as grey shaded distributions (representing the 1$\sigma$ uncertainty region), while the overimposed model expectations are shown as coloured lines (same colours as in Figure \ref{authorf_fig:cntmir}).}  
\label{authorf_fig:lzdist15}
\end{figure}
The evolution functions described by these parameters and characterising the five IR populations of our model are shown in Figure \ref{authorf_fig:evol}. To allow direct comparisons, in limited redshift ranges we can approximate our evolution curves to those more commonly used, with the form (1+$z$)$^k$. We note that the {\tt spiral} galaxies show low evolutionary rates, both in luminosity and density, increasing up to $z$$\sim$0.3 ($\propto$$(1+z)^{1.5}$ in luminosity and $\propto$$(1+z)^{0.8}$ in density), then slowly decreasing towards the higher $z$'s. {\tt Starburst} galaxies evolve fast ($\propto$$(1+z)^{3.5}$ in luminosity and $\propto$$(1+z)^{2.3}$ in density) up to $z$$\sim$1, with the luminosity and density remaining approximatively constant between $z$$=$1 and $z$$=$2, then decreasing at higher redshifts. {\tt LLAGN} show luminosity evolution (L($z$)$\propto$$(1+z)^{3.7}$) similar to and density evolution ($\rho$($z$)$\propto$$(1+z)^{2.8}$) higher than   those of starburst galaxies, but with a more pronounced peak at $z$$\simeq$1.2$\div$1.4, followed by a faster decrease at $z$$>$1.5. {\tt AGN1} evolve in luminosity at a rate $\propto$$(1+z)^{3.3}$ up to $z$$\sim$1.5, then their luminosity remains almost constant between $z$$\simeq$1.5 and $z$$\simeq$2.5, while the evolution of the {\tt AGN2} objects shows a flattening at even higher redshift (i.e. between $z$$=$2 and $z$$=$3), increasing towards the peak at a rate of $\approx$$(1+z)^{2.7}$ in luminosity and $\approx$$(1+z)^{2.2}$ in density and decreasing faster at $z$$>$3. 

\section{Results}
\label{authorf_sec:results}

\subsection{The 15-$\mu$m Data}
\label{authorf_sec:15mic}

In Figures~\ref{authorf_fig:cntmir} and \ref{authorf_fig:lzdist15} we show the results of the 15-$\mu$m data and model comparison, plotted in terms of differential source counts as function of flux density and of redshift and luminosity distributions (global and separately for each population). Note that when comparing the single populations source counts, redshift and luminosity distributions from the model to the observed ones, we consider and apply to the model the same optical and 15-$\mu$m limits and spectroscopic incompleteness of the data. The model, constrained just up to the data limits, is then extrapolated to fainter optical magnitudes and infrared flux densities. 
The model is able to reproduce the number density, shape and peak location of all the distributions. A discrepancy is observed in the source counts, with the model overpredicting the data at flux densities of a few mJy. However, we must stress that the highly statistically significant S1 survey, which constitutes the only data comparison that we used for redshifts and luminosities at 15 $\mu$m, in the few mJy range seems somewhat under-dense with respect to other, less statistically significant, surveys from smaller fields (see Figure~\ref{authorf_fig:cntmir}). 
Another little discrepancy between data and model consists of a high-$z$, high-$L$ tail in the {\tt starburst} modelled redshift and luminosity distributions that is not observed in the data. We note, however, that we are dealing with very small numbers of objects in these tails: $\sim$2 expected, versus 0 observed. \\
The 15-$\mu$m counts are dominated by {\tt spiral} galaxies and {\tt AGN1} at S$_{15 \mu m}$$\gsimeq$1.5 mJy, while at fainter flux densities the {\tt LLAGN} constitute the main population responsible for the peak at S$_{15 \mu m}$$\simeq$0.3--0.4 mJy observed in the differential counts. The {\tt starburst} and {\tt AGN2} populations are never dominant at 15 $\mu$m, though their contribution is significant at sub-mJy level.

\subsection{The 24-$\mu$m Data}
\label{authorf_sec:24mic}

\begin{figure*}
  \begin{center}
    \includegraphics[width=7 cm]{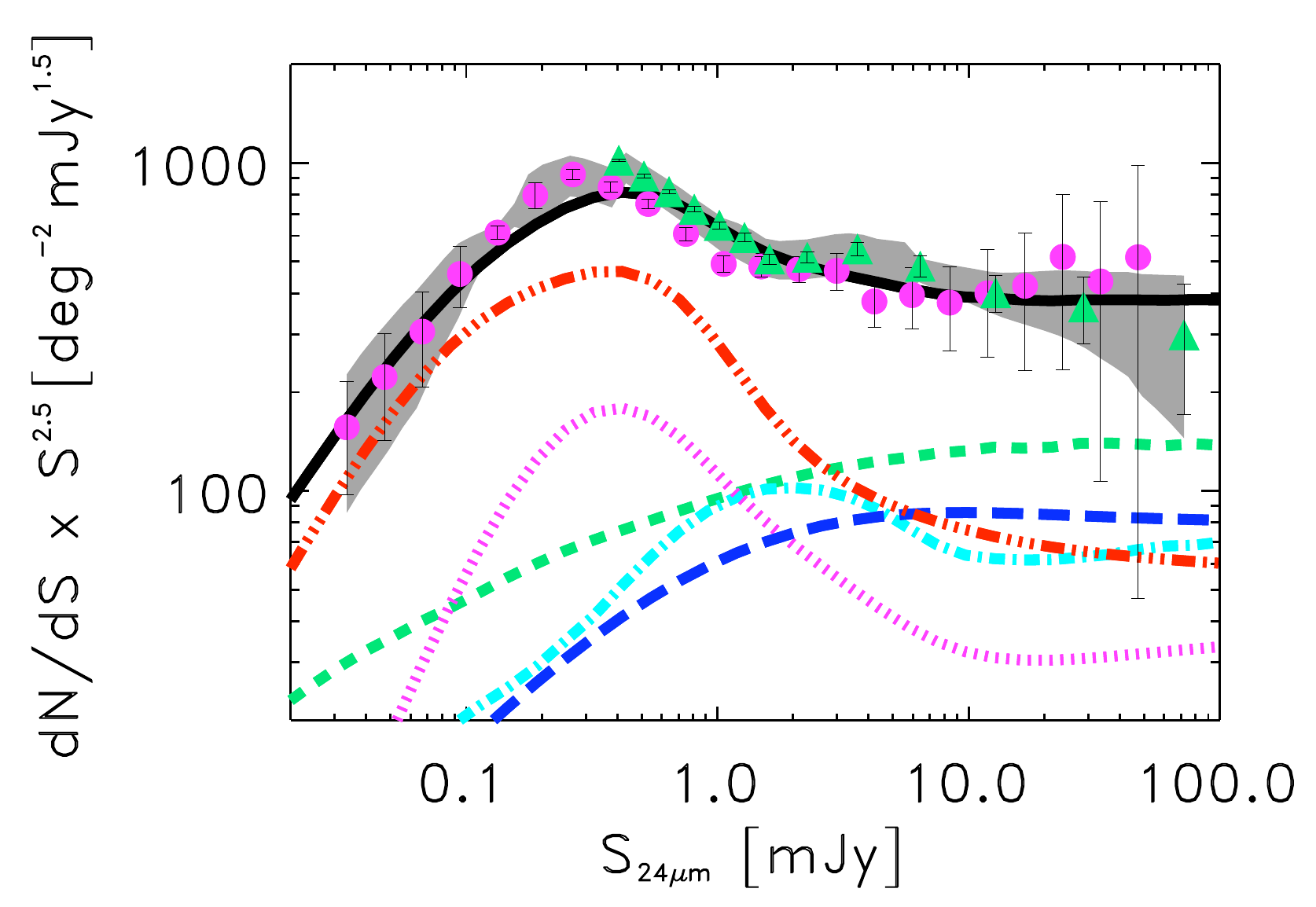}        
\includegraphics[width=6 cm,height=4.5cm]{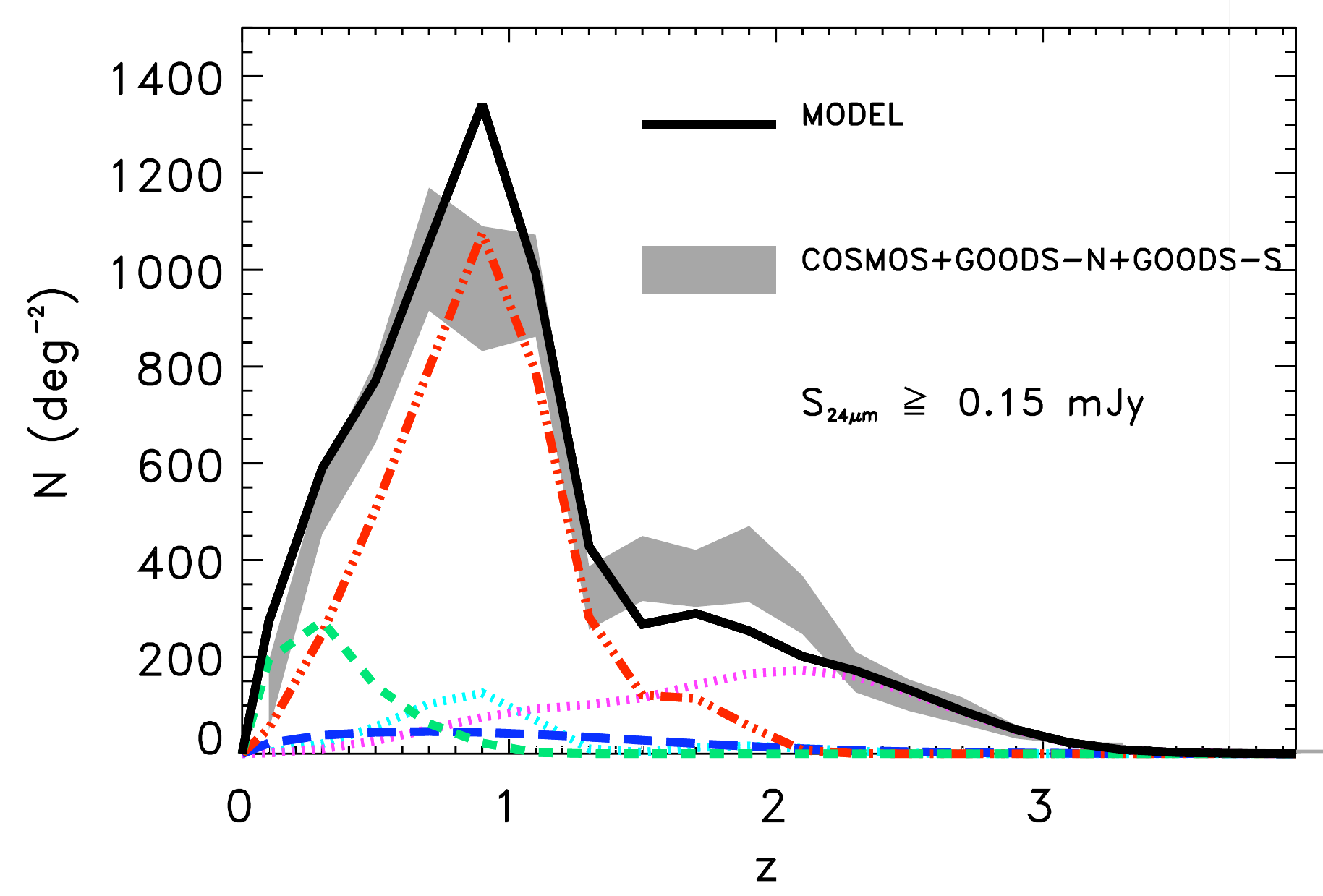} 
\includegraphics[width=13 cm,height=7cm]{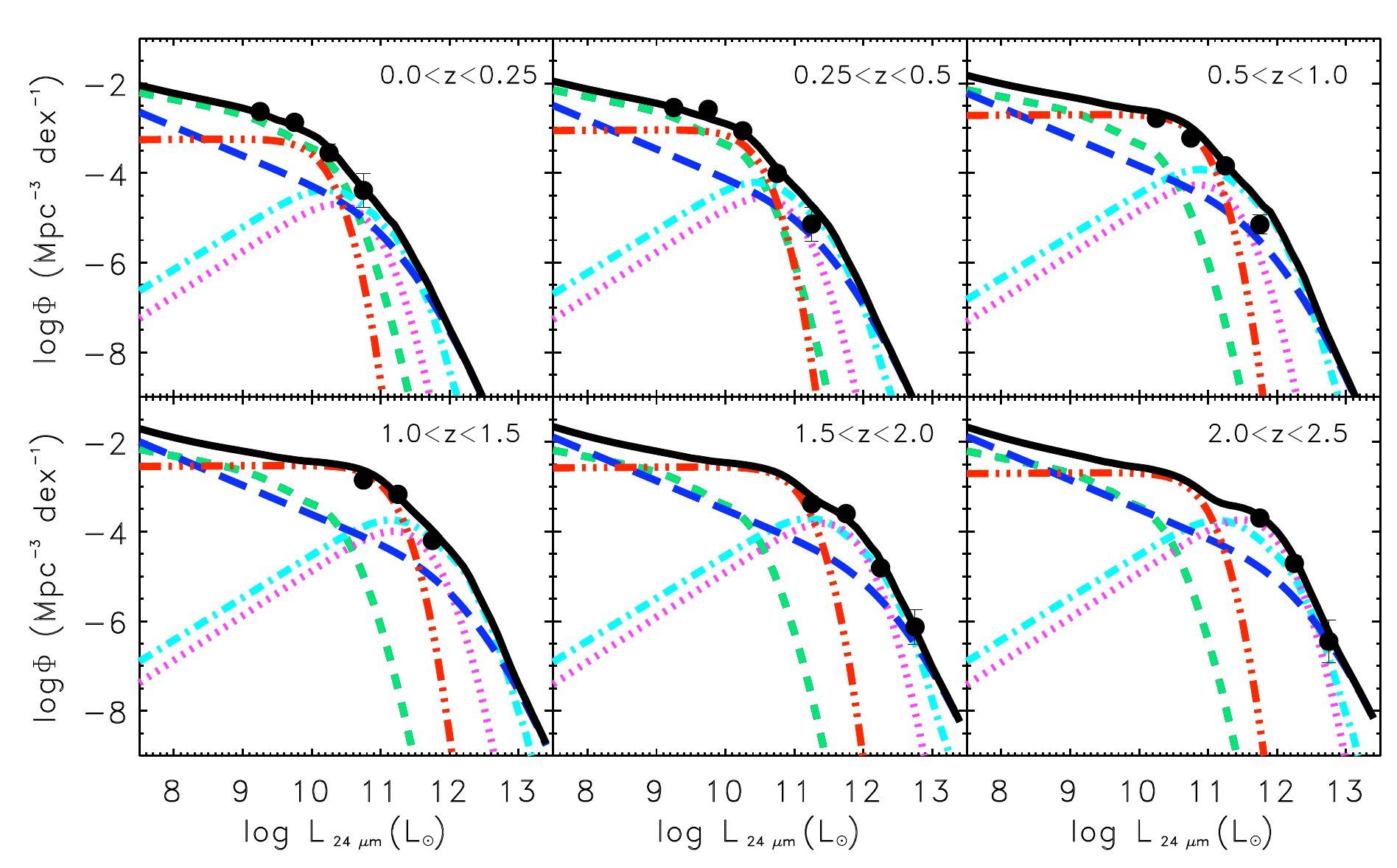} 
       \end{center}
  \caption{{\em Upper left panel}: 24 $\mu$m differential source counts normalised to Euclidean obtained from our model (coloured curves, as in Fig.~\ref{authorf_fig:cntmir}) and observed by {\em Spitzer} (magenta filled circles: Papovich et al. 2004; green filled triangles: Shupe et al. 2008; grey shaded area: uncertainty region of the weighted averaged counts from the two surveys.). {\em Upper right panel:} redshift distribution of the 24-$\mu$m sources to a flux density S$_{24\mu m}$$=$0.15 mJy: total (black line) and decomposed into populations, compared to the combined observed data (grey shaded area) from the COSMOS  (Le Floc'h et al. 2009), GOODS-S and GOODS-N fields (Rodighiero et al 2010). {\em Lower panels:} Rest-frame LF at 24 $\mu$m in different redshift intervals: model expectations (coloured curves as in �Fig.~\ref{authorf_fig:cntmir}) compared to observed data in the VVDS+GOODS fields (black circles; Rodighiero et al. 2010). }  
\label{authorf_fig:lfzdist24}
\end{figure*}

As shown in Figure~\ref{authorf_fig:lfzdist24}, the agreement between data and model is remarkably good also at 24 $\mu$m, since the model is able to reproduce well the source counts, the $z$-distributions and the luminosity function in different redshift intervals (data from Rodighiero et al. 2010). Note that, according to our model, the peak observed in the counts at S$_{24 \mu m}\simeq$0.3 mJy is made mainly by galaxies containing a {\tt LLAGN}, with a minor contribution also from the {\tt AGN2} population. At flux densities $\gsimeq$1 mJy {\tt spiral} galaxies dominate the counts, while {\tt starburst} galaxies and {\tt AGN1} contribute almost equally and similarly to {\tt LLAGN}. The redshift distribution down to S$_{24 \mu m}$$=$0.15 mJy shows a peak at $z$$\sim$0.8--0.9 mainly due to {\tt LLAGN} and a higher-$z$ tail due to the {\tt AGN2} objects. {\tt Spiral} galaxies contribute only to very low redshifts, while {\tt starburst} galaxies peak at similar $z$ to that of {\tt LLAGN}, though the number of ``pure'' {\tt starburst} is much lower. The redshift distribution shown in the plot is obtained from the combination of the redshift distributions in three different surveys (COSMOS, Le Floc'h et al. 2009 and GOODS-N and -S, Rodighiero et al. 2010), which were all in reasonably good agreement with our model.  

\subsection{The 100- and 160-$\mu$m {\em Herschel} Data}
\label{authorf_sec:100mic}

\begin{figure*}
  \begin{center}
    \includegraphics[width=8 cm]{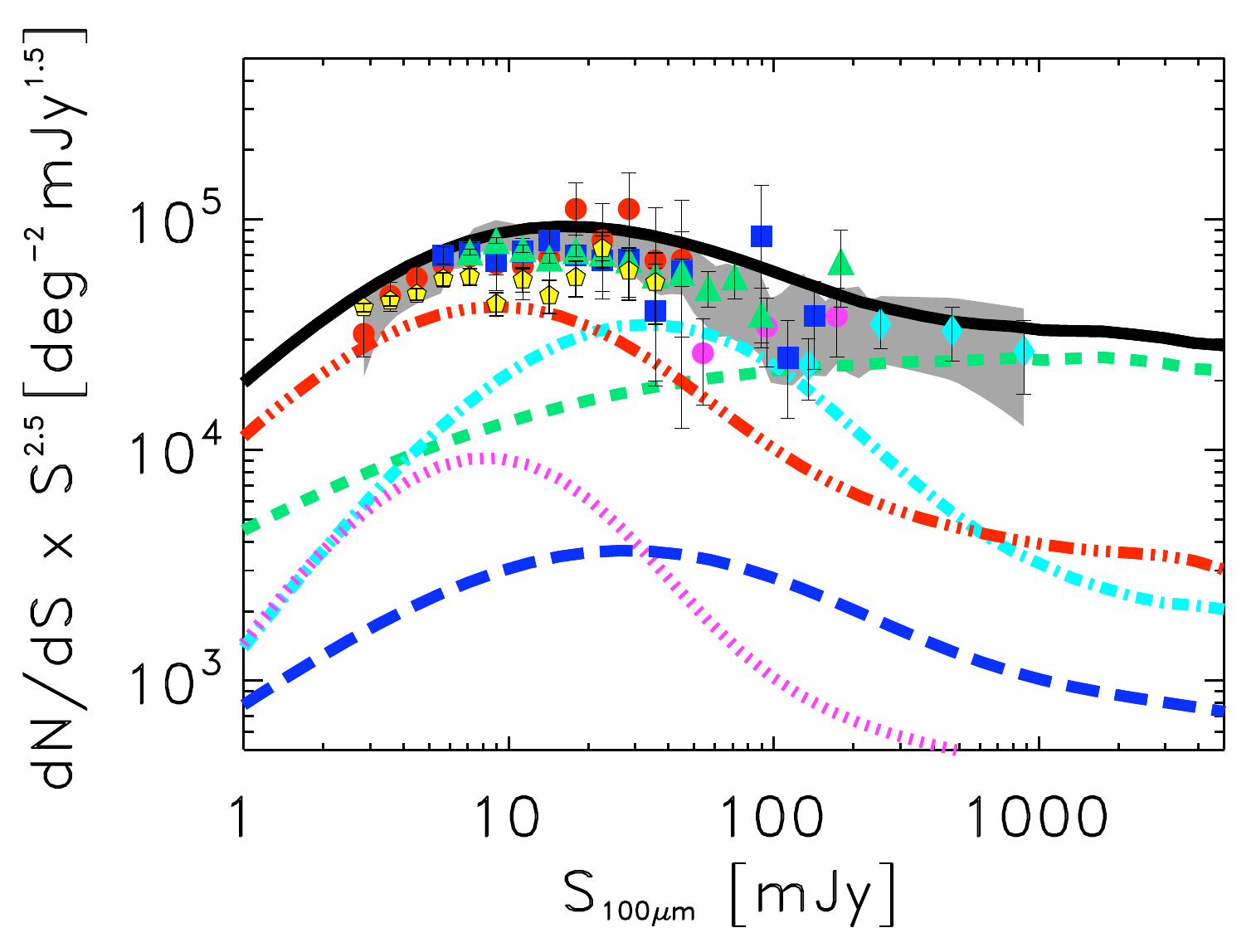}
    \includegraphics[width=8. cm]{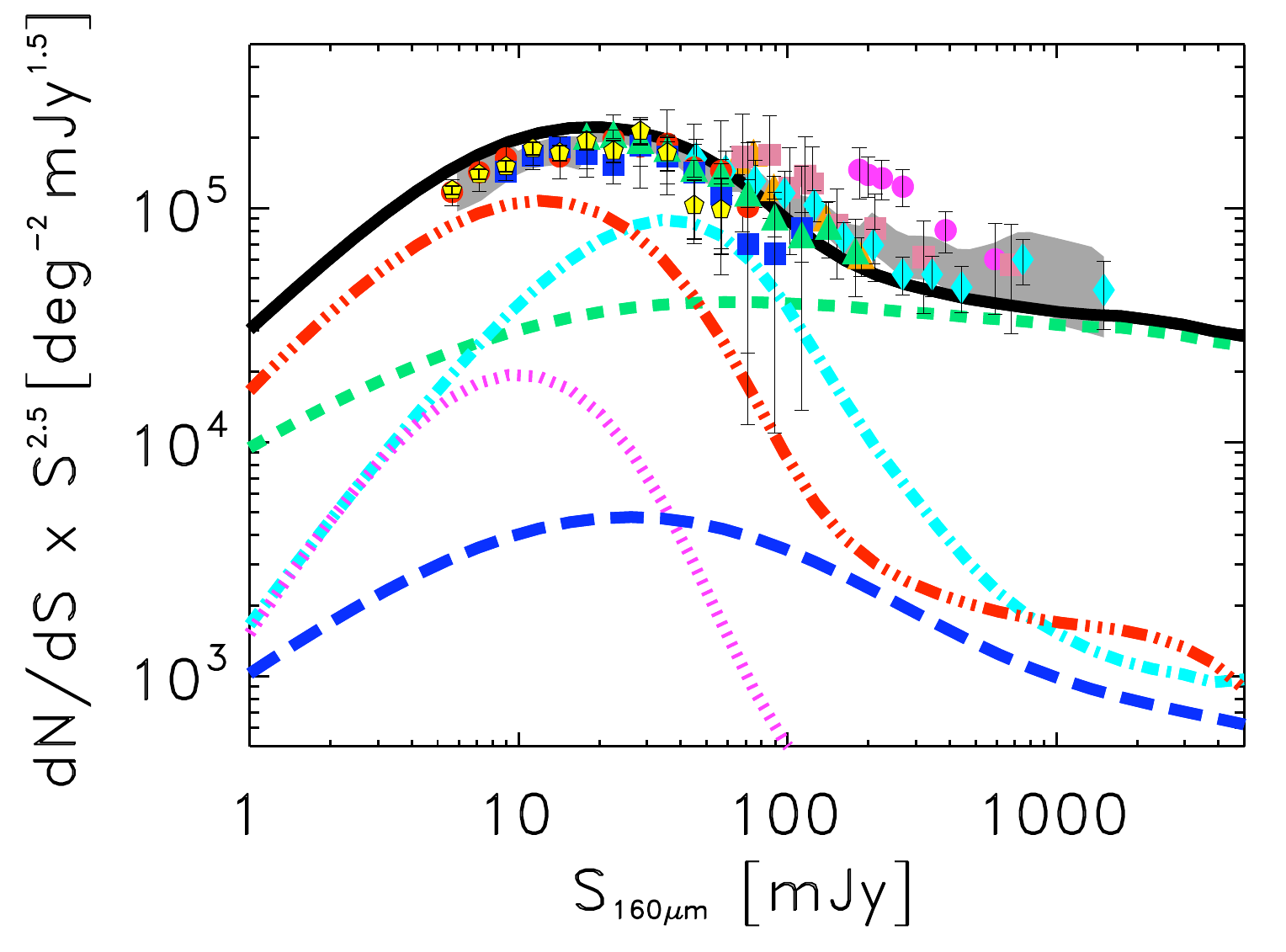} 
    \includegraphics[width=8. cm]{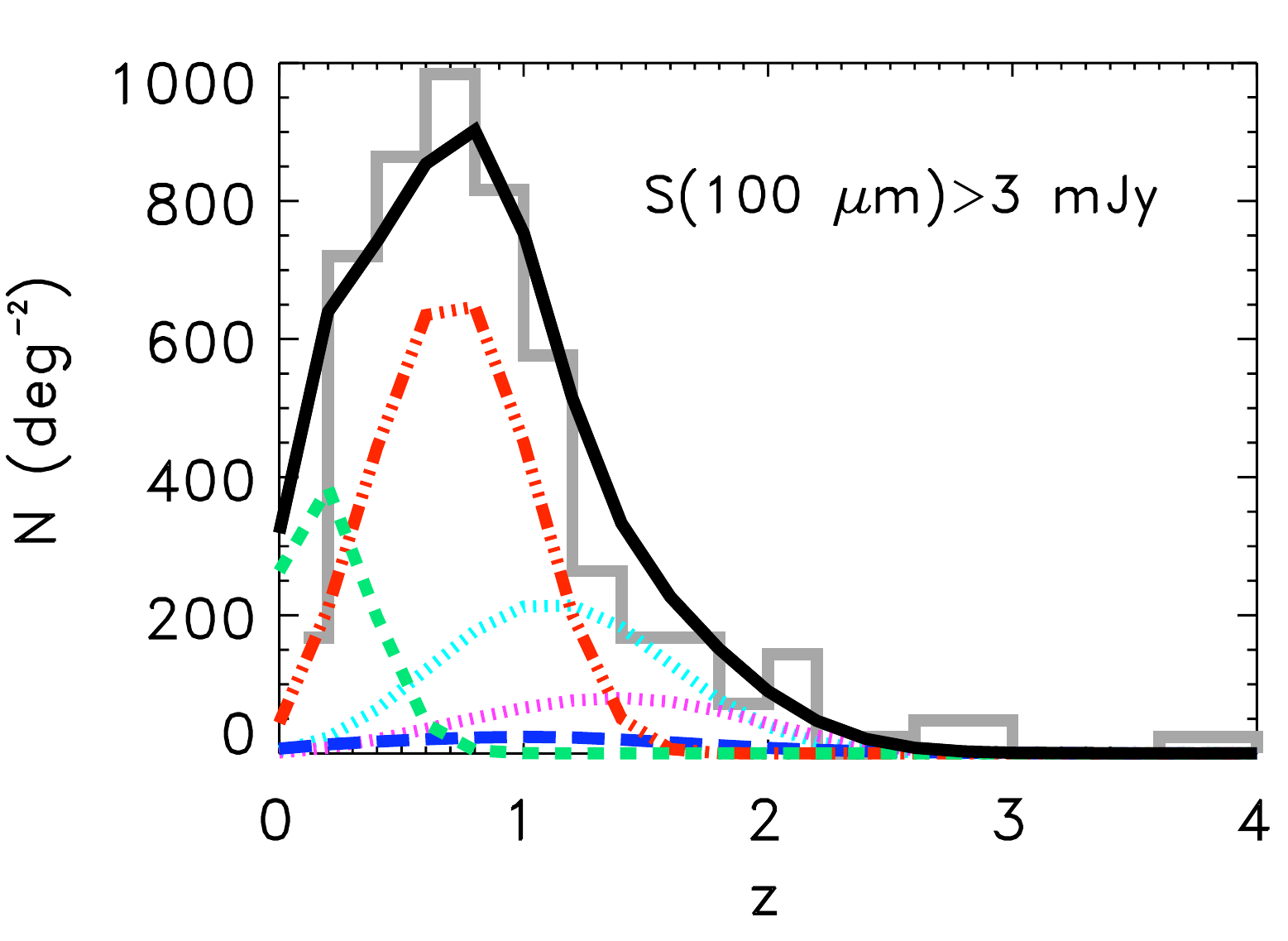}
    \includegraphics[width=8. cm]{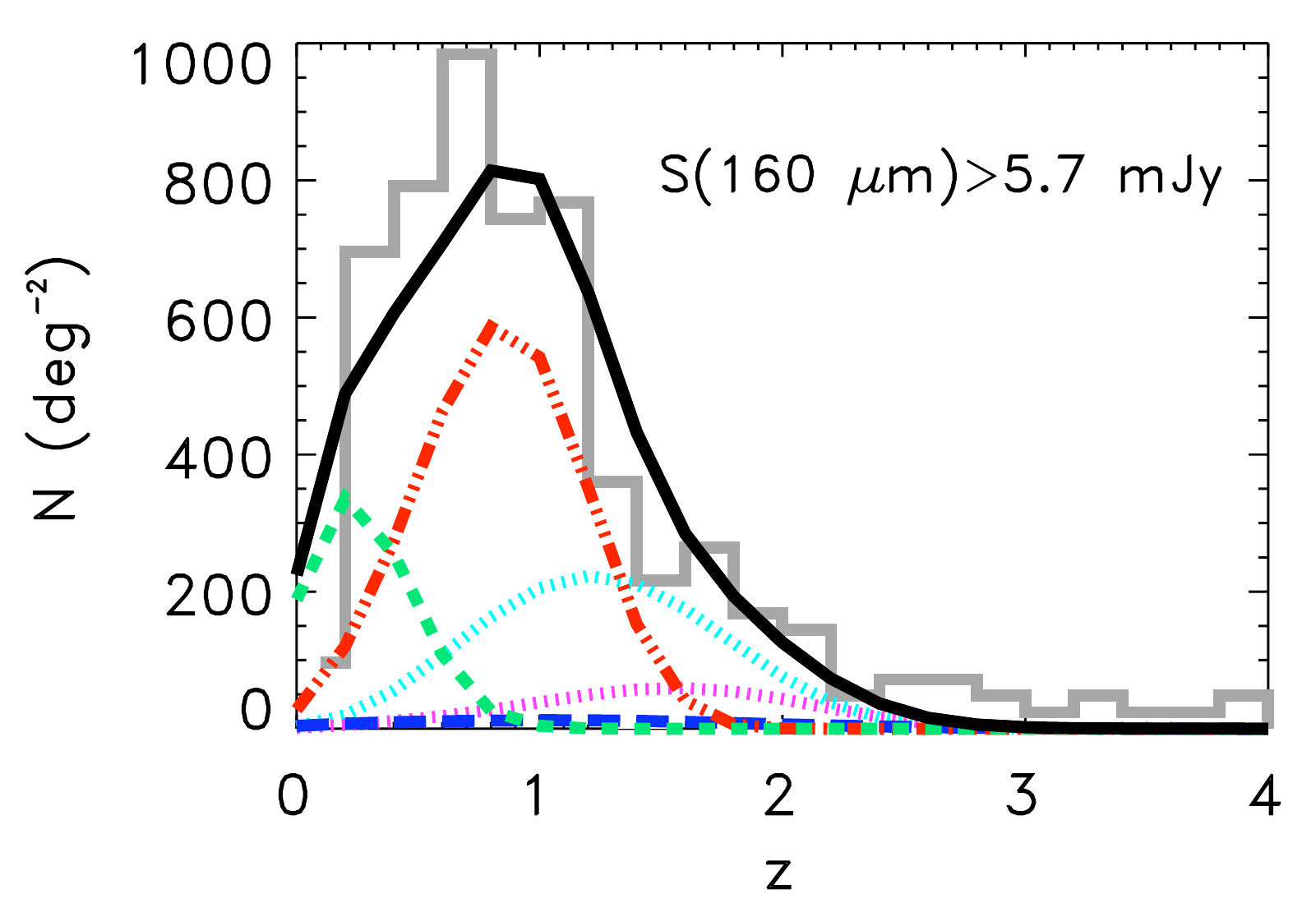} 
      \end{center}
  \caption{{\em Top:} Differential source counts normalised to Euclidean at 100 $\mu$m ({\em left}) and 160 $\mu$m ({\em right}). {\em Herschel} data from PEP (Berta et al. 2010) are represented by the red filled circles (GOODS-N), the blue filled squares (Lockman Hole) and the green triangles (COSMOS) in both plots. The other symbols show the previous results from {\em ISO} (100-$\mu$m: magenta filled circles from Rodighiero \& Franceschini 2004, cyan filled diamonds from Heraudeau et al. 2004; 160-$\mu$m: magenta filled circles from Dole et al. 2001) and {\em Spitzer} (160-$\mu$m: pink filled squares from Dole et al. 2004, orange filled triangles from Frayer et al. 2006, cyan filled diamonds from Bethermin et al. 2010a). The grey shaded area represents the uncertainty region of the weighted averaged counts from all the different surveys. {\em Bottom:} Redshift distribution of the PEP sources at 100 $\mu$m ({\em left}) and 160 $\mu$m ({\em right}) cut at a flux density corresponding to the 3$\sigma$ limit (grey histogram), compared to model expectations to the same limit. The model curves representing the different populations are the same as in Figures~\ref{authorf_fig:llf}, \ref{authorf_fig:cntmir} and \ref{authorf_fig:lfzdist24}.}  
\label{authorf_fig:cntpacs}
\end{figure*}

The recent results from {\em Herschel} have just provided (and will furtherly provide) stringent constraints to evolution models in the FIR, with large samples of sources with extensive multiwavelength information, crucial not only for statistical studies, such as source counts, but also for detailed SED-fitting analysis and photometric redshift derivation. 
The {\em Herschel} data, spanning from 70 to 500 $\mu$m, will systematically cover the FIR/sub-mm spectrum of thousands of galaxies, from $z$$=$0 up to $z$$=$3--4. 
In particular, the PEP extragalactic survey samples 4 different layers with {\em Herschel-PACS}: from the wide and shallow COSMOS field, through medium size areas like Lockman Hole, to the very deep GOODS-N and -S areas and beyond, through gravitational lensing in galaxy clusters (e.g. Abell 2218, Altieri et al. 2010). The first PEP data available, the Science Demonstration Phase (SDP) observations dedicated to the GOODS-N field, allowed us to perform a detailed broad-band SED-fitting analysis and to classify the 100-$\mu$m and 160-$\mu$m sources using the same technique used for the ELAIS-S1 sources, consistently with the model populations. The PEP observations provide a strong constraint to the SED shape of the IR sources, since for the first time the peak of the FIR bump has been sampled for large numbers of galaxies at cosmological distance. As discussed by Gruppioni et al. (2010), the template SEDs of Polletta et al. (2007) generally  provide very good fits to the SEDs of the PEP sources. However, in $\sim$10--12\% of cases the observed SEDs are very well reproduced by the templates over the entire UV/optical/NIR/MIR range, while they are systematically lower than the data in the FIR range. In these cases, the PEP 100- and 160-$\mu$m fluxes are higher by up to a factor of $\sim$4 than the template at the same wavelengths. This happens mainly for the {\tt LLAGN} templates (for about 40\% of the PEP sources fitted by the Seyfert 2/Seyfert 1.8 SEDs) and in a smaller fraction of cases also for the {\tt spiral} ones. We therefore constructed three new templates with a rest-frame 0.1--15 $\mu$m spectrum similar to that of Polletta et al. (2007), but with a higher FIR bump, by averaging together (in wavelength-bins) the observed rest-frame SEDs (normalised to $K_s$ band) exhibiting an excess in the FIR and fitted by the same template (Seyfert2/1.8/Sdm). We have therefore taken into account this result in our model, by considering both the Seyfert 2 template and the modified-Seyfert 2 one as representative of the {\tt LLAGN} population. Since no trend with either luminosity or redshift has been observed in the PEP SDP sample for the {\tt LLAGN} SEDs, in the model we have considered the relative contribution of the original and modified Polletta et al. (2007) template as given by the relative proportion of sources fitted by the two templates. 
The Luminosity Function and its evolution have then been studied for the five populations separately, by means of the SDP observations, and compared to our model expectations, showing a very good agreement between data and model (see Gruppioni et al. 2010). The PEP SDP data in the GOODS-N have been supplemented with the COSMOS and Lockman Hole wider and shallower surveys to build galaxy number counts at 100 and 160 $\mu$m (Berta et al. 2010), considered as constraints to our model in the FIR. 
In Figure~\ref{authorf_fig:cntpacs} we show the results of the comparison between the observed PEP data -- in terms of source counts and redshift distributions -- and our model expectations at 100 $\mu$m and 160 $\mu$m.
\begin{figure*}
  \begin{center}
    \includegraphics[width=7.5cm]{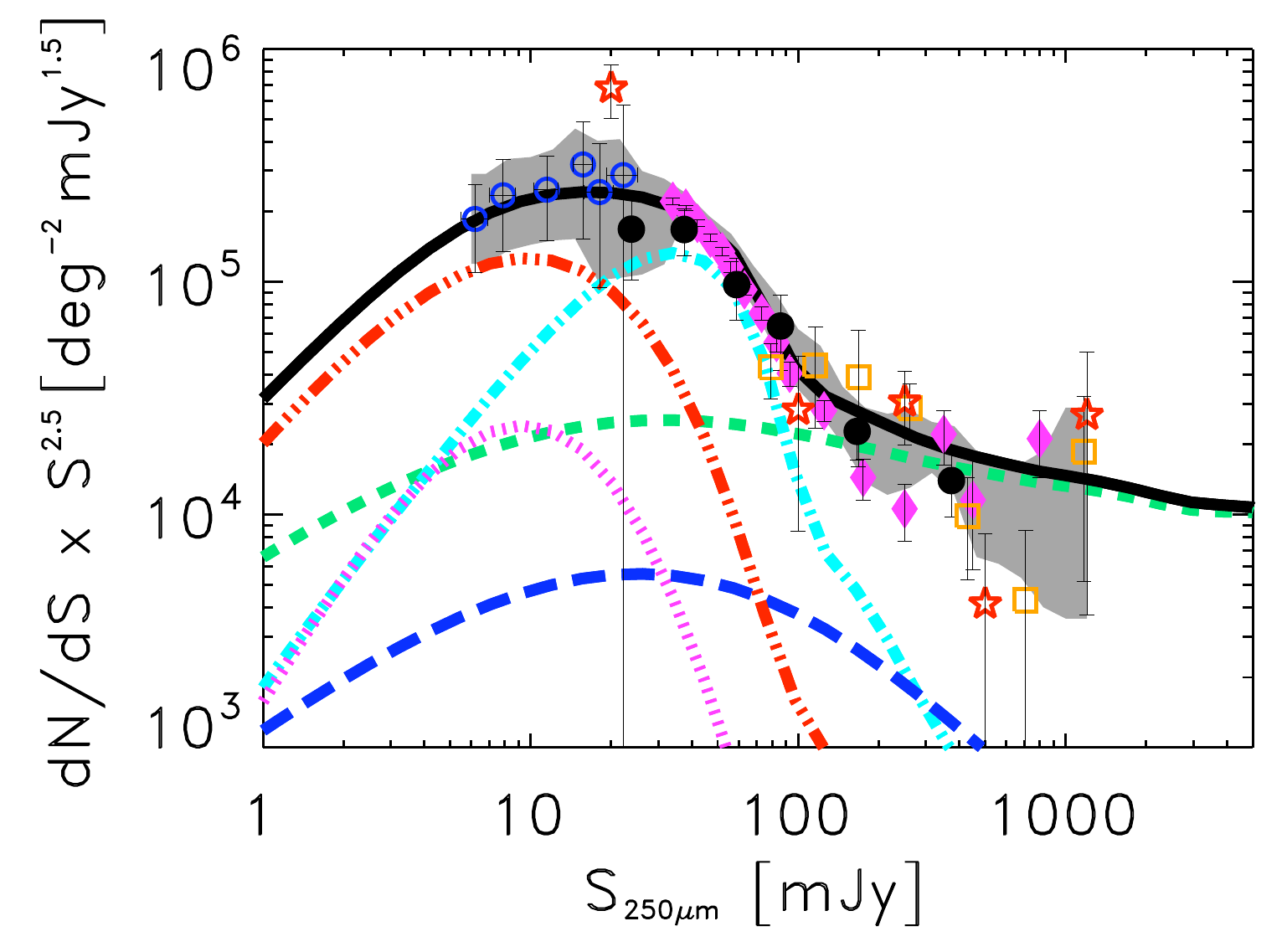}        
    \includegraphics[width=7cm,height=5.2cm]{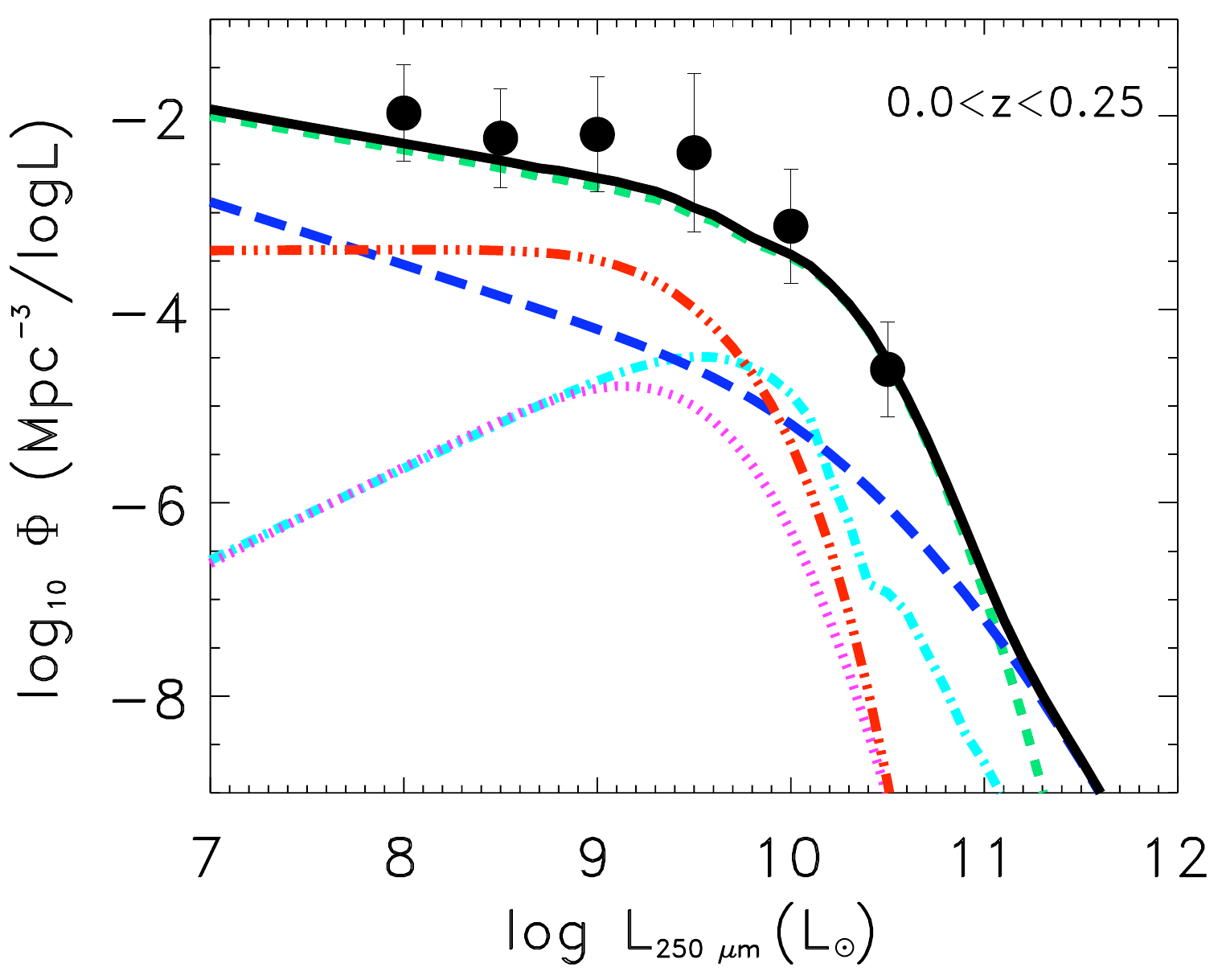}
\includegraphics[width=7.5cm]{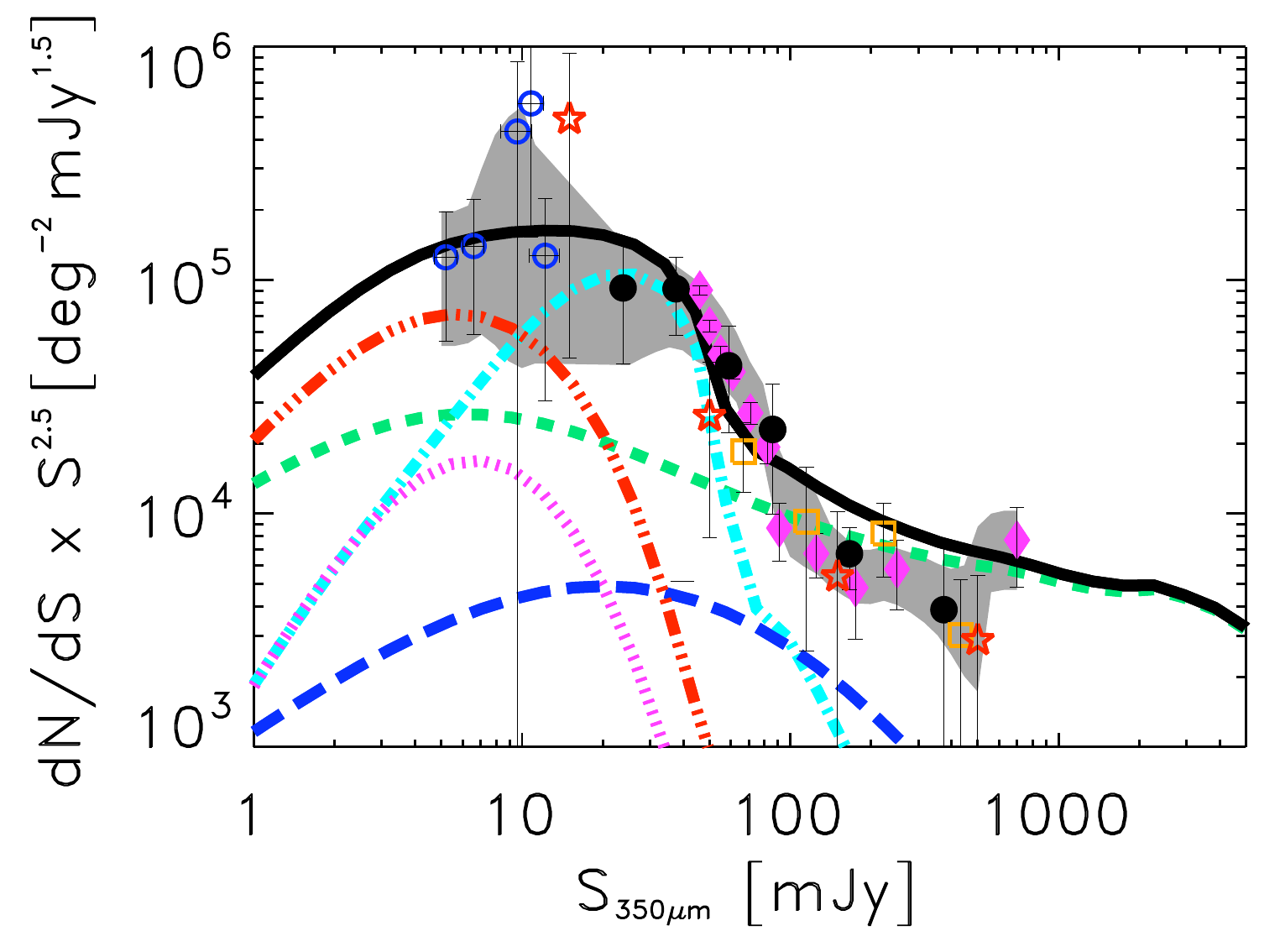} 
    \includegraphics[width=7cm,height=5.2cm]{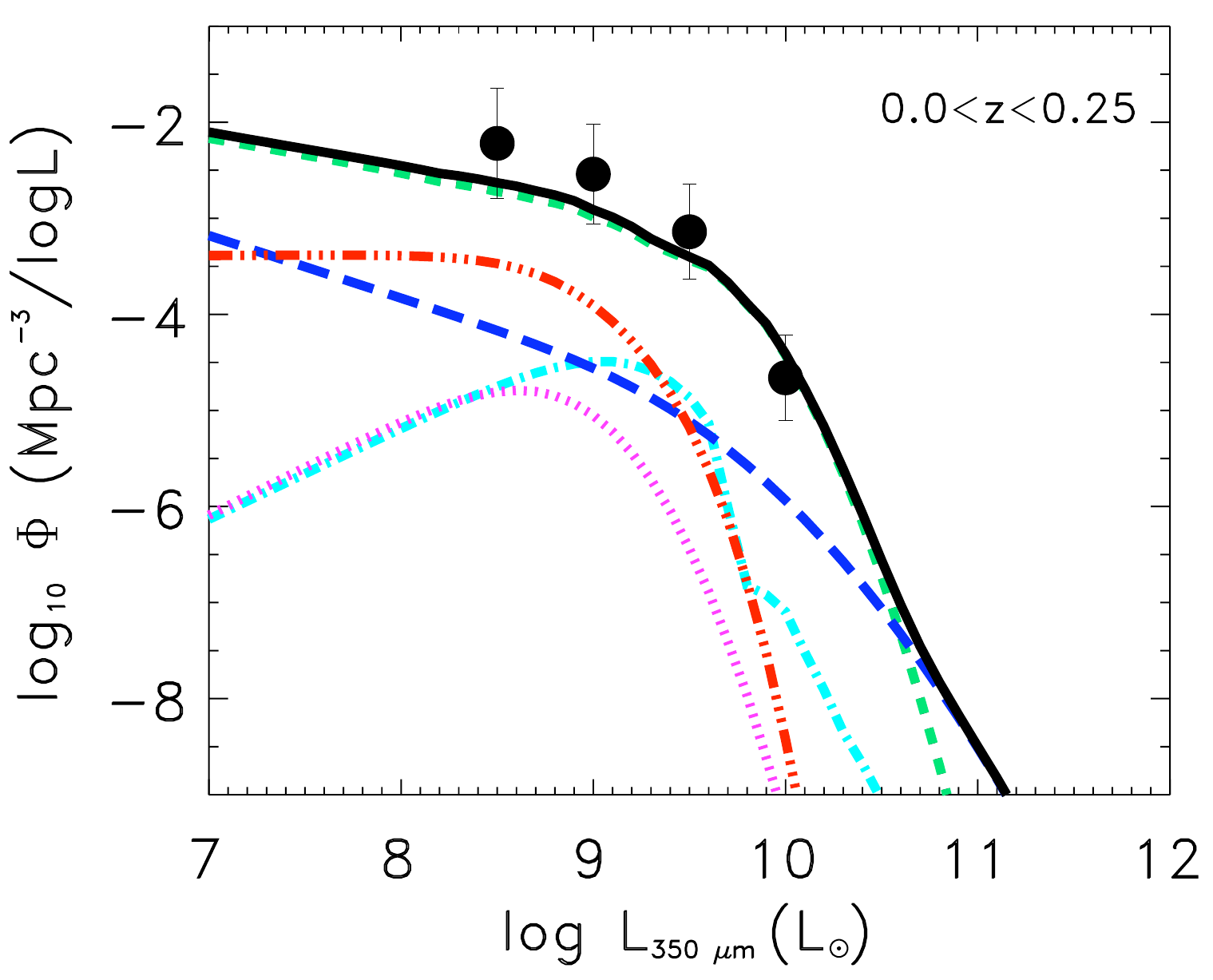}
\includegraphics[width=7.5cm]{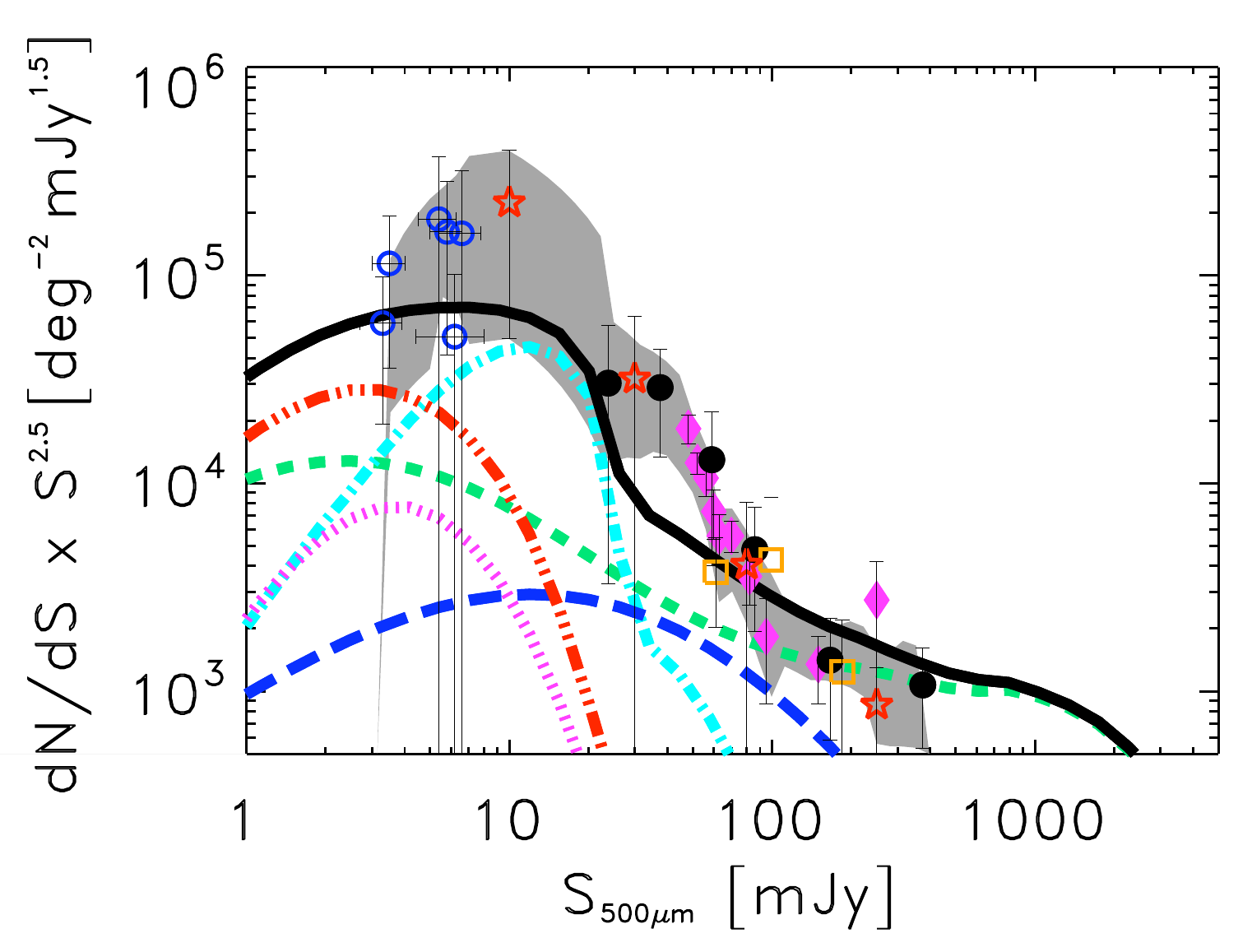} 
    \includegraphics[width=7cm,height=5.2cm]{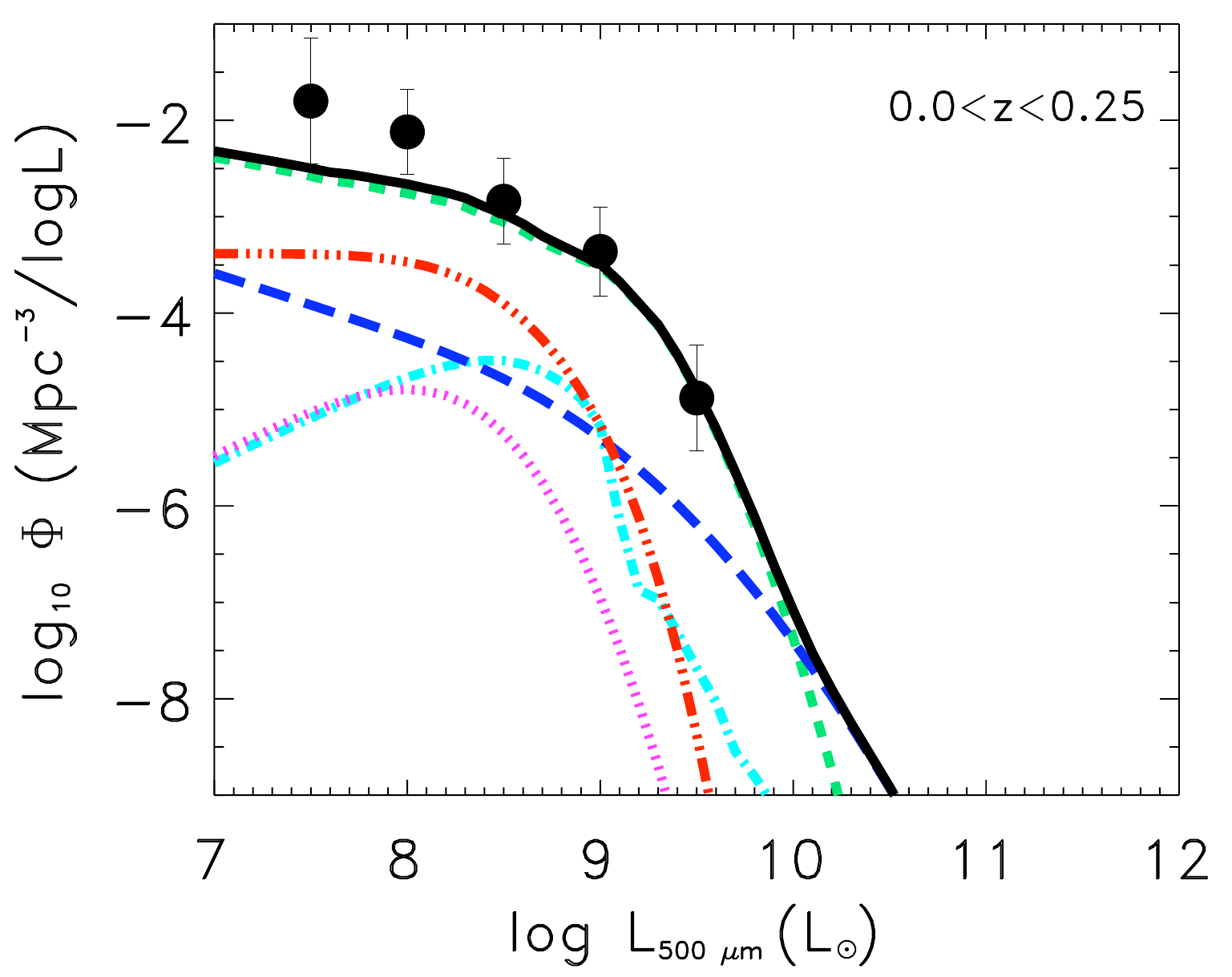}
      \end{center}
  \caption{{Differential source counts normalised to Euclidean ({\em left}) and local luminosity functions (\em right}) at the {\em Herschel-SPIRE} wavelengths: 250 $\mu$m ({\em top}), 350 $\mu$m ({\em middle}) and 500 $\mu$m ({\em bottom}). The black filled circles are data from the HerMES Survey (counts: Oliver et al. 2010; LLFs: Vaccari et al. 2010), while the magenta filled diamonds are counts from the H-ATLAS Survey (Clements et al. 2010). The open symbols are source counts derived from {\em BLAST} data by Bethermin et al. (2010b; orange squares: deep field with and without 24-$\mu$m priors; blue circles: counts computed with a stacking analysis on 24-$\mu$m positions) and Patanchon et al. (2009; red stars: P(D) analysis). The grey shaded area represents the uncertainty region of the weighted averaged counts from all the different surveys.}  
\label{authorf_fig:cntspire}
\end{figure*}

\subsection{The 250-, 350- and 500-$\mu$m {\em Herschel} Data}
\label{authorf_sec:250mic}

In the sub-mm band the instrument {\em BLAST} performed the first wide and deep survey in the 250--500 $\mu$m range (Devlin et al. 2009) before {\em Herschel}, producing source counts through a P(D) fluctuation analysys (Patanchon et al. 2009) and through other three different methods (e.g. 1) blind extraction using a particular algorithm called FASTPHOT; 2) extraction using the {\em Spitzer} 24-$\mu$m sources as a prior; 3) a stacking analysis of the {\em Spitzer} 24-$\mu$m galaxies in the {\em BLAST} data; Bethermin et al. 2010b).
The HerMES (Oliver et al. 2010) and the H-ATLAS Surveys (Eales et al. 2010) have already observed a number of fields of different areas and sensitivity (HerMES SDP: GOODS-N, Lockman North, Lockman SWIRE and FLS; H-ATLAS: 14 sq. deg. in the GAMA9 field) using {\em SPIRE} on {\em Herschel}, providing the galaxy number counts down to $\sim$20 mJy at 250 $\mu$m, 350 $\mu$m and 500 $\mu$m. Oliver et al. (2010) compared the observed counts with eight models available in the literature, showing that many of them cannot fit the bright end ($>$100 mJy) and none but one (Valiante et al. 2009) can fit the steep rise at 20$<$S$<$100 mJy observed in the HerMES data. 
Similarly, Clements et al. (2010) find that none of the current models is an ideal fit to the H-ATLAS source counts, the best performing model with respect to the shape of the counts being the Negrello et al. (2007), but only at 500 $\mu$m, since this model overpredicts the effect of the bump at 250 and 350 $\mu$m. 
In Figure~\ref{authorf_fig:cntspire} ({\em left panels}) we show the comparison between our model
and the sub-mm observed source counts. The model is able to reproduce very well the 250-$\mu$m and 350-$\mu$m data over the whole flux density range, fitting the bright end, the steep rise and the peak of the source counts (note the particularly good fit to the rise of the 250-$\mu$m counts, sampled with very high statistical significance by the H-ATLAS data). The HerMES local luminosity functions (LLFs) of Vaccari et al. (2010) are also well reproduced by our model, as shown in the {\em right panels} of 
Figure~\ref{authorf_fig:cntspire}. At 500 $\mu$m, we notice a slight discrepancy between data and model, mainly in the source counts, where the model tends to underestimate the data at 30$\lsimeq$S$_{500}$$\lsimeq$100 mJy. In particular, the steep rise of the source counts starts at brighter flux densities ($\sim$100 mJy) in the data than in the model, the latter predicting a smoother rise up to $\sim$20--30 mJy. The model LLF at 500 $\mu$m is flatter than the data in the two lower luminosity bins, though the knee and the bright tail of the local luminosity function is well reproduced. 
The discrepancy between data and model observed at 500 $\mu$m could be due to the presence (even locally) of a ``cold'' population, whose SEDs are not well represented by the templates considered by our model (which are just extrapolated in the sub-mm regime, not using the {\em SPIRE} and {\em BLAST} data as constraints). 
Indeed, $z$$<$1 analogues of the ``cold'' $z$$>$1 sub-mm galaxies have already been found by Symeonidis et al. (2009) before {\em Herschel}, while Elbaz et al. (2010) find lower dust temperatures than previously inferred (due to the lack of constraints at FIR wavelengths before {\em Herschel}) in $z$$<$1--1.5 galaxies detected by {\em PACS} and {\em SPIRE}. Moreover, they also find SEDs ``colder'' than those of their local analogues for a significant fraction (10--20\%) of $z$$\sim$1 LIRGs and $z$$\sim$1.6 ULIRGs. On the other hand, Shultz et al. (2010),
by analysing the colours of the HerMES sources, discovered a population of red bright objects that may consist mostly of ``colder'' SEDs, but with a fraction ($>$12\%) of distant lensed ones. We would therefore need to further investigate the FIR/sub-mm SEDs of the IR populations: as soon as we could have access to the {\em SPIRE} data for large samples of galaxies, we will be able to properly fit the observed SEDs up to 500 $\mu$m, comparing, adjusting and modifying our templates to better reproduce the real Universe. In addition, since a considerable fraction of sub-mm bright sources are expected to be lensed by foreground galaxies (e.g. according to Negrello et al. 2007 all the 500-$\mu$m sources brighter than 100 mJy and with 2$<$$z$$<$3 are lensed; see also the recent H-ATLAS results of Negrello et al. 2010), the effect of lensing should  be properly taken into account when performing statistical studies (with both data and models) like source counts and luminosity functions.

\section{The AGN Contribution to the Infrared Emission}
\label{authorf_sec:agn}

As shown in the previous sections, from our model we expect a significant
contribution to the source counts and luminosity functions from objects
containing an AGN. However, AGNs and starbursts often co-exist in the same
object, while in the IR populations defined by our different SEDs, the real
AGN contribution is not disentangled from that due to star-formation. Here we
try to identify -- in a very simplified way -- the AGN contribution to the
monochromatic and total IR luminosity emitted by the different populations
considered in our model. To this purpose, we have decomposed, by mean of a
standard $\chi^{2}$ minimisation technique, the template SEDs 
into three distinct components: a stellar component emitting most of its power
in the optical/NIR, an AGN component due to hot dust heated by the central
black hole, and peaking in the MIR, and a starburst (SB) component representing the
major contribution to the FIR spectrum. The algorithm combines synthesis stellar
models built using the Padova evolutionary tracks (Bertelli et al. 1994), AGN dusty tori models from Fritz et al. (2006) and 6 empirical starburst SEDs
to reproduce the observed broad-band spectra. We refer to Pozzi et al. 2010 (see also
Hatziminaoglou et al. 2008, 2010; Vignali et al. 2009) for a detailed description of the method and of the individual model components.
In particular, we run the algorithm on the template SEDs representative of our AGN
populations ({\tt LLAGN}, {\tt AGN2}and {\tt AGN1}): Seyfert 2, Markarian
231 and TQSO1 of Polletta et al. (2007), respectively. 

\begin{figure}
  \begin{center}
    \includegraphics[width=8.5 cm,height=2.5cm]{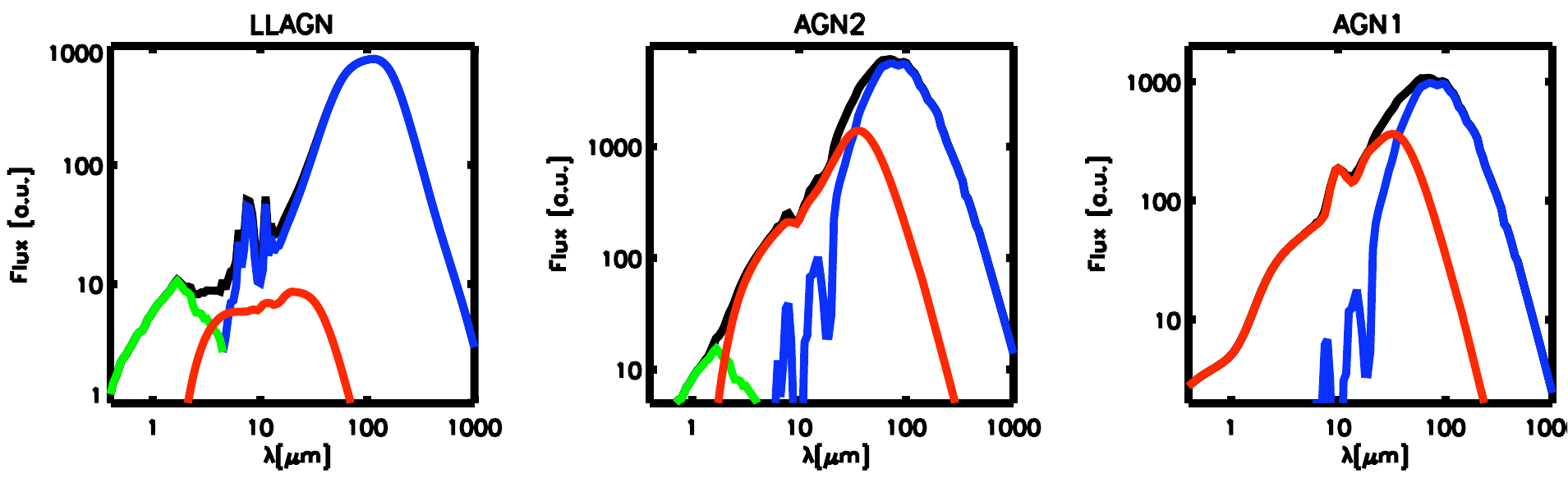}        
       \end{center}
  \caption{Templates representative of the three populations containing an AGN ({\em left}: {\tt LLAGN}; {\em middle}: {\tt AGN2}; {\em right}: {\tt AGN1}) decomposed into three different contributions: a stellar component emitting most of its power
in the optical/NIR (green), an AGN component due to hot dust heated by the central
black hole, and peaking in the MIR (red), and a starburst representing the
major contribution to the FIR spectrum (blue). The total SED is represented by the black solid line.}  
\label{authorf_fig:agntempl}
\end{figure}

In Figure~\ref{authorf_fig:agntempl} we show the results of the template
decomposition, with the AGN contribution highlighted in red. Note that the optical/NIR emission in the {\tt LLAGN} spectrum
is almost exclusively due to the stellar component from the host galaxy, while in the {\tt AGN1} spectrum 
the emission from the accretion disk is dominant in that wavelength range. The {\tt AGN2} SED shows an intermediate situation between {\tt LLAGN} and {\tt AGN1}, though apparently closer to that of the {\tt AGN1}, with the torus emission dominating in the whole MIR range. However, we note that the torus emission in the {\tt AGN2} spectrum, at odds with the {\tt AGN1} one, must be of ``type 2'' (i.e. edge-on), such as to reproduce the strong silicate absorption at 9.7 $\mu$m (typical of heavily obscured sources) observed in the {\em IRS} spectrum of our prototypical  {\tt AGN2} object Markarian 231 (Weedman et al. 2005). Though Markarian 231 is a well known heavily obscured BAL QSO, the Mrk231 ({\tt AGN2}) template reproduces the observed SED of  ``obscured'' AGNs regardless of their optical spectra (i.e. both type 1 and type 2 in the optical; Gruppioni et al. 2008) and also those of many Dust Obscured Galaxies (DOGs; Dey et al. 2008; Lanzuisi et al. 2009), with extreme MIR-to-optical ratios (Gruppioni, Vignali, Fritz et al. in preparation), likely to harbour obscured AGNs.
The modelled 9.7-$\mu$m feature is in absorption in the {\tt LLAGN}  SED, while it is in emission in the {\tt AGN1} one, as commonly observed in type 2 and 1 AGN spectra respectively (but see
i.e. Mason et al. 2009 for examples of silicate emission in type 2 AGNs).
Using these SED decompositions, we have derived the fractional contributions of AGN, starburst (SB) and evolved stars to the
monochromatic and bolometric (both 8--1000 $\mu$m and 1--1000 $\mu$m) IR luminosities  for the three representative template SEDs. 
These fractions, reported in Table~\ref{authorf_tab:agnsb}, have been used to disentangle the contribution to the total IR
luminosity function due to SB and AGN activity.
The evolved stars component in the decomposition of the {\tt AGN2} SED was necessary to reproduce the optical/NIR part of the spectrum, since the ``type 2'' torus needed for modelling the silicate feature in absorption, in the Fritz et al. (2006) model does not account for emission observed in these bands. Different tori models (i.e. the ``clumpy'' one of Nenkova et al. 2008) can explain the optical spectra of type 2 AGNs in terms of scattering contribution with no need for a stellar contribution. However, whether there is or not a stellar component in the {\tt AGN2} template is not relevant for our further analysis and conclusions and is beyond the scope of this work.
\begin{table}
 \caption{AGN, SB and stellar contribution to the IR luminosity}
\begin{tabular}{|l|c|c|c|c|}
\hline \hline
 SED &  $\lambda$ & L$_{\lambda}^{AGN}/$L$_{\lambda}$  & L$_{\lambda}^{SB}/$L$_{\lambda}$  &  L$_{\lambda}^{stars}/$L$_{\lambda}$ \\ 
        &   ($\mu$m)  &   (\%)  &  (\%)  &  (\%)  \\   \hline \hline 
           &  3.6   & 50 & 0 & 50 \\
            &  8.0  &  18 & 82 & 0 \\                      
 {\tt LLAGN}   & 24   &    12 & 88 & 0      \\
            &  100    &   0 & 100 & 0 \\ 
            & 8--1000 &   4 & 96 & 0 \\                   
            &  1--1000  &  9 & 68 & 23 \\  \hline                    
           &  3.6    & 93 & 0 & 7 \\
            &  8.0    & 91 & 8 & 1 \\                      
{\tt AGN2}   & 24   &  66 & 34 & 0   \\
            &  100    &  3 & 97 & 0 \\                      
            & 8--1000 &   36 & 64 & 0 \\                   
            &  1--1000    &  44 & 52 & 4 \\ \hline
          &  3.6    & 100 & 0 & 0 \\
           &  8.0    & 97 & 3 & 0   \\                      
 {\tt AGN1}  & 24   &   79 & 21 & 0       \\
            &  100    &  4 & 96 & 0 \\                      
            & 8--1000 &   54 & 46 & 0 \\                   
           &  1--1000    &  69 & 31 & 0 \\                      
 \hline \hline
\end{tabular}
\label{authorf_tab:agnsb}
\end{table}

\subsection{Star-Formation and Accretion History}
\label{authorf_sec:sfh}
We have used the LFs provided by our model, and the SED decomposition described
above, to estimate the cosmic evolution of the star-formation density (SFD), $\rho_{SFR}(z)$ (derived from the total comoving IR luminosity density due to star-formation, $\rho_{IR}^{SF}(z)$), and of the SMBH accretion density, $\Psi_{BHAR}(z)$, as a function of 
redshift. The comparison between these two quantities is a crucial tool for understanding
galaxy and AGN evolution and the role played by the AGNs in the formation of
galaxies. 
\begin{figure}
  \begin{center}
    \includegraphics[width=8cm]{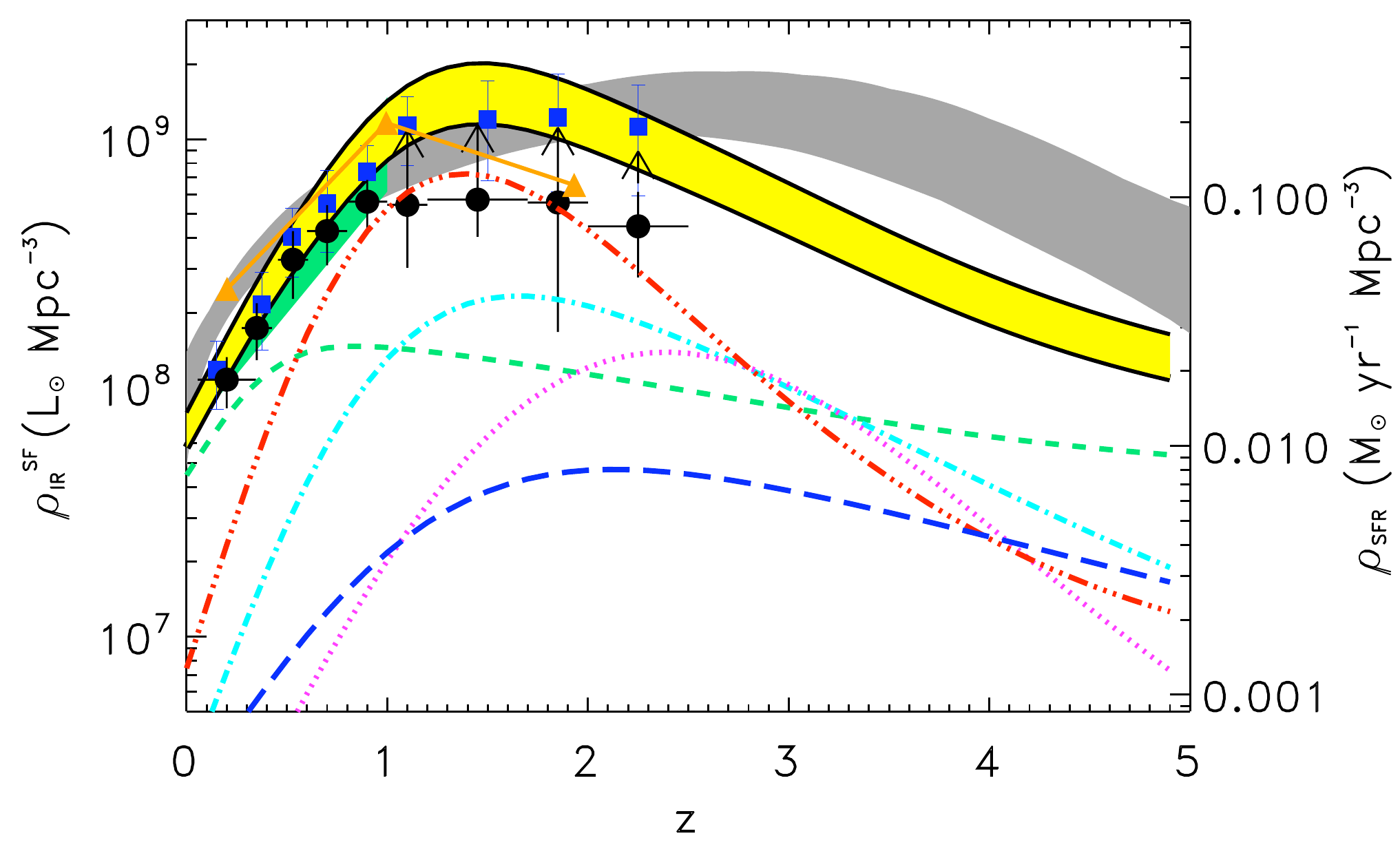}        
       \end{center}
  \caption{Redshift evolution of the total IR luminosity density $\rho_{IR}^{SF}(z)$ (and SFD, $\rho_{SFR}$) as expected from our model (yellow shaded area) and compared with estimates from IR surveys (green shaded area: Le Floc'h et al. 2005; orange filled triangles: Caputi et al. 2007; blue filled squares: Rodighiero et al. 2010; black filled circles: Gruppioni et al. 2010, with the arrows standing for lower limits due to luminosity incompleteness). 
 The contributions to $\rho_{IR}^{SF}$ (or $\rho_{SFR}$) from the different populations are shown as green dashed line, cyan dot-dashed line, red dot-dot-dot-dashed lines,  magenta dotted line and blue long-dashed line for the {\tt spiral}, {\tt starburst}, {\tt LLAGN}, {\tt AGN2} and {\tt AGN1} class respectively.
We have calculated the upper and lower limits of our total SFD expectation by considering as AGNs only those spectroscopically confirmed and all the SED-fitting derived ones respectively (see Gruppioni et al. 2008). On the right y-axis we report the SFD obtained by means of the Kennicutt (1998) conversion and, for comparison, we plot (as grey shaded area) the evolution of the SFR density $\rho_{SFR}$ obtained from a large compilation of extinction-corrected UV/optical and H$\alpha$ data by Hopkins \& Beacom (2006). Given the lack of data constraints at high-$z$, our model expectations are highly uncertain at
$z$$>$2.5.}  
\label{authorf_fig:sfdz}
\end{figure}
\noindent We have estimated $\rho_{IR}^{SF}(z)$  as follows:
\begin{equation}
\rho_{IR}^{SF}(z)=\int_{0}^{\infty}{L_{8-1000}^{SB}  \phi(L_{8-1000}) dlog L_{8-1000} }\label{equation:rhoir}
\end{equation}
\noindent where ${L_{8-1000}^{SB}}$ is the 8--1000 ${\mu}m$ IR luminosity due to star-formation.
The SFD has been derived from the IR luminosity, according to the conversion
of Kennicutt (1998): $\rho_{SFR}$$=$1.7${\times}10^{-10}$ $\rho_{IR}$ [$M_{\odot}$ yr$^{-1}$ Mpc$^{-3}$] (for a Salpeter IMF).

In Figure~\ref{authorf_fig:sfdz} we show $\rho_{IR}^{SF}(z)$ as expected from our model and compared with estimates from different IR surveys
(i.e. at 24 $\mu$m: Le Floc'h et al. 2005; Caputi et al. 2007; Rodighiero et al. 2010; at 100 $\mu$m and 160 $\mu$m: Gruppioni et al. 2010). 
To derive the uncertainties of our calculations, we have considered as upper limit for the AGNs the number of AGNs derived from the SED fitting analysis and as lower limit the fraction of AGNs spectroscopically confirmed through optical emission lines (see La Franca et al. 2004; Gruppioni et al. 2008). By considering these values we obtain the lower and the upper limit to the estimate of $\rho_{IR}^{SF}(z)$ respectively.
We have also compared the SFD evolution obtained from the IR to that derived from optical/UV and H$\alpha$ observations, presented in a
large and homogenised collection from different surveys by Hopkins \& Beacom (2006). 
Our model expectation is in very good agreement with all the IR data estimates, confirming the rapid increase of $\rho_{IR}^{SF}$ up to $z$$\sim$1. The increase of the IR luminosity density is followed by a peak at 1$<$$z$$<$2 and by a decrease from $z$$\sim$2--2.5 towards the higher redshifts. At the lower redshifts ($z$$<$0.3) the IR luminosity density is dominated by the {\tt spiral} population, while in the 0.3$<$$z$$<$2--2.5 range, the principal contributors to $\rho_{IR}^{SF}$ are galaxies with a {\tt LLAGN} SED. Pure {\tt starburst} galaxies are also important in the same redshift interval, but are never dominant at any $z$: at $z$$>$2.5 the {\tt AGN2} SED objects start dominating up to $z$$>$4, when they are overtaken by the {\tt AGN1} population.   
Therefore, galaxies likely to contain a low-luminosity AGN and galaxies powered by a starburst are responsible for the peak of the IR luminosity density at $z$$\sim$1--2, then galaxies hosting increasingly powerful AGNs become increasingly important towards the higher $z$'s. 
  
The total IR emissivity predicted by our model at $z$$>$2 is lower and shows a faster convergence than previously published results based on UV/optical or H$\alpha$ observations (i.e. Hopkins \& Beacom, 2006), which are subject to large extinction corrections, or on 24-$\mu$m data (i.e. Perez-Gonzalez 2005), which require large spectral extrapolations to derive the total IR luminosity, especially at high redshifts. Similar high-$z$ convergence of the total IR luminosity density has been found recently by Franceschini et al. (2010) with a model based on the analysis of a large IR database of high-redshift galaxies at long wavelengths.  However, up to now the total emissivity of IR galaxies at high redshifts is poorly constrained, due to the scarcity of {\em Spitzer} galaxies at $z$$>$2 and the incomplete information on
the $z$-distribution of sub-mm sources (Chapman et al. 2005). At high redshifts also our model expectations are very uncertain, since no constraint from data at $z$$>$2--2.5 are
available from pre-{\em Herschel} Surveys. The large numbers of high-$z$ galaxies provided by the deep Surveys performed with {\em Herschel}
(and furtherly with {\em SPICA}) will be crucial to assess galaxy and AGN evolution in the IR at $z$$>$2.5.

The SMBH accretion rate can be derived once the bolometric luminosity
function $\Phi(L_{bol}^{AGN}, z)$ is known, where $L_{bol}^{AGN}={\epsilon}_{rad}{\dot M} c^{2}$ is the intrinsic
bolometric luminosity produced by a SMBH accreting at a rate of ${\dot M}$
with a radiative efficiency ${\epsilon}_{rad}$. Crucial factors in the determination of  $\Phi(L_{bol}^{AGN}, z)$ (Hopkins, Richards \& Hernquist 2007; Merloni \& Heinz 2008) are the completeness of any AGN survey and a suitable
correction to estimate, from observations in one band, the AGN bolometric
luminosity (i.e. bolometric correction $BC$). Observations in the hard X-ray band
are commonly used, since the X-ray surveys provide a relatively unbiased census
of AGN in the Universe. 
\noindent Here we estimate, for the first time, the SMBH accretion from the
IR emission originating from the circumnuclear dusty material that intercepts a fraction of the
inner optical-UV thermal accretion disk emission. While the nature of the
dusty material is still a matter of debate (i.e. smooth vs. clumpy distribution),
dust around the BH is observed in almost all the AGNs (see the review of Elitzur 2008 and references therein). 
\begin{figure}
  \begin{center}
    \includegraphics[width=8 cm]{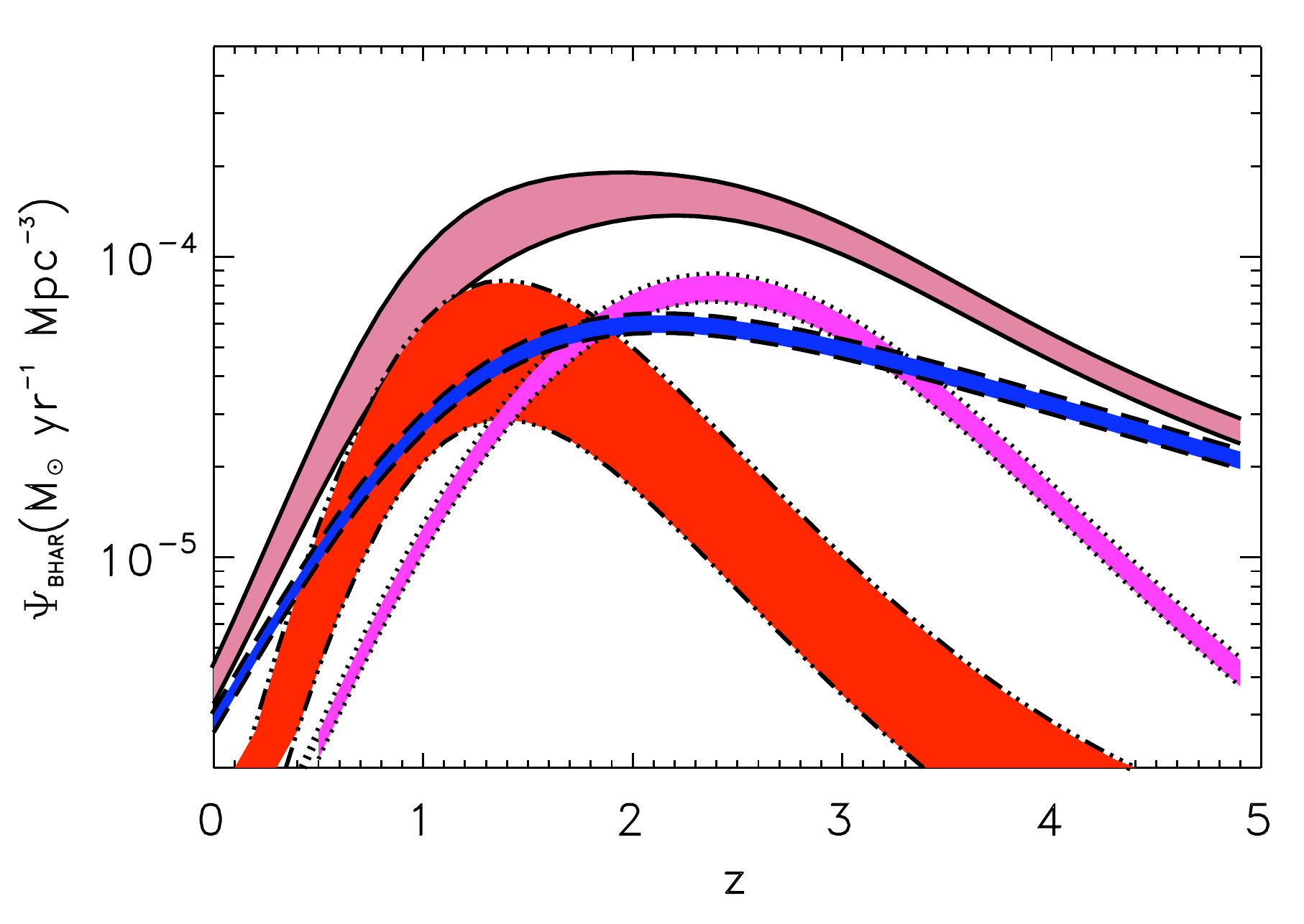}        
       \end{center}
  \caption{Redshift evolution of the BHAR density, $\Psi_{BHAR}(z)$, estimated from our IR analysis. 
  Pink shaded area (solid lines) shows the total accretion rate, while the red (dot-dot-dot-dashed lines), blue (long dashed lines)
  and magenta (dotted lines) show the accretion rate due to {\tt LLAGN}, {\tt AGN1} and {\tt AGN2} objects respectively. 
  The uncertainty regions for each population have been obtained by considering AGNs all the SED-fitting derived ones and only those spectroscopically confirmed (see Gruppioni et al. 2008) in the calculation of the upper and lower envelopes respectively. 
}  
\label{authorf_fig:bharz}
\end{figure}
\noindent As bolometric correction $BC$ we used a value of $\sim$1.5 for the {\tt AGN1}
and {\tt AGN2} templates, and $BC$$\sim$2 for the {\tt LLAGN} template.
These bolometric corrections are first-order empirical estimates derived in
Pozzi et al. (2007), where the broad-band SEDs of a sample of X-ray
selected AGN have been studied. As
explained in Pozzi et al. (2007), these corrections take into
account the geometry of the torus (i.e. covering factor $f$) and the effects of
orientation (i.e. effective optical depth $\tau_{9.7{\mu}m}$ of the dust along the line of sight). The geometry correction is based on statistical arguments, employing the
ratio between obscured and unobscured quasars as required by recent X-ray background
synthesis models (Gilli et al. 2007) to infer a typical torus covering
factor of $f\approx0.67$ (hence $BC$$=$1.5). As reported by Vasudevan et
al. (2010), this value is also consistent with the covering
factor obtained from recent detailed clumpy torus models
(i.e. Nenkova et al. 2008) for typical torus parameters (e.g. number of
line-of-sight clouds $\sim5$, opening angle $\sim$30-45$^{\circ}$).
The effect of orientation was computed from the ratio between the luminosities of a
face-on and an edge-on AGN using the AGN SEDs of Silva et al. (2004).
\noindent Consistent results were derived in Pozzi et al.(2010) using a more
sophisticated approach, i.e. computing the bolometric corrections using the best-fitting torus models,
as the ratio between the accretion disk luminosity given as input to the radiative transfer
model and the observed reprocessed infrared luminosity in output (see Fig. 4 from the pioneering
work of Pier \& Krolik 1992). Since the latter approach implies an exploitation of the degeneracy of the torus
solutions which is beyond the scope of the present work, we prefer the first, simpler
and straightforward, method. 

\noindent The expression used for estimating the SMBH accretion (BHAR) is:
\begin{equation}
\Psi_{BHAR}(z)=\int_{0}^{\infty}{ \frac{(1-\epsilon_{rad})~ BC ~ L_{1-1000}^{AGN}}{\epsilon_{rad} c^2}
   \phi(L_{1-1000}) dlog L_{1-1000}}
\end{equation}
\noindent where $BC$ is the bolometric correction to the 1--1000 ${\mu}$m IR
luminosity depending on the SED type and  ${L_{1-1000}^{AGN}}$  indicates the
1--1000 ${\mu}m$ IR luminosity due to the AGN. For $\epsilon_{rad}$ we have assumed the
canonical value of 0.1 (see e.g. Hopkins, Richards \& Hernquist 2007), but changing this value would simply result in a change of the overall normalization.
\begin{figure}
  \begin{center}
    \includegraphics[width=8 cm]{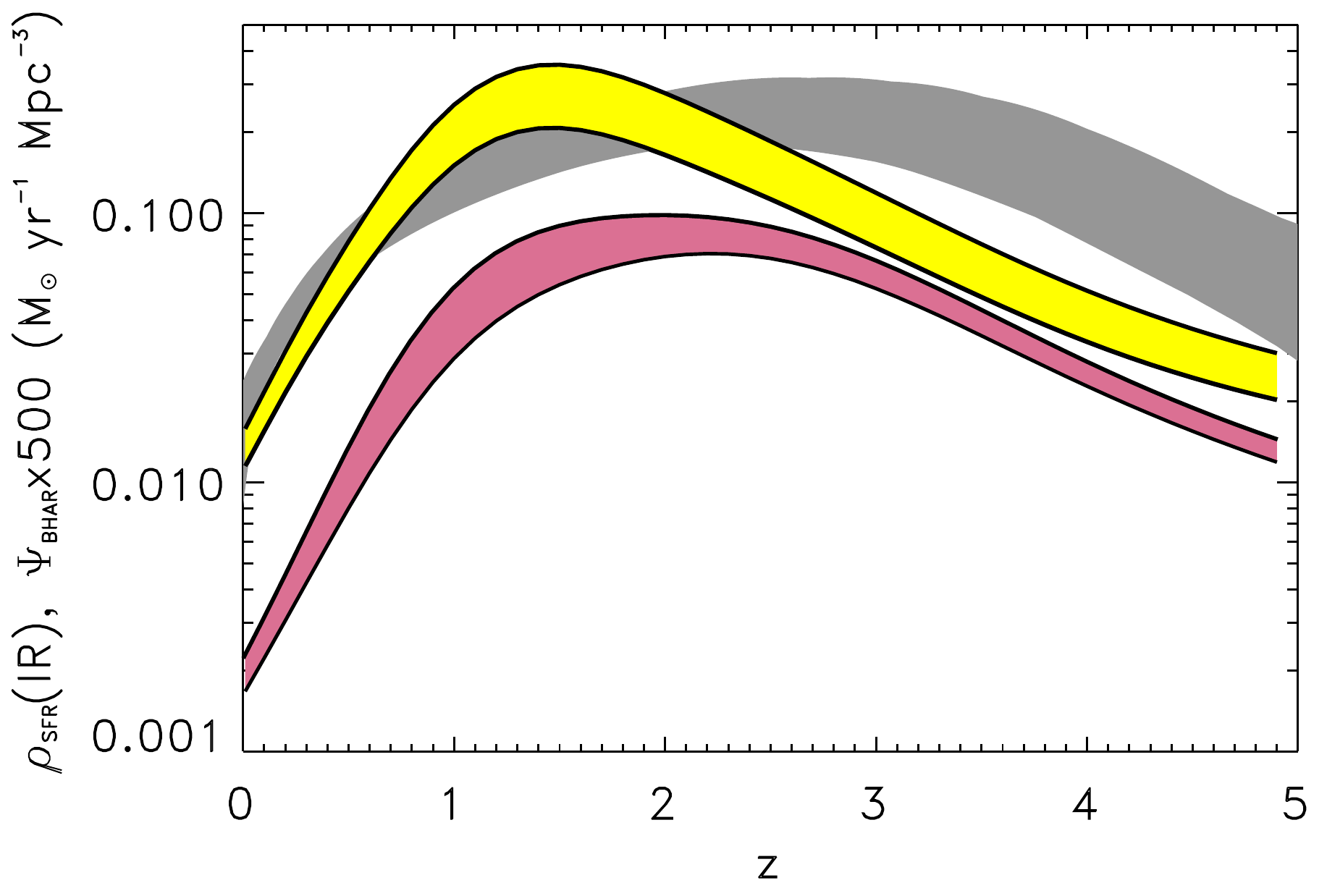}        
       \end{center}
  \caption{Redshift evolution of the BHAR density $\Psi_{BHAR}(z)$, multiplied by a constant factor (equal to 500) is plotted 
as a pink shaded area alongside to our SFD derivation shown in Figure~\ref{authorf_fig:sfdz} (yellow shaded area) and to the best fit to a large compilation of optical/UV data for the SFD (grey shaded area, from Hopkins and Beacom 2006). 
}  
\label{authorf_fig:sfd_bhar}
\end{figure}
In Figure~\ref{authorf_fig:bharz} we show the predicted $\Psi_{BHAR}(z)$, with the contribution from the different AGN populations.
The uncertainty region of the single contributions and of the total estimate have been obtained by considering, as for the
SFD, the SED-fitting and spectroscopic fractions of the different AGN types respectively as upper and lower limits in our calculations. We note that for the {\tt AGN1} and {\tt AGN2} SED objects the difference between the numbers derived by SED-fitting and by optical spectroscopy is small, while for the {\tt LLAGN} the fraction of spectroscopically confirmed AGNs is only about 30\% of the SED-fitting one (see Gruppioni et al. 2008). However, even considering the lower limits for objects containing an AGN, which conservatively takes into account the possibility that a significant fraction of our {\tt LLAGN} are not AGNs, but starburst galaxies not well reproduced by our limited set of templates, the global result of our analysis does not change significantly.\\
The {\tt AGN1} contribution dominates the BHAR, especially at low ($z$$<$0.3) and high redshifts ($z$$>$3), with the
{\tt LLAGN} and the {\tt AGN2} population's accretion density peaking at 1$<$$z$$<$2 and 2$<$$z$$<$3 respectively and reaching the
{\tt AGN1} accretion values just in these redshift ranges. The BHAR obtained from our IR estimate is
reasonably consistent in shape with previous derivations from X-rays (i.e. Merloni, Rudnick \& di Matteo 2004; Merloni \& Heinz 2008), though it is about a factor of $\sim$2 higher at 1$\lsimeq$$z$$\lsimeq$3, where the BHAR peaks. Note that the derivation of BHAR from IR luminosity is completely independent from that obtained from X-ray data and requires bolometric corrections at least a factor of $\sim$10 lower and probably less uncertain than those required from observations in the X-ray band ($\sim$1.5--2 against $\sim$10--40 in the X-rays). Moreover, to
derive the BHAR from X-rays, substantial assumptions need to be made regarding the number and redshift distribution of compton-thick AGNs missed by the X-ray surveys. These obscured objects should instead be observable in the IR and are, in principle, already included in our calculations. The main source of uncertainty in our approach is due to the identification of sources containing a {\tt LLAGN}: in our calculations we have considered as lower value that obtained by considering as {\tt AGN} only those spectroscopically confirmed. Only future space missions with high resolution spectrometers in the MIR
and FIR range, like {\em MIRI} (Wright et al. 2004) on {\em JWST} and {\em SAFARI} (Swinyard et al. 2009) on {\em SPICA}, will be able to definitely confirm or reject the presence of low-luminosity AGNs inside objects classified as {\tt LLAGN} on the basis of their SEDs. 

The value of local BH accreted mass that we obtain by integrating our $\Psi_{BHAR}(z)$ over time
(i.e. $\rho_{BH,0}$=$\int_{0}^{\infty}{\Psi_{BHAR}(z) (dt'/dz) dz}$, with $dt/dz$$=$1$/[H_0 (1+z) \sqrt{\Omega_m (1+z)^3+\Omega_{\Lambda}}$] being the differential cosmic time as a function of redshift) is $\rho_{BH,0}$$=$6.5--9.2$\times$10$^5$ M$_{\odot}$ Mpc$^{-3}$. Despite the large uncertainties and the simplifying assumptions in our calculation, we find that our value of $\rho_{BH,0}$ is in broad agreement with (although somewhat on the high side of) previous derivations from X-rays, of 9$\times$10$^5$ M$_{\odot}$ Mpc$^{-3}$ ($0.1/\epsilon_{rad}$) (Barger et al. 2001), of (7.5--16.8)$\times$10$^5$($0.1/\epsilon_{rad}$) M$_{\odot}$ Mpc$^{-3}$  (Elvis, Risaliti \& Zamorani 2002), 4.81$^{+1.24}_{-0.99}$$\times$10$^5$($0.1/\epsilon_{rad}$) M$_{\odot}$ Mpc$^{-3}$ (Hopkins, Richards \& Hernquist 2007) and (3.2--5.4)$\times$10$^5$ M$_{\odot}$ Mpc$^{-3}$ (Shankar, Weinberg \& Miralda-Escud\'e 2009).

\section{Predictions for Future Surveys with {\em JWST-MIRI} and {\em SPICA-SAFARI}}
\label{authorf_sec:pred}
\begin{figure*}
  \begin{center}
    \includegraphics[width=7.5cm]{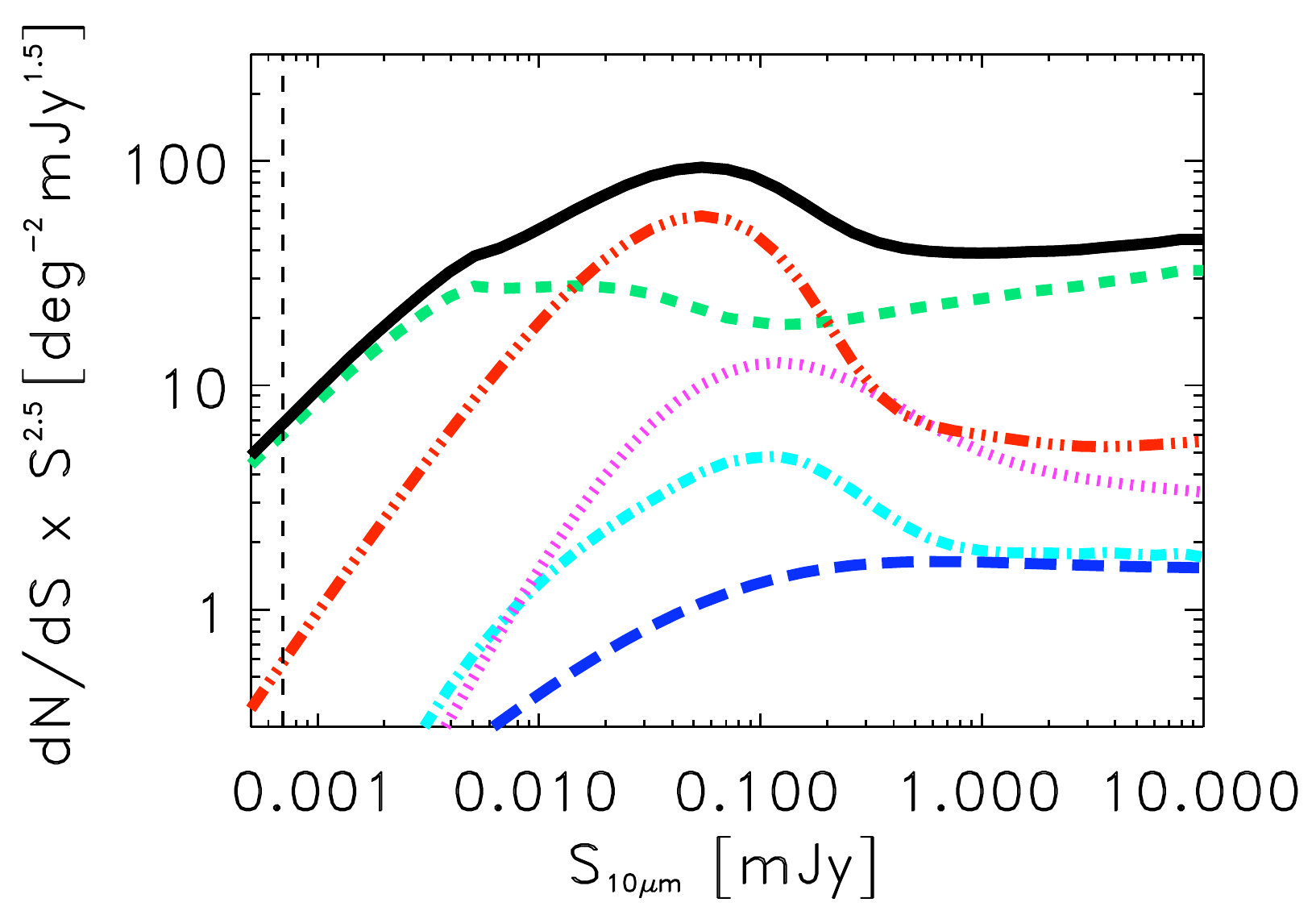}
    \includegraphics[width=7.cm,height=5cm]{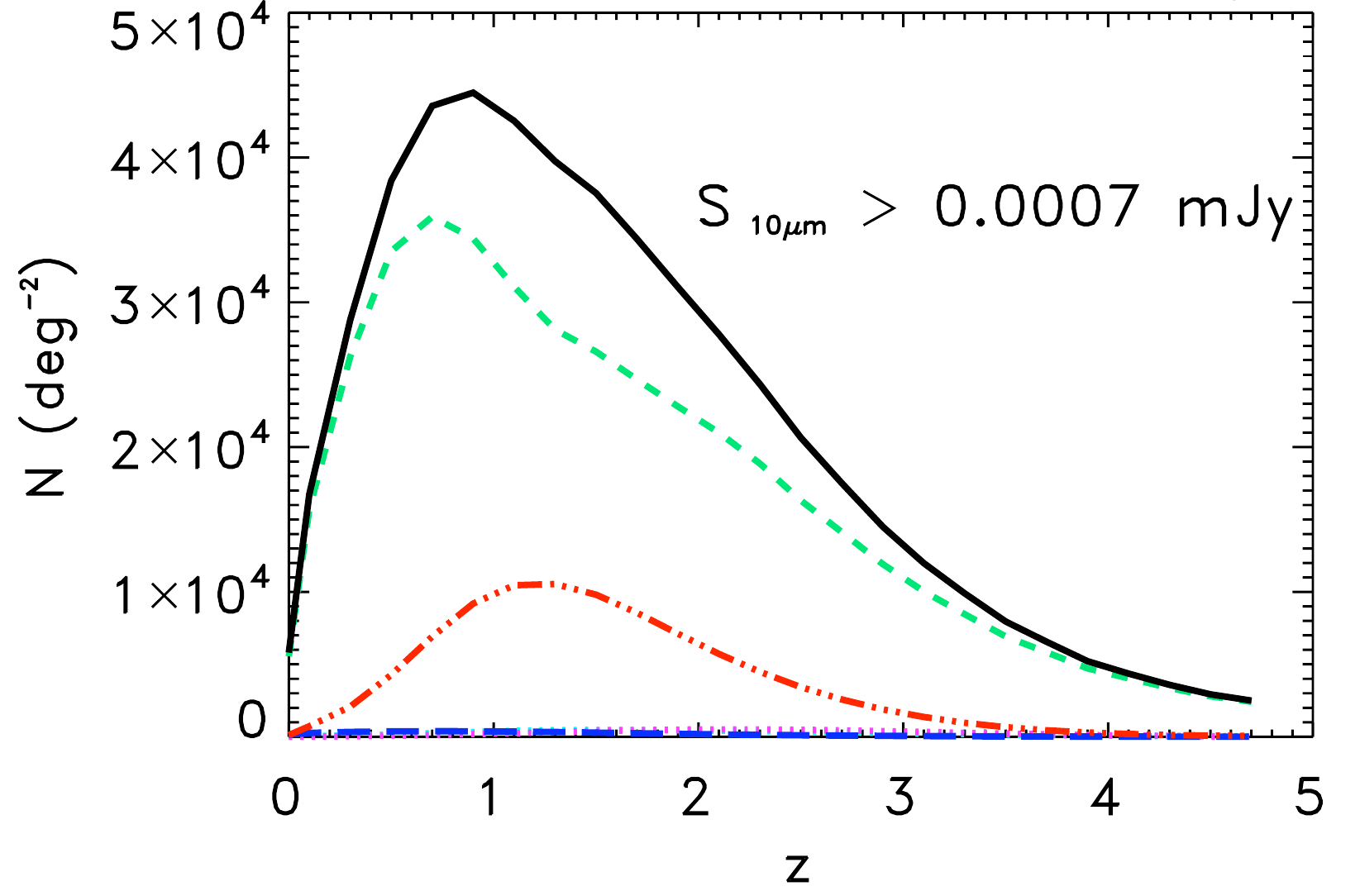}   
    \includegraphics[width=7.5cm]{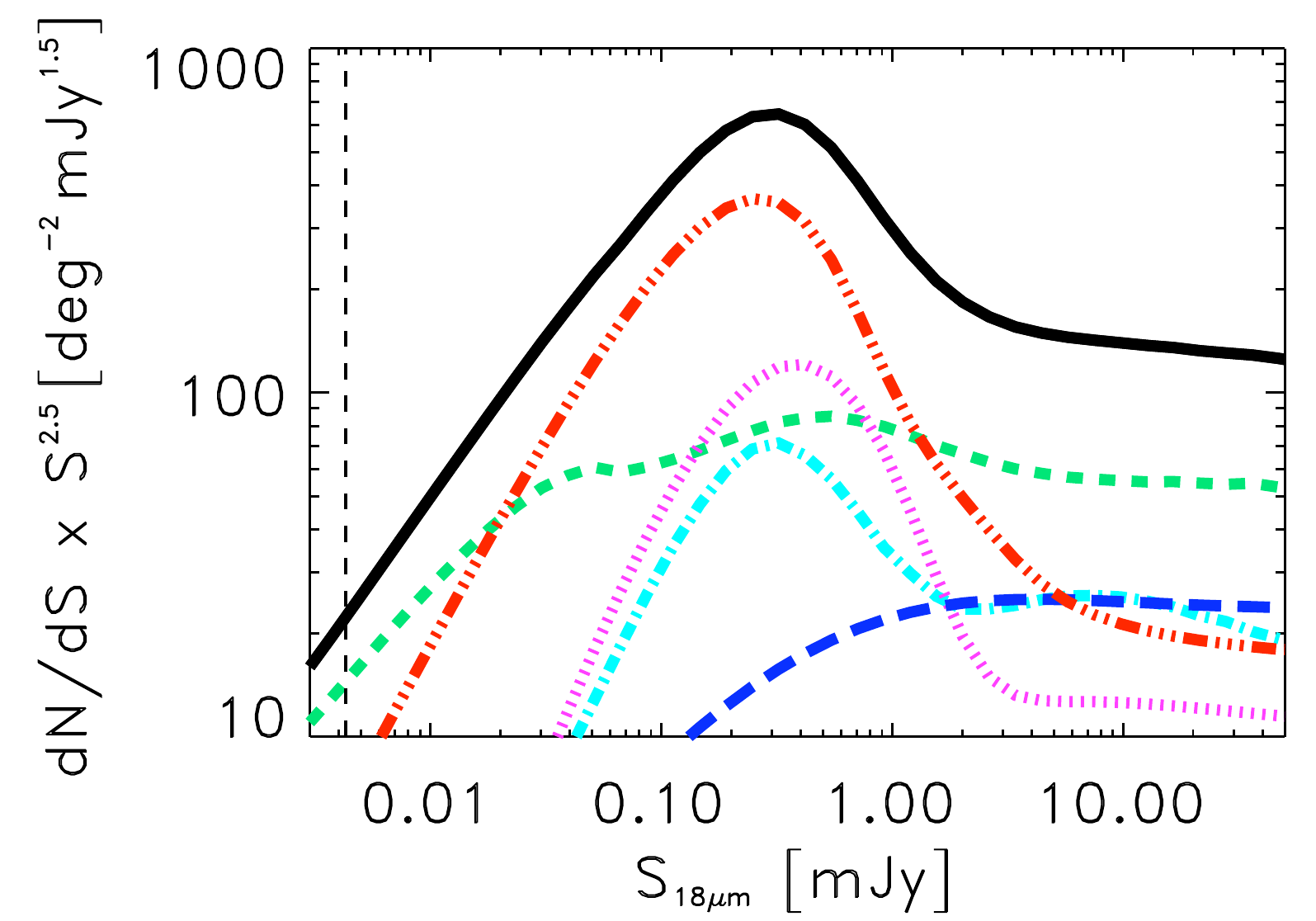}
    \includegraphics[width=7.2cm,height=5cm]{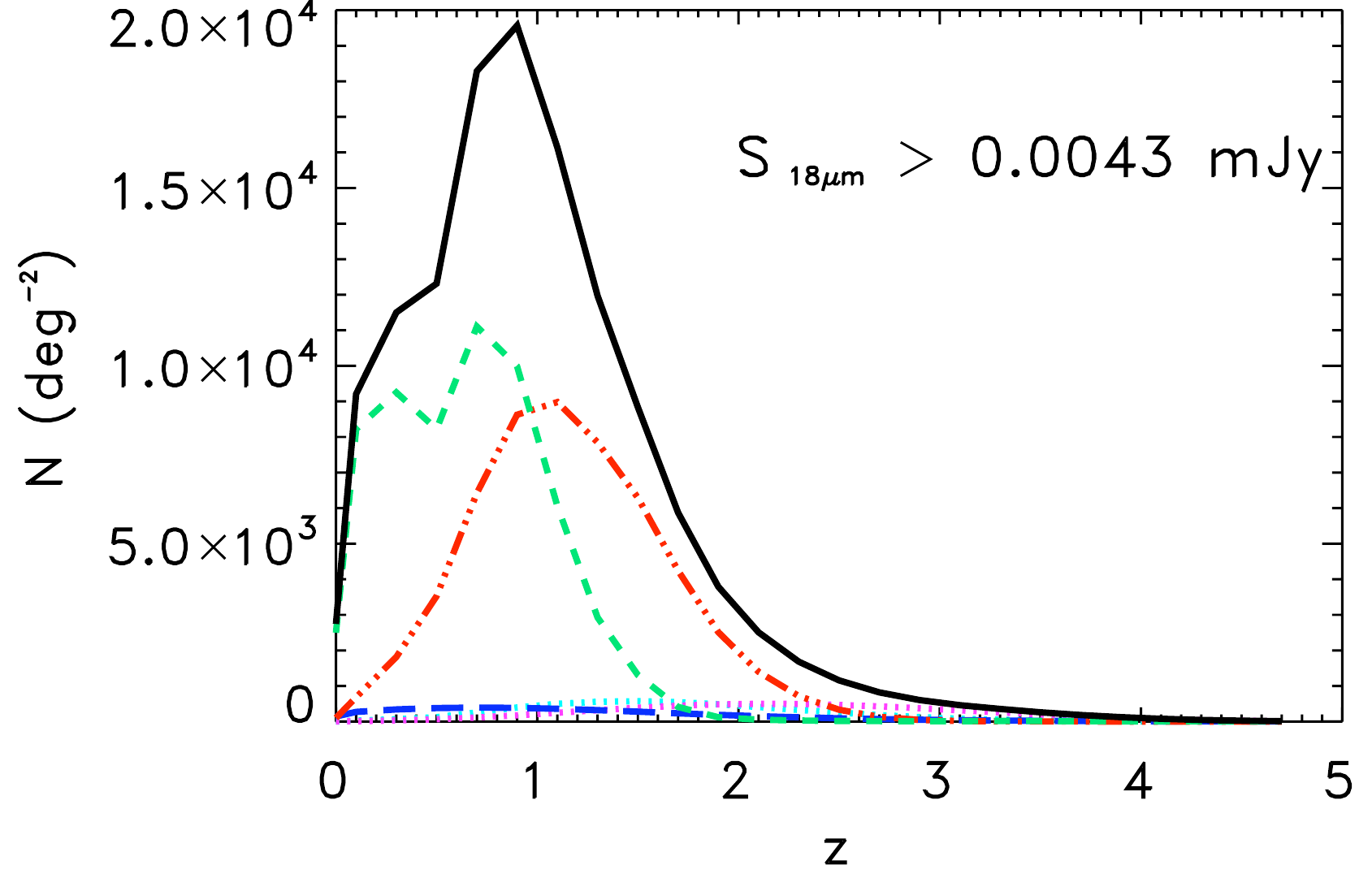}
    \includegraphics[width=7.5cm]{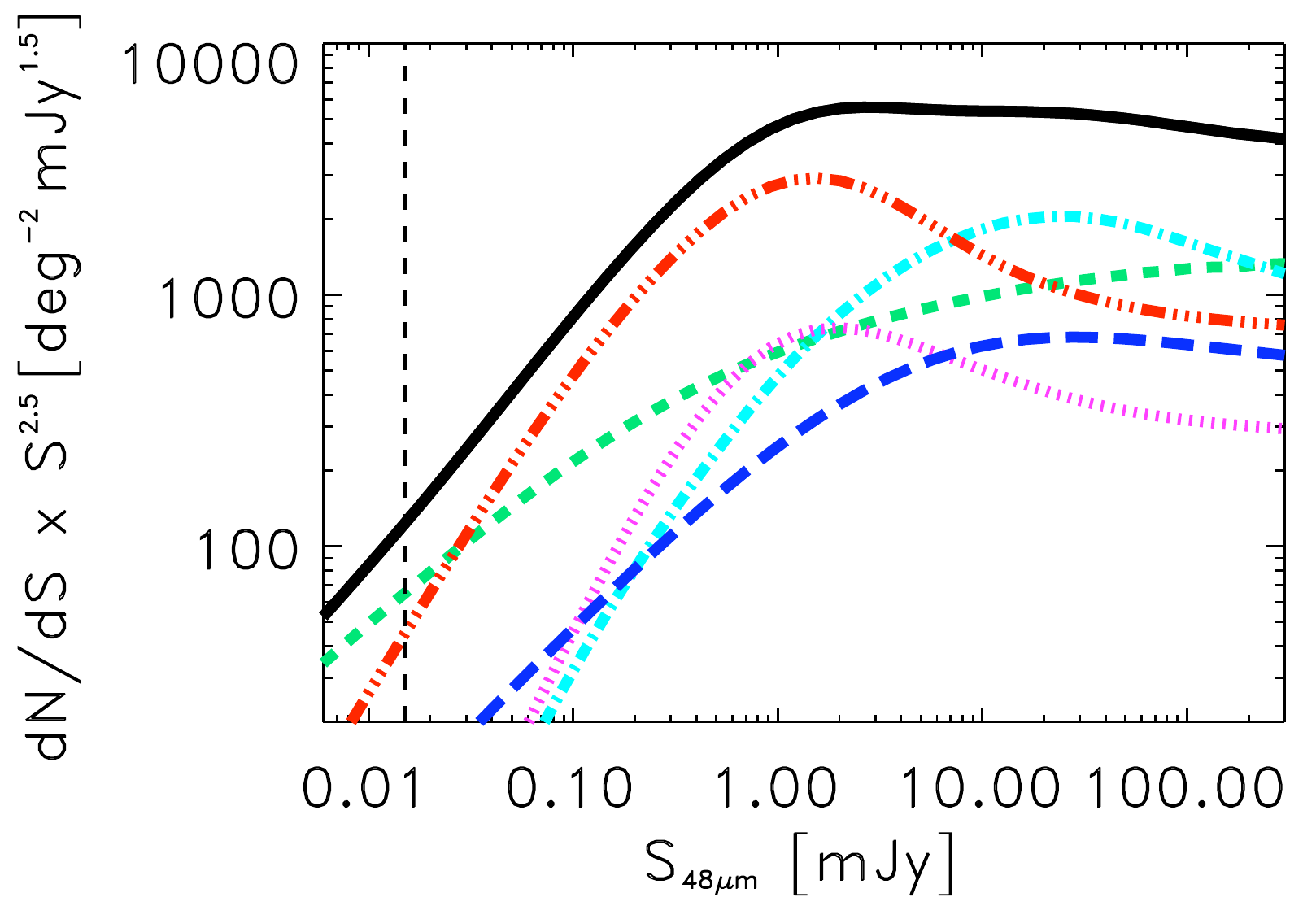}
    \includegraphics[width=7.2cm,height=5cm]{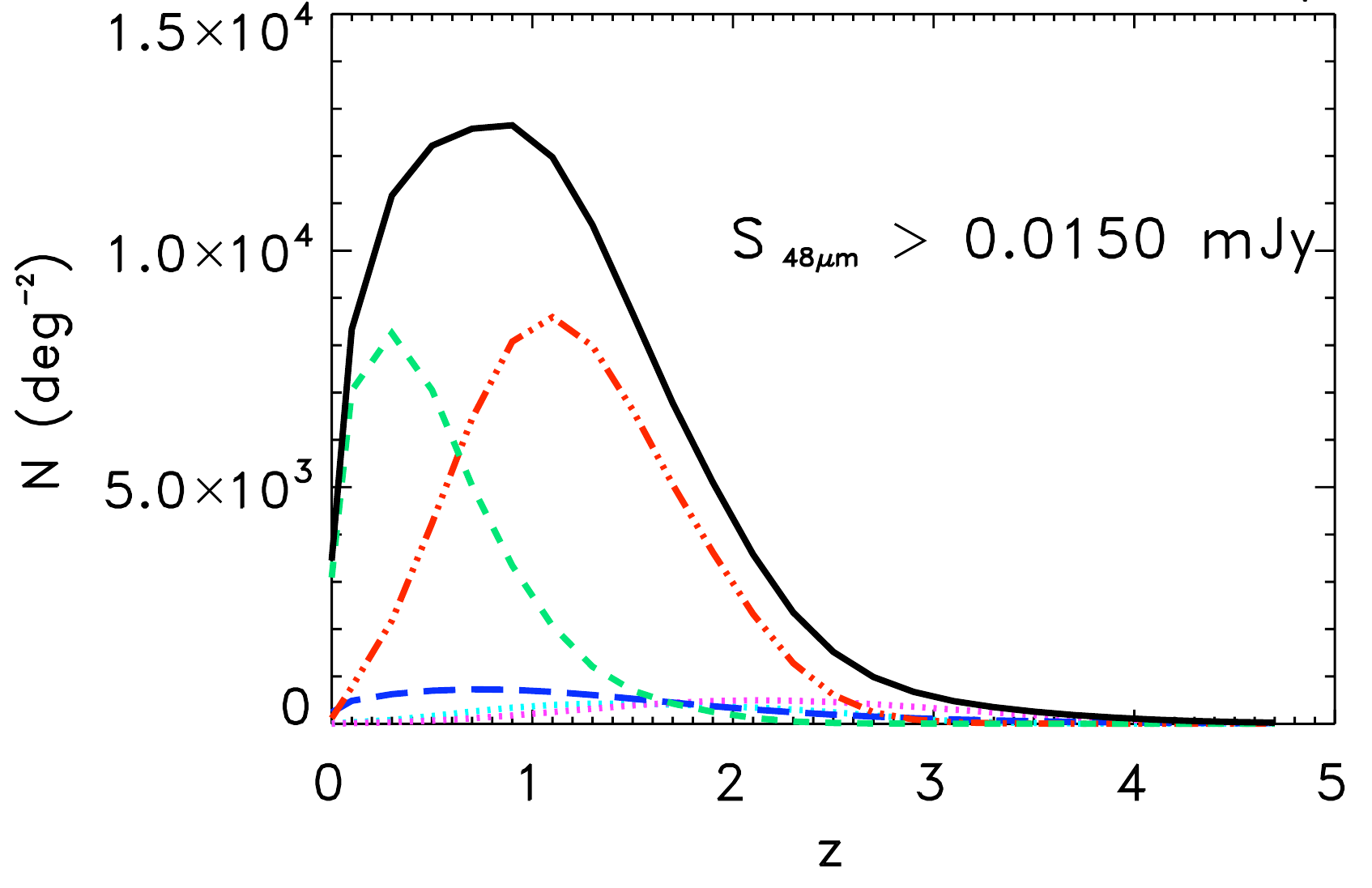}
    \includegraphics[width=7.2cm]{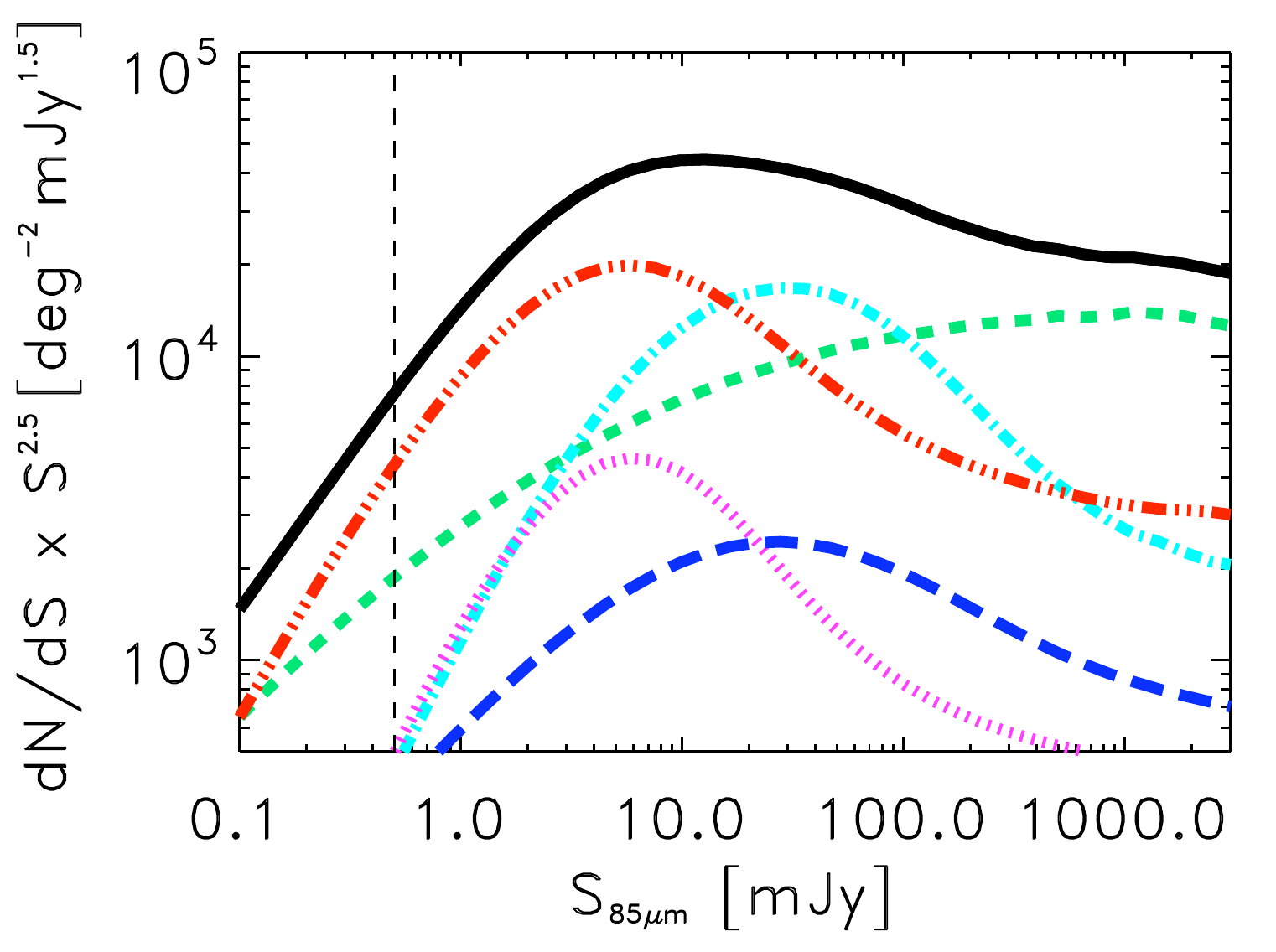}
    \includegraphics[width=7.1cm,height=5cm]{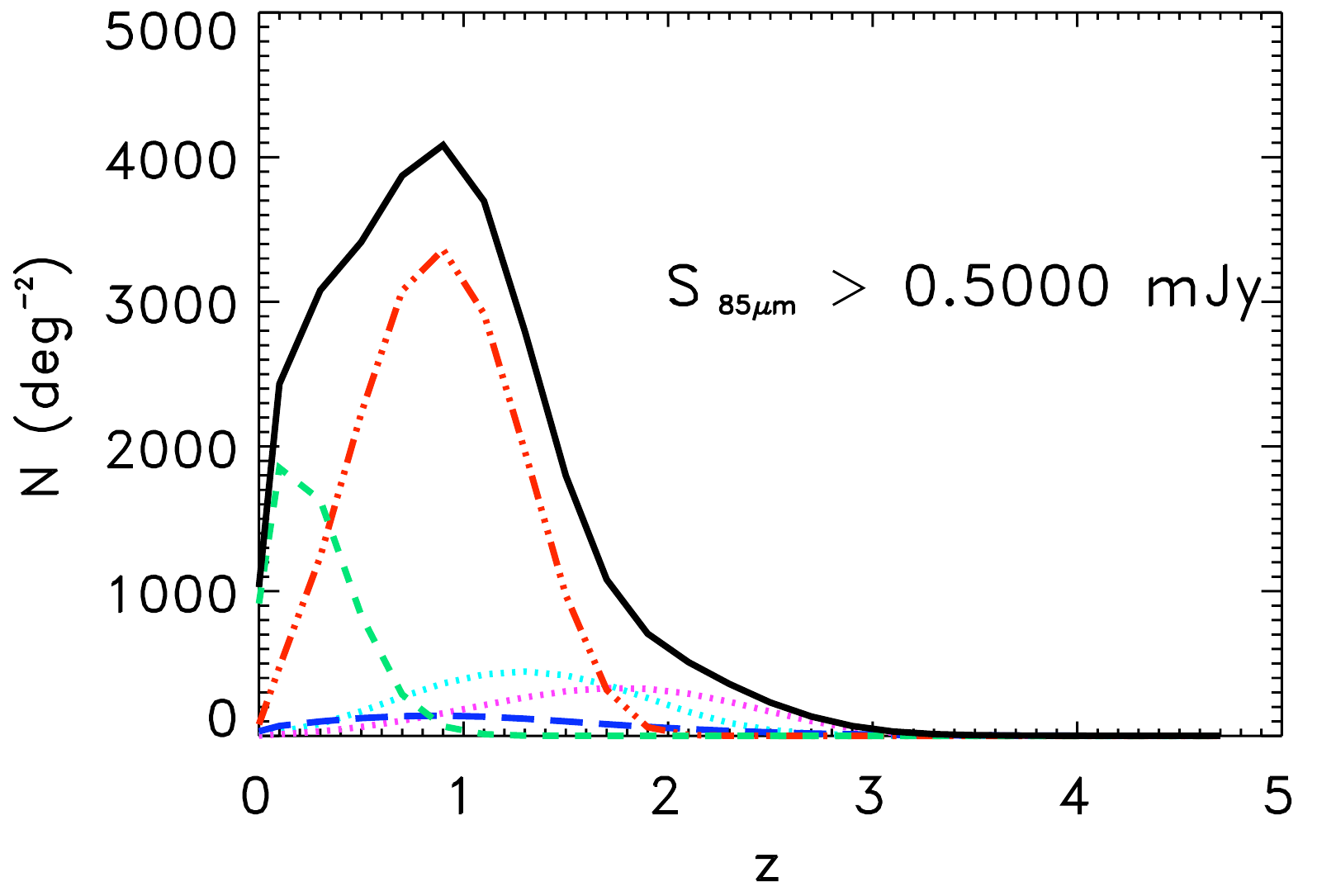}    
     \end{center}
  \caption{Differential extragalactic source counts normalized to the Euclidean slope ($left$) and redshift distributions ($right$) as predicted by our model in four selected bands that will be covered by future Surveys with {\em MIRI} (i.e. 10 and 18 $\mu$m) and  {\em SAFARI} (i.e. 48 and 85 $\mu$m). The redshift distributions have been simulated to the estimated point source 10$\sigma$ detection level for a 10$^4$ seconds exposure with {\em MIRI} at 10 and 18 $\mu$m (provided by the {\em MIRI} science team at the official {\em MIRI} web pages) and close to the estimated {\em SAFARI} confusion limits (given the specification of the {\em SPICA} Study Team Collaboration, 2009, and the recent proposed rescope of the telescope) at 48 and 85 $\mu$m. These limits are shown by vertical dashed lines in the {\em left} panels.}
  \label{authorf_fig:cntmirisaf}
\end{figure*}

We have used the model described above to make predictions for surveys to be performed with the {\em MIRI} and {\em SAFARI} instruments on board of the {\em JWST} and of the JAXA-ESA satellite {\em SPICA} respectively. For our simulations, we chose two representative bands of {\em MIRI} (i.e. 10 and 18 $\mu$m) and two of {\em SAFARI} (i.e. 48 and 85 $\mu$m). In Figure~\ref{authorf_fig:cntmirisaf} we show the extragalactic differential source counts and the redshift distributions expected in the four bands. The adopted flux limits in our simulations are the estimated 10$\sigma$ detection level for a 10$^4$ seconds exposure with {\em MIRI} at 10 and 18 $\mu$m (0.7 $\mu$Jy and 4.3 $\mu$Jy respectively, as provided by the {\em MIRI} science team at the official {\em MIRI} web pages) and close to the estimated {\em SAFARI} confusion limits at 48 and 85 $\mu$m ($\sim$0.015 mJy and $\sim$0.5 mJy respectively, estimated by means of our model and based on the telescope specification given by the {\em SPICA} Study Team Collaboration, 2009, and on the recent proposed rescope of the {\em SPICA} telescope, as the change of the primary mirror diameter from 3.5 to 3.2 m).
To these limits, we expect to detect $\sim$6$\times$10$^5$ sources/deg$^{2}$ at 10 $\mu$m and $\sim$1.5$\times10^5$ sources/deg$^{2}$ at 18 $\mu$m with {\em MIRI}, while $\sim$2$\times10^5$ and $\sim$8$\times10^4$ sources/deg$^{2}$ with {\em SAFARI} at 48 and 85 $\mu$m respectively.\\
The deep Surveys that will be performed with SAFARI at 85 $\mu$m (i.e. 60-110 $\mu$m band) and 160 $\mu$m (i.e. 110-210 $\mu$m band) will easily reach the confusion limits. In particular, given the nominal field of view of SAFARI of 2$^{\prime}\times2^{\prime}$, at 85 $\mu$m it will be possible to observe down to confusion a field as large as one of the GOODS areas ($\sim10^{\prime}\times15^{\prime}$) in just 2.5 minutes and a COSMOS-like field (2 deg$^2$) in $\sim$2 hours (not considering the overheads). On the other hand, at 48 $\mu$m (i.e. 35-60 $\mu$m band) about 26 hours would be needed to cover one field like GOODS and $\geq 1000$ hours for a larger field like COSMOS to 0.015 mJy (timely prohibitive). These confusion limited surveys with SAFARI will be $> 3-10$ times deeper than the deepest GTO Survey with Herschel PACS (i.e. PEP 100 $\mu$m Survey in the GOODS-S to 1.7 mJy), detecting, in a GOODS-size field, hundreds of $L_{IR}$$<$10$^{10}$ L$_{\odot}$ galaxies at $z$$\leq$1 and few hundreds of LIGs at $z$$\sim$2.  In addition, a confusion limited survey in the 35-60 $\mu$m band will detect also of the order of 150 AGNs (all with $L_{IR}$$\gsimeq$10$^{11}$ L$_{\odot}$) at $z$$>$3. The 110-210 $\mu$m will get confusion limited very soon (estimated S$_{conf}$(5$\sigma$)$\sim$13 mJy), therefore this band is more suited for large and moderately shallow surveys to detect rare and luminous objects.

\begin{figure}
\begin{center}
\includegraphics[width=8cm]{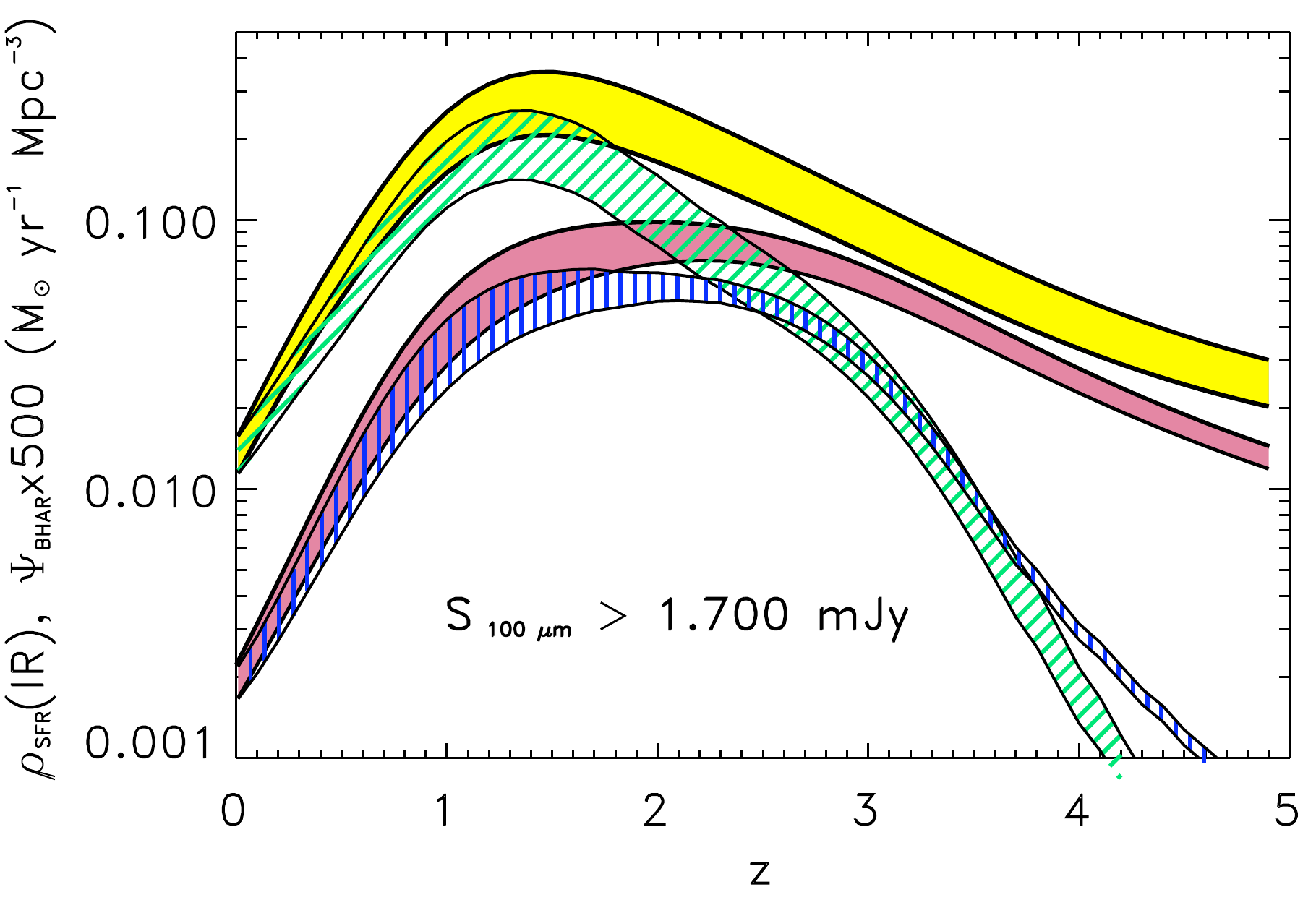}
\includegraphics[width=8cm]{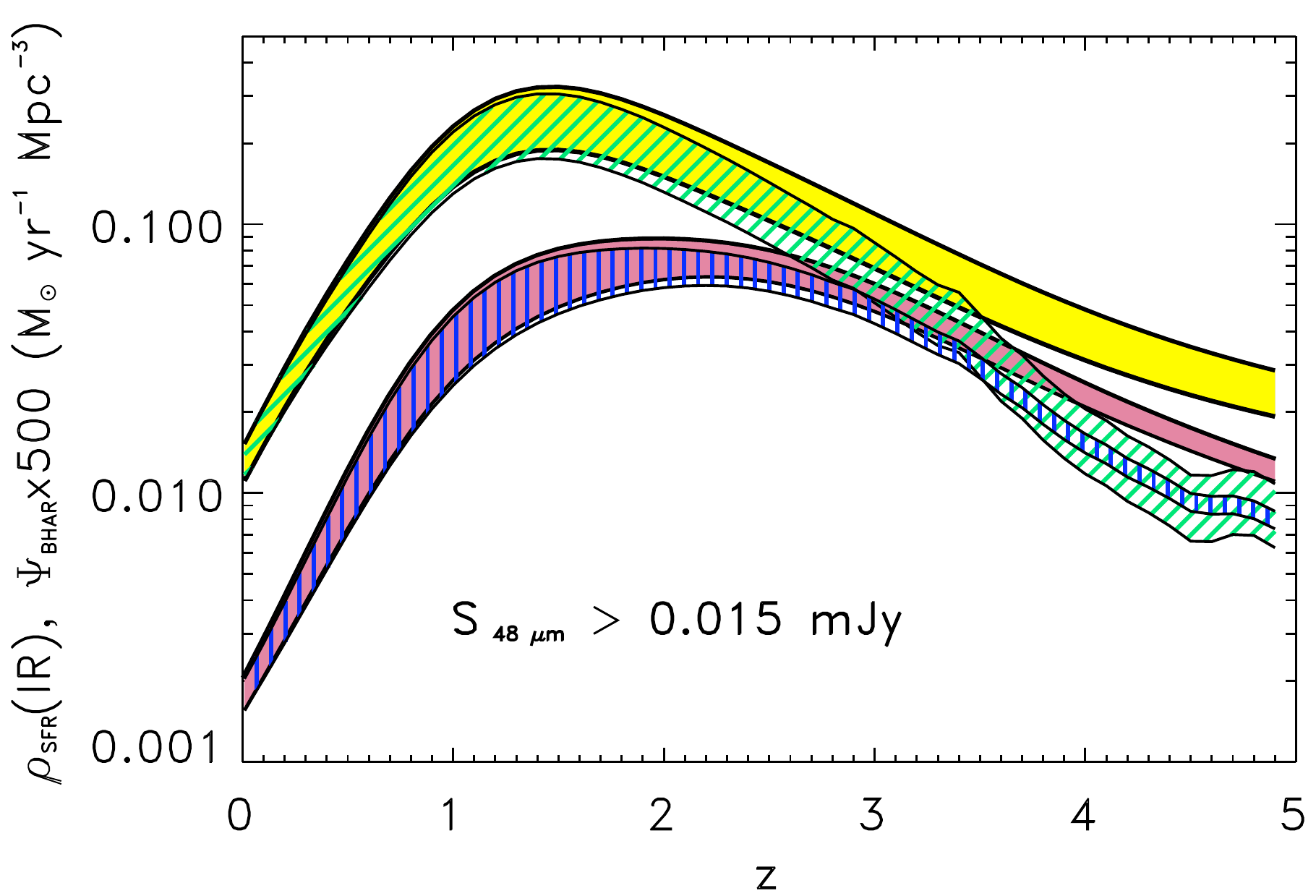}
\end{center}
\caption{BHAR density, $\Psi_{BHAR}$ (blue vertical-dashed area), and SFD, $\rho_{SFR}$ (green diagonal-dashed area), as predicted by our model for the deepest PEP {\em Herschel} Survey at 100 $\mu$m ($top$) and for a confusion limited Survey with {\em SAFARI} at 48 $\mu$m ($bottom$) and compared to the "total" model expectations reported in Fig.~\ref{authorf_fig:sfd_bhar}}
\label{authorf_fig:sfh_lim}
\end{figure}
From Fig.~\ref{authorf_fig:cntmirisaf} we can notice how the dominating populations are expected to change with increasing wavelength, with the redshift distribution at 10 $\mu$m being largely dominated by spiral galaxies and that at 85 $\mu$m being dominated by Seyfert 2-SED objects, with a major contribution of spiral galaxies at lower redshifts ($z$$\leq$0.5).  At 10 $\mu$m we expect a significant fraction of spiral galaxies also at high-$z$ (between $z$$\sim$2 and 4), not present in the other bands.

By considering the limiting fluxes discussed above, we have computed the limiting luminosities for the prospected Surveys with {\em MIRI} and {\em SAFARI} and estimated the evolution of the SFD and BHAR that we expect to obtain with real data from these surveys (using the same procedure described in Section~\ref{authorf_sec:sfh}, but integrating the total IR LF only down to the limiting luminosities). In Figure~\ref{authorf_fig:sfh_lim}, we show the SFD and BHAR (multiplied by a factor of 500) that we expect to obtain with data from the confusion limited Survey with {\em SAFARI} at 48 $\mu$m, compared to the predicted SFD and BHAR from our model, corresponding to an "ideal" FIR survey covering all the luminosities at all redshifts (the same curves shown in Fig.~\ref{authorf_fig:sfd_bhar}). For comparison, in the $upper$ panel of Figure~\ref{authorf_fig:sfh_lim}, we show the same simulation with the deepest PEP Survey in the GOODS-S, reaching 1.7 mJy at 100 $\mu$m (the GOODS-Herschel Survey will reach $\sim$0.6 mJy, but just in a very small area of 50 arcmin$^2$). In agreement with the results of Gruppioni et al. (2010; see also Fig.~\ref{authorf_fig:sfdz}) in the GOODS-N, our predictions show that PEP is complete in SFD up to $z$$\sim$1--1.5, becoming more and more incomplete with increasing redshift (i.e. at $z$=3 the incompleteness in SFD due to the survey flux limit is about a factor of 3 or more). The BHAR that we could measure with PEP is complete up to $z$$\sim$1.5--2, then decreasing less rapidly than the SFD up to $z$=3 and dropping down at higher $z$'s. With the {\em SAFARI} Survey we expect to be able to measure almost all the SFD to $z$$\sim$2 and most of it to $z$$\sim$3--3.5, and almost all the BHAR to $z$$\sim$3. Moreover, as mentioned in the previous Section, the high resolution spectrometer of {\em SAFARI} will be crucial in identifying AGNs and separating the SB from the AGN contribution, to measure with great precision the SFD and BHAR in the high redshift Universe.

\section{Conclusions}
\label{authorf_sec:concl}
We have presented a new model for galaxy and AGN evolution in the IR, in which the evolutionary properties of the different IR populations, defined by means of a detailed SED-fitting analysis (see Gruppioni et al. 2008; 2010), are separately studied and constrained by all the available results from MIR and FIR surveys. The model identifies five main SED classes: three containing an AGN at different levels of dominance ({\tt AGN1} where the unobscured AGN dominates up to the MIR domain, {\tt LLAGN} with a galaxy SED but showing a flattening at MIR wavelengths explainable only by the presence of a dusty torus, and {\tt AGN2} where an obscured AGN and starburst co-exhist) and two powered by star-formation only (normal {\tt spiral} and moderate-to-extreme {\tt starburst} galaxies).

The main results of this work are summarised as follows:
\begin{itemize}
\item[$\bullet$] The {\tt spiral} galaxies show low evolutionary rates, both in luminosity and density, increasing up to $z$$\sim$0.3 ($\propto$$(1+z)^{1.5}$ in luminosity and $\propto$$(1+z)^{0.8}$ in density), then slowly decreasing towards the higher $z$'s. {\tt Starburst} galaxies evolve fast ($\propto$$(1+z)^{3.5}$ in luminosity and $\propto$$(1+z)^{2.3}$ in density) up to $z$$\sim$1, with the luminosity and density remaining approximatively constant between $z$$=$1 and $z$$=$2, then decreasing at higher redshifts. {\tt LLAGN} show luminosity evolution (L($z$)$\propto$$(1+z)^{3.7}$) similar to, and density evolution $(\rho$($z$)$\propto$$(1+z)^{2.8}$) higher than, those of starburst galaxies, but with a more pronounced peak at $z$$\simeq$1.2$\div$1.4, followed by a faster decrease 
at $z$$>$1.5. The {\tt AGN1} luminosity evolves as $\propto$$(1+z)^{3.3}$ up to $z$$\sim$1.5, then remains almost constant between $z$$\simeq$1.5 and $z$$\simeq$2.5, while the evolution of the {\tt AGN2} objects shows a flattening at even higher redshift (i.e. between $z$$=$2 and $z$$=$3), increasing towards the peak at a rate of $\approx$$(1+z)^{2.7}$ in luminosity and $\approx$$(1+z)^{2.2}$ in density and decreasing faster at $z$$>$3. With our backward evolution model we are able to reproduce well the source counts, redshift distributions and luminosity functions in the MIR and FIR bands, provided by all the main IR space observatories (i.e. {\em ISO}, {\em Spitzer} and {\em Herschel}). 
\item[$\bullet$] We have decomposed the template SEDs representative of the populations containing an AGN into three distinct components: a stellar component emitting most of its power
in the optical/NIR, an AGN component due to hot dust heated by the central
black hole, and peaking in the MIR, and a SB component representing the
major contribution to the FIR spectrum. In this way, we have estimated -- although in a very simplified way -- the AGN contribution to the
monochromatic and total IR luminosity emitted by the different populations
considered in our model. 
\item[$\bullet$] Using the LFs given by our model and the total IR luminosity due to SF derived from our template SED decomposition, we have estimated the cosmic evolution of the 
total IR luminosity density due to star-formation, $\rho_{IR}^{SF}(z)$, as a function of redshift. Our model expectation is in very good agreement with all the IR data estimates, confirming the rapid increase of $\rho_{IR}^{SF}$ up to $z$$\sim$1. The IR luminosity density shows a peak at 1$<$$z$$<$2 and a decrease from $z$$\sim$2--2.5 towards the higher redshifts. At $z$$<$0.3 $\rho_{IR}^{SF}$ is dominated by the {\tt spiral} population, while in the 0.3$\leq$$z$$\lsimeq$2--2.5 range the principal contributors are galaxies with a {\tt LLAGN} SED. {\tt Starburst} galaxies are also important in the same redshift interval, but are never dominant at any $z$. The {\tt AGN2} SED objects start dominating at $z$$>$2.5 up to $z$$\sim$4, when they are overtaken by the {\tt AGN1} population. Star-forming galaxies containing or not a low-luminosity AGN (with LIRG luminosities) are therefore responsible for the peak of the IR luminosity density at $z$$\sim$1--2, then galaxies hosting increasingly powerful AGNs (in the ULIRG luminosity range) become increasingly important towards the higher $z$'s. 
\item[$\bullet$] For the first time, the SMBH Accretion Density $\Psi_{BHAR}(z)$ as a function of redshift has been derived from IR rather than X-ray data. The BHAR obtained from our IR estimate is
reasonably consistent in shape with previous derivations from X-rays (i.e. Merloni, Rudnick \& Di Matteo, 2004; Merloni \& Heinz 2008), though it is slightly higher at 1$\lsimeq$$z$$\lsimeq$3, where the BHAR peaks. The Compton-thick AGNs are already included in our calculations, while substantial assumptions need to be made regarding their number and redshift distribution in the X-rays. The {\tt AGN1} contribution dominates the BHAR, especially at low ($z$$<$0.3) and high redshifts ($z$$>$3), with the {\tt LLAGN} and the {\tt AGN2} population's accretion density peaking at 1$<$$z$$<$2 and 2$<$$z$$<$3 respectively and reaching the {\tt AGN1} accretion values just in these redshift ranges.  
\item[$\bullet$] We have simulated source counts, redshift distributions and SFD and BHAR that we expect to obtain with the future cosmological Surveys in the MIR/FIR that will be performed with {\em JWST-MIRI} and {\em SPICA-SAFARI}.
\end{itemize}

\section*{Acknowledgments}

The authors acknowledge financial contribution from the contracts PRIN-INAF 1.06.09.05 and ASI-INAF I/057/08/0. We thank an anonymous referee for helpful comments 
and J. Fritz and A. Feltre for useful suggestions about SED decomposition.

\label{lastpage}

\end{document}